\documentclass[longauth]{aa}  

\usepackage{graphicx}
\usepackage{txfonts}
\usepackage{color}
\usepackage{array,multirow}
\usepackage{tabularx}
\usepackage{booktabs}
\usepackage[table]{xcolor}
\usepackage{booktabs}

\begin{document}


\title {The Galaxy Activity, Torus and Outflow Survey (GATOS):}
   \subtitle{II. Torus and polar dust emission in nearby Seyfert galaxies }

\author{A. Alonso-Herrero\inst{\ref{inst1}}
         \and
         S. Garc\'{\i}a-Burillo\inst{\ref{inst2}}
          \and
          S. F. H\"onig\inst{\ref{inst3}}
          \and
          I. Garc\'{\i}a-Bernete\inst{\ref{inst4}}
          \and
          C. Ramos Almeida\inst{\ref{inst5}, \ref{inst6}}
          \and
          O. Gonz\'alez-Mart\'{\i}n\inst{\ref{inst7}}
          \and
          E. L\'opez-Rodr\'{\i}guez\inst{\ref{inst8}}
          \and
          P. G. Boorman\inst{\ref{inst9}}
          \and
          A. J. Bunker\inst{\ref{inst4}}
          \and
          L. Burtscher\inst{\ref{inst10}}
          \and
          F. Combes\inst{\ref{inst11}}
          \and
          R. Davies\inst{\ref{inst12}}
          \and
          T. D\'{\i}az-Santos\inst{\ref{inst13}}
          \and
          P. Gandhi\inst{\ref{inst3}}
          \and
          B. Garc\'{\i}a-Lorenzo\inst{\ref{inst5}, \ref{inst6}}
          \and
          E. K. S. Hicks\inst{\ref{inst14}}
          \and
          L. K. Hunt\inst{\ref{inst15}}
          \and
          K. Ichikawa\inst{\ref{inst12}, \ref{inst16}, \ref{inst17} }
          \and
          M. Imanishi\inst{\ref{inst18}, \ref{inst19}, \ref{inst20}}
          \and
          T. Izumi\inst{\ref{inst21}}
          \and
          A. Labiano\inst{\ref{inst1}}
          \and
          N. A. Levenson\inst{\ref{inst23}}
          \and
          C. Packham\inst{\ref{inst24}, \ref{inst18}}
          \and
          M. Pereira-Santaella\inst{\ref{inst22}}
          \and
          C. Ricci\inst{\ref{inst25}, \ref{inst26}}
          \and
          D. Rigopoulou\inst{\ref{inst4}}
          \and
          P. Roche\inst{\ref{inst4}}
          \and
          D. J. Rosario\inst{\ref{inst27}}
          \and
          D. Rouan\inst{\ref{inst28}}
          \and
          T. Shimizu\inst{\ref{inst12}}
          \and
          M. Stalevski\inst{\ref{inst29}, \ref{inst30}}
          \and
          K. Wada\inst{\ref{inst31}, \ref{inst32}, \ref{inst33}}
          \and
          D. Williamson\inst{\ref{inst3}}
        }

\institute{Centro de Astrobiolog\'{\i}a (CAB, CSIC-INTA), ESAC
     Campus, E-28692 Villanueva de la Ca\~nada, Madrid,
     Spain\\
     \email{aalonso@cab.inta-csic.es}\label{inst1}
   \and
      Observatorio de Madrid, OAN-IGN, Alfonso XII, 3, E-28014 Madrid, Spain\label{inst2}
   \and
   Department of Physics \& Astronomy, University of Southampton, Hampshire SO17 1BJ, Southampton, UK\label{inst3}
  \and
   Department of Physics, University of Oxford, Oxford OX1 3RH, UK\label{inst4}
   \and
   Instituto de Astrof\'{\i}sica de Canarias, Calle v\'{\i}a L\'actea, s/n, 38205 La Laguna, Tenerife, Spain\label{inst5}
 \and
   Departamento de Astrof\'{\i}sica, Universidad de La Laguna, 38205 La Laguna, Tenerife, Spain\label{inst6}
   \and
   Instituto de Radioastronom\'{\i}a y Astrof\'{\i}sica (IRyA-UNAM), 3-72 (Xangari), 8701, Morelia, Mexico\label{inst7}
   \and
   Kavli Institute for Particle Astrophysics and Cosmology (KIPAC),
   Stanford University, Stanford, CA 94305, USA \label{inst8}
   \and
   Astronomical Institute, Academy of Sciences, B\^ocn\'{\i} II 1401, CZ-
   14131 Prague, Czechia\label{inst9}
   \and
   Leiden Observatory, PO Box 9513, 2300 RA Leiden, The
   Netherlands\label{inst10}
   \and
   LERMA, Observatoire de Paris, Coll\`ege de France, PSL University,
   CNRS, Sorbonne University, Paris, France\label{inst11}
   \and
   Max-Planck-Institut f\"ur Extraterrestrische Physik,
   Postfach 1312, D-85741 Garching, Germany\label{inst12}
   \and
   Institute of Astrophysics, Foundation for Research and Technology-
   Hellas, GR-71110, Heraklion, Greece\label{inst13}
   \and
   Department of Physics \& Astronomy, University of Alaska Anchorage,
   Anchorage, AK 99508-4664, USA\label{inst14}
   \and
   INAF - Osservatorio Astrofisico di Arcetri, Largo Enrico Fermi 5,
   50125 Firenze, Italy\label{inst15}
   \and
   Frontier Research Institute for Interdisciplinary Sciences, Tohoku
   University, Sendai 980-8578, Japan\label{inst16}
   \and
   Astronomical Institute, Tohoku University 6-3 Aramaki, Aoba-ku,
   Sendai, 980-8578 Japan\label{inst17}
   \and
   National Astronomical Observatory of Japan, National Institutes of
Natural Sciences (NINS), 2-21-1 Osawa, Mitaka, Tokyo 181-8588,
Japan\label{inst18}
\and
Department of Astronomy, School of Science, The Graduate University
for Advanced Studies, SOKENDAI, Mitaka, Tokyo 181-8588,
Japan\label{inst19}
\and
National Astronomical Observatory of Japan, 2-21-1 Osawa, Mitaka,
Tokyo 181-8588, Japan\label{inst20}
\and
Department of Astronomical Science, The Graduate University for
Advanced Studies, SOKENDAI, 2-21-1 Osawa, Mitaka, Tokyo 181-
8588, Japan\label{inst21}
\and
Centro de Astrobiolog\'{\i}a (CAB, CSIC-INTA), Carretera de Ajalvir, 28850
  Torrej\'on de Ardoz, Madrid, Spain\label{inst22}
  \and
  Space Telescope Science Institute, Baltimore, MD 21218,
  USA\label{inst23}
  \and
  The University of Texas at San Antonio, One UTSA Circle, San
  Antonio, TX 78249, USA\label{inst24}
  \and
  N\'ucleo de Astronom\'{\i}a de la Facultad de Ingenier\'{\i}a, Universidad
Diego Portales, Av. Ej\'ercito Libertador 441, Santiago,
Chile\label{inst25}
\and
Kavli Institute for Astronomy and Astrophysics, Peking University,
Beijing 100871, China\label{inst26}
\and
Centre for Extragalactic Astronomy, Department of Physics,
Durham University, South Road, Durham DH1 3LE, UK\label{inst27}
\and
LESIA, Observatoire de Paris, PSL Research University, CNRS,
Sorbonne Universit\'es, UPMC Univ. Paris 06, Univ. Paris Diderot,
Sorbonne Paris Cit\'e, 5 place Jules Janssen, 92190 Meudon, France\label{inst28}
\and
Astronomical Observatory, Volgina 7, 11060 Belgrade,
Serbia\label{inst29}
\and
Sterrenkundig Observatorium, Universiteit Ghent, Krijgslaan 281-S9, Ghent, B-9000, Belgium\label{inst30}
\and
Kagoshima University, Graduate School of Science and Engineering,
Kagoshima 890-0065, Japan\label{inst31}
\and
Ehime University, Research Center for Space and Cosmic Evolution,
Matsuyama 790-8577, Japan\label{inst32}
Hokkaido University, Faculty of Science, Sapporo 060-0810, Japan\label{inst33}}
   \date{Received:  --; accepted ---}

 
  \abstract
  {We compare high angular resolution mid-infrared (mid-IR) and
    ALMA  far-infrared  (far-IR) images of twelve
    nearby (median 21\,Mpc) Seyfert galaxies selected from the Galaxy Activity Torus
    and Outflow Survey (GATOS). The mid-IR  unresolved emission contributes
more than 60\% of the nuclear (diameters of 1.5\arcsec$\sim$150\,pc) emission in most galaxies. By
    contrast,  the ALMA $870\,\mu$m continuum emission is mostly
    resolved with a median diameter of 42\,pc and typically along the equatorial
    direction of the torus \citep[Paper I of the series][]{GarciaBurillo2021}.  The Eddington
ratios and nuclear hydrogen column densities ($N_{\rm H}$) of half the sample are
 favorable to launching polar and/or equatorial dusty winds, according
 to numerical simulations. Six of
these show  mid-IR extended emission approximately in the polar direction as traced
by the narrow line region and 
perpendicular to the ALMA emission.  In a few galaxies, the nuclear $N_{\rm H}$
might be too high to uplift large quantities of dusty material
along the polar direction. Five galaxies have low $N_{\rm H}$
and/or Eddington ratios and thus 
polar dusty winds are not likely. We generate new radiative transfer CAT3D-WIND disk-wind
models and model images at 8, 12, and $700\,\mu$m. We  tailor these
models to the properties of the 
GATOS Seyferts in this work.
At low wind-to-disk cloud ratios the far-IR model images have disk-
and ring-like morphologies. The characteristic
“X”-shape associated with dusty winds is seen better in the far-IR at
intermediate-high inclinations for the extended-wind configurations. In most of the
explored models, the mid-IR emission comes mainly from the inner part of the disk/cone.
Extended bi-conical and one-sided polar mid-IR emission is seen in extended-wind configurations and high
wind-to-disk cloud ratios. When convolved to the typical angular resolution
of our observations, the CAT3D-WIND model images reproduce qualitative
aspects of  the observed mid-
and far-IR morphologies. However, low to intermediate values of the wind-to-disk ratio
are required 
to account for the observed large fractions
of unresolved mid-IR emission in our sample. This work and Paper I
provide observational support for 
the torus+wind scenario. The wind
component is more relevant at high Eddington ratios and/or AGN
luminosities, and polar dust emission is predicted at  nuclear
column densities of up to  $\sim 10^{24}\,{\rm
  cm}^{-2}$. The torus/disk component, on the other hand,  prevails at low luminosities
and/or Eddington ratios.

}  


   \maketitle

%


 \section{Introduction}

 The fundamental component of the Unified
Model for Active Galactic Nuclei (AGN) is an obscuring torus or
disk\footnote{Throughout this work we will use the terms torus and
 nuclear disk interchangeably. In particular, unless otherwise indicated, the term torus does not
  necessarily refer to a geometrically thick one, which is defined as
  having 
  a height over radial size ratio of $H/R=1$.}
made of dust and molecular gas \citep[see][for 
reviews]{Antonucci1993, UrryPadovani1995, Netzer2015}. In the
classical scenario, the torus obscures the view of the broad line region (BLR) along certain lines of
sight and the nuclei are classified as type 2. Those nuclei observed along or
near the
polar direction of the torus have a direct view of
the BLR and are classified as type 1. Initially, \cite{PierKrolik1993}
derived a compact size (a few parsecs) for 
the torus of the archetypical Seyfert 2 galaxy
NGC~1068,  from the fit of the infrared (IR) spectral
energy distribution (SED) with the torus models of 
\cite{PierKrolik1992}. Subsequent modelling of a sample of Seyfert 1s
by \cite{Granato1994} however,
required tori extending for up to a few hundred parsecs.
The narrow line region (NLR) of
Seyfert galaxies extends
on much larger scales (hundreds of parsecs up to a $\sim$kpc) than the dusty molecular torus and is thus seen
in both type 1 and type 2 AGN.

\begin{table*}
\caption{The sample.}             
\label{tab:Sample}      
\centering                          
\begin{tabular}{c c c c c c c c}        
\hline\hline                 
Galaxy      &  Dist    & Type   & log $L$(2-10keV) & X-ray log $N_{\rm H}$ &
                                                                   $\log
                                                                   \lambda_{\rm
  Edd}$ & ALMA log $N_{\rm H2}$ \\
            &   (Mpc)&  & (erg s$^{-1}$) & (cm$^{-2}$) & & (cm$^{-2}$)
\\
\hline
NGC~1365    & 18.3 & Sy1.8 & 42.09 & 22.2 & $-1.6$ & 22.3  \\
NGC~3227    &23.0  & Sy1.5 & 42.37 & 21.0 & $-1.2$ & 22.7  \\
 NGC~4388    &18.1  & Sy1.9 & 42.45 & 23.5 & $-1.2$ &22.3  \\
NGC~4941    & 20.5 & Sy2 &   41.40  & 23.7 & $-2.4$ & 21.9 \\
NGC~5506    & 26.4 & Sy1.9 & 42.98 & 22.4 & $-2.3$ & 22.6 \\
NGC~5643    & 16.9 & Sy2    & 42.41 & 25.4 & $-1.3$ & 23.6  \\
NGC~6300    & 14.0 & Sy2    & 41.73 & 23.5 & $-1.9$ & 23.4  \\
NGC~6814    & 22.8 & Sy1.5 & 42.24 & 21.0 & $-1.6$ & $\le$21.8 \\
NGC~7213    & 22.0 & Sy1.5 & 41.85 & 20.0 & $-3.0$ & $\le$22.0 \\
NGC~7314    & 17.4 & Sy1.9 & 42.18 & 21.6 & $-1.2$ & 22.1  \\
NGC~7465    & 27.2 & Sy2    & 41.93 & 21.5 & $-2.2$ & 22.7 \\
NGC~7582    & 22.5 & Sy2    & 43.49 & 24.3 & $-1.7$ & 22.6 \\
\hline
\end{tabular}
\tablefoot{The distances are 
  from the NASA/IPAC Extragalactic Database (NED) for $H_0 = 73\,{\rm
    km\,s}^{-1}\,{\rm Mpc}^{-1}$, $\Omega_m=0.27$, and $\Omega_\Lambda=0.73$. The Seyfert types
  are taken from \cite{VeronCetty2006}, except for NGC 7213, which is taken from
\cite{Phillips1979}. The absorption
  corrected $2-10\,$keV luminosities and X-ray column densities are
  from \cite{Ricci2017}. The Eddington ratios ($\lambda_{\rm Edd}$)
  are from \cite{Koss2017} except for NGC~1365       that is
    from \cite{Vasudevan2010}.
    The H$_2$  column densities
    are  based on ALMA CO(3-2)  estimates  at
    the AGN position from   GB21.
}
\end{table*}

The angular resolutions needed  in the IR to resolve the obscuring structures of
nearby AGN  have not been available until recently. 
In the mid-infrared\footnote{Throughout this work, mid-IR
  refers to the $\sim 7-26\,\mu$m spectral range that can be
  observed from the ground.} (mid-IR), interferometric observations with the
Very Large Telescopes Interferometer (VLTI) of nearby
Seyferts are generally modeled with an unresolved source and a resolved
source. Both show compact sizes
\citep[$\sim 1-10\,$pc, ][]{Burtscher2013, LopezGonzaga2016}. Some of
the resolved model components are elongated in the polar direction, with
this component accounting
for most of the mid-IR emission on these
scales \citep{Hoenig2013, Tristram2014, LopezGonzaga2014,
  LopezGonzaga2016, Leftley2019}. This polar dust
emission appears to be related
to the large scale (up to a few hundred parsec) emission
detected in the mid-IR \citep{Cameron1993, Tomono2001, Radomski2003, Packham2005,
  Asmus2014, Asmus2019, GarciaBernete2016} and with SOFIA at $30\,\mu$m
\citep[see][]{Fuller2019}.
In many local Seyferts that is spatially coincident with that of the
NLR and/or ionization cones. In the near-infrared
(near-IR) \cite{Pfuhl2020}
reconstructed VLT/GRAVITY  $K$-band
observations of NGC~1068  with a ring-like structure with
a radius of 0.24\,pc. This emission is believed to be associated with the dust
sublimation region but, according to these authors, the geometry is not
consistent with that expected from a geometrically thick
torus. Furthermore, for a sample of Seyfert 1s, 
\cite{Dexter2020} resolved the radii of the hot
dust continuum emission and showed 
they follow the  luminosity-size relation \citep{Suganuma2006, Kishimoto2007}, as expected for
the dust sublimation region.

Observations with  the Atacama Large Millimeter/submillimeter
Array (ALMA)  of NGC~1068 resolved the far-infrared
(far-IR) or sub-millimeter continuum  emission of the torus \citep{GarciaBurillo2016, GarciaBurillo2019, 
  LopezRodriguez2020}. At $432\,\mu$m the torus diameter is
7-10\,pc. The derived (sub)millimeter spectral indices at the AGN
position of NGC~1068
indicate the presence of cold dust but with an important contribution
from synchrotron emission at these wavelengths
\citep{GarciaBurillo2019, Pasetto2019}. The torus was also detected in a variety  
of molecular gas transitions that probe a range of gas
densities with the torus diameter reaching $\sim 30\,$pc in the low
density tracers \citep[see][]{GarciaBurillo2016, GarciaBurillo2019,
  Gallimore2016, Impellizzeri2019, Imanishi2020}. 

Molecular tori with
diameters of up to 50\,pc are now routinely observed with ALMA in other nearby Seyfert galaxies
and low luminosity AGN. The tori are sometimes morphologically and kinematically
decoupled from the host galaxy. However, in most cases the tori are
connected to reservoirs of
molecular gas on scales of $\sim 100\,$pc that are associated with dynamical resonances
\citep[see][and also below]{Izumi2018, AlonsoHerrero2018, AlonsoHerrero2019,
  Combes2019}. These reservoirs  are likely related to  the 100\,pc-torus  invoked by 
\cite{MaiolinoRieke1995}  to explain the properties of 1.8 and 1.9
Seyfert nuclei. These would be type 1s seen through these large scale
dust structures that are coplanar with the host galaxy disk. Finally there is evidence that in some cases the torus itself is not only
 rotating but also outflowing
 \citep{Gallimore2016, AlonsoHerrero2018, GarciaBurillo2019}. 

In  parallel with the new observational constraints on the torus properties,
theoretical models are continuously evolving. In the first static
models  the dust was homogeneously 
distributed \citep{PierKrolik1992, Granato1994, Efstathiou1995,
  Fritz2006}. Subsequently, the dust was distributed in clouds in the
so-called clumpy torus models 
\citep{Nenkova2008I, Nenkova2008II, Schartmann2008,
  Hoenig2010model} and in two phases \citep{Stalevski2012,
  Siebenmorgen2015}. Torus models  also incorporated an additional
polar dust  component to 
account for the mid-IR imaging and interferometric
observations of some Seyfert galaxies \citep{Efstathiou1995b, Gallagher2015, Hoenig2017,
  Stalevski2017, Isbell2021}. These models reproduced satisfactorily the
observed nuclear IR emission of samples of nearby AGN \citep[see
e.g.,][]{RamosAlmeida2009, RamosAlmeida2011, AlonsoHerrero2011,
  Ichikawa2015, GarciaGonzalez2017,
  GarciaBernete2019,GonzalezMartin2019}. \cite{Elitzur2006}
put forward a scenario where the torus is part of a clumpy outflow and 
recently,
\cite{Venanzi2020} demonstrated theoretically that dusty  
winds can be launched at the inner walls of the torus.
Radiation hydrodynamical
models \citep{Schartmann2014, Wada2016, Williamson2019, Williamson2020} incorporated
predictions for the dust IR emission and the
molecular gas emission and kinematics. 

\cite{Hoenig2019} assembled the information gathered from the analysis
of IR and sub-mm data of nearby and bright AGN and
proposed a new paradigm for the obscuring structures
around radio-quiet
AGN \citep[see also][]{RamosAlmeida2017, Lyu2021}. The {\it torus} is now
envisioned as a multi-component multi-phase structure \citep[see
figure~1 and figure~ 4 of][for schematic pictures,
respectively]{Izumi2018, Hoenig2019}. In short, the hot innermost part of
the equatorial disk/torus is close to the sublimation radius  on
sub-pc scales in Seyferts and emits mostly in the near-IR.
The dusty
inner molecular torus as well as the wind region are traced by the
mid-IR  emission as well as hot and relatively warm molecular gas on scales
of a few pc to probably tens of parsecs. Both dust components (that
is, the inner molecular disk and the wind) are
likely to contribute to the AGN obscuration. 
Finally, the cold outer part of the equatorial disk is probed by the
cold molecular gas and dust
emission and extends on scales from
5\,pc out to tens of parsecs.

This is the second paper in a series aimed at understanding the nuclear
activity and its connection with  the host galaxy in nearby Seyfert
galaxies. In the first paper of the series, \cite{GarciaBurillo2021},
GB21 from now on,
obtained ALMA observations of a volume-limited and complete sample of  Seyfert galaxies
to study their torus properties. The galaxies are part of 
the Galactic Activity, Torus, and Outflow Survey (GATOS). We
 drew the GATOS
galaxies from the 70 Month Swift-BAT All-sky Hard X-Ray Survey
\citep{Baumgartner2013}. The Swift-BAT 14-195 keV  energy range ensures a
nearly complete
selection for nearby AGN
at $L_{\rm AGN} (14-150\,{\rm  keV})> 10^{42}\,{\rm erg
  \,s}^{-1}$ \citep{Ricci2017, Koss2017}. GB21 selected galaxies in
the southern hemisphere and with distances in 
the $10-40\,$Mpc range. We summarize the main
properties of the sample of 12 Seyfert galaxies in
Table~\ref{tab:Sample}.  The median galaxy distance is  21\,Mpc and
the median value of the intrinsic (absorption corrected)
2-10\,keV luminosity  is $2\times 10^{42}\,{\rm erg \, s}^{-1}$.
The Eddington ratios ($\lambda_{\rm Edd}$) vary from 0.001 to 0.06.
The sample probes a range of X-ray column densities and includes two Compton-thick objects. 

In this paper we characterize
for the first time the  
torus and polar dust emission components  in  nearby Seyferts,
using mid and far-IR observations with physical resolutions 
$7-50$\,pc.
The paper is organized as follows. We describe the existing
mid-IR observations with angular resolutions $0.2-0.4\arcsec$ in
Sect.~\ref{sec:midIRobservations}. We summarize in
Sect.~\ref{sec:ALMAobservations} the ALMA observations
analyzed and discussed in detail by GB21.
In Sect.~\ref{sec:PSFsub} we derive the
extended mid-IR emission of our Seyfert galaxies, and  compare it with
ALMA and NLR observations. In  Sect.~\ref{sec:columnEddington} we
investigate the mid-IR morphology dependence on 
AGN properties. In
Sect.~\ref{sec:comparisonmodelCAT3D-WIND}
we generate mid- and far-IR torus model images using the disk-wind
 models of \cite{Hoenig2017}. In
Sect.~\ref{sec:comparisondatamodel}  we use these model images to
simulate  our observations.
Sections~\ref{sec:discussion} and \ref{sec:conclusions} present the
discussion and summary, respectively.

\section{Observations}\label{sec:observations}
The observations used in this work have already been  presented in
the literature. In what follows we describe them briefly.

\subsection{Mid-IR observations}\label{sec:midIRobservations}
We used  fully-reduced mid-IR imaging observations taken
with 8-10\,m class telescopes and already
published in the literature (see references in Table~\ref{tab:midIR}). The instruments included 
VISIR \citep{Lagage2004} and the upgraded VISIR \citep{Kaeufl2015,
  Kerber2016} on the VLT, T-ReCS \citep{Telesco1998} on Gemini-South
(Gemini-S), and
CanariCam \citep{Telesco2003, Packham2005GTC} on
the Gran Telescopio Canarias (GTC). We used observations taken with
various filters in the
atmospheric $N$-band listed in Table~\ref{tab:midIR}.

For each galaxy we also obtained fully-reduced
images of standard stars taken with the same filter and observed
close in time. All the observations have angular resolutions, as
measured from the full width at half maximum (FWHM) of the stars, close to the
theoretical diffraction limit of the corresponding telescopes in the
mid-IR
($0.2-0.4\arcsec$). At the distances of our galaxies, these correspond
to physical resolutions between 17\,pc for NGC~6300 and 50\,pc for NGC~7465.
The pixel sizes are as follows, 0.0453\arcsec \, for the upgraded
VISIR, $0.075\arcsec$ for VISIR, $0.0798\arcsec$ for CanariCam and
$0.09\arcsec$ for T-ReCS.

We note that the observations from the atlas of
\cite{Asmus2014} were already at the usual orientation of north up,
east to the left. The T-ReCS and CanariCam original images had
different orientations and we kept them during the Point Spread
Function (PSF) subtraction
analysis (see Sect.~\ref{sec:PSFsub}), and only rotated them to the
usual orientation after the PSF subtraction.  We refer the reader to
the works listed in the notes of
Table~\ref{tab:midIR}
for full details on the observations and data reduction.

\begin{table*}
\caption{Mid-IR observations and analysis.}             
\label{tab:midIR}      
\centering                          
\begin{tabular}{c c c c c c c c c c c c}        
\hline\hline                 
Galaxy      &  Tel/Inst & Filter & $\lambda_{\rm c}$ &
                                                       \multicolumn{2}{c}{FWHM}
  & Ref &  Unresolv. & PA$_{\rm MIR-ext}$ & Morphology\\
            &   & & ($\mu$m) & (\arcsec) & (pc) & &  (\%) & ($^\circ$) \\
\hline
NGC~1365    & Gemini-S/T-ReCS & Si-2 & 8.74 & 0.34 & 30 &1 & 57& $-65$
                                                                 to
                                                                 $-$70
            & unresolv. + two-sided polar\\
NGC~3227    & GTC/CanariCam   & Si-2 & 8.74 & 0.31 & 35 & 2 & 69  & 30
                                                                    to
                                                                    45
  & unresolv. + one-sided polar\\
NGC~4388    & GTC/CanariCam   & Si-2 & 8.74 & 0.39 & 34 & 2 & 72 & 30

            & unresolv.+ one-sided polar\\
NGC~4941    & VLT/VISIR            & NeII-1 & 12.27 & 0.35 & 35 & 3
            & $\sim$100 &\dots & unresolved\\
NGC~5506    & VLT/VISIR            & NeII-1 & 12.27 & 0.33 &  42 & 3 &
                                                                       61&
                                                                            30
                                                                            to
                                                                            90
            & unresolv. + polar + equatorial/host\\
NGC~5643    & VLT/VISIR(u)            & B12.4  & 12.47   & 0.34 & 28 &
                                                                      4 & 53 & 48 to
                                                                70 & unresolv. + polar + equatorial/host\\
NGC~6300    & VLT/VISIR            & PAH1  & 8.59    & 0.25 & 17 & 3 &
                                                                       36 &$-65$
                                                                to
                                                                $-76$
            & unresolv. + equatorial/host\\
NGC~6814    & VLT/VISIR            & PAH2  & 11.25   & 0.35 & 39 & 3 &
                                                                       78
            & \dots & unresolved\\
NGC~7213    & VLT/VISIR(u)            & PAH1  & 8.59  & 0.26 & 28& 5
            & $\sim$100 & \dots & unresolved\\
NGC~7314    & Gemini-S/T-ReCS & Si-2    & 8.74  & 0.37 & 31& 6 & 67 &
                                                                     \dots  & unresolved\\
NGC~7465    &  GTC/CanariCam   & Si-2 & 8.74 & 0.38 &  50& 2 & 80
                                                        &\dots & unresolved\\
NGC~7582                    & VLT/VISIR(u)            & B12.4  & 12.47   & 0.36
  &  40& 4 & 70 & 45 to 55& unresolv. + one-sided polar\\
  \hline
\end{tabular}

\tablefoot{The FWHM of the observations are measured by fitting a 2D
  Gaussian to the standard
  star images taken close in time to the galaxy observations.  Fully reduced
images are from: 1. \cite{AlonsoHerrero2012}.  2. \cite{AlonsoHerrero2016}.
  3. \cite{Asmus2014}. 4. \cite{Asmus2019}. 5. \cite{Leftley2019}. 6. \cite{GarciaBernete2016}.  The
  unresolved flux fraction is measured within $1.5-1.6\arcsec$ diameter
  apertures. When two
values for PA$_{\rm MIR-ext}$ (see Section~\ref{sec:PSFsub}) are given, the first one corresponds to
the inner regions and the second to the outer regions within
approximately 1$\arcsec$. }
\end{table*}

\subsection{ALMA CO(3-2) and continuum $870\,\mu$m observations}\label{sec:ALMAobservations}
We obtained ALMA band 7 (frequency range $275-373\,$GHz,
wavelength range $0.8-1.1\,$mm)  observations of the GATOS core sample of 10
Seyferts in Cycles
6 and 7.  We targeted the CO(3-2) and HCO$^+$(4-3) transitions as well
as the adjacent continuum at $870\,\mu$m with a common angular
resolution of 0.1\arcsec, which translates into physical resolutions
in the range $7-10\,$pc for our sample. GB21 presented the
observations, data reduction, and analysis. For this work we use the 
fully reduced maps of the $870\,\mu$m continuum and integrated intensity CO(3-2)  (see Sect.~\ref{sec:comparisonALMA}) 
generated from their
moderate spatial resolution  (MSR) data sets as well as the torus
properties derived from the modelling of the MSR $870\,\mu$m continuum
(see Sect.~\ref {sec:comparisonALMA}).  
The ALMA band 7 observations of the other two galaxies in our sample,
namely NGC~1365
and NGC~3227, were part of other ALMA programs and were published by
\cite{Combes2019} and \cite{AlonsoHerrero2019}, respectively.


\section{Extended mid-IR emission}\label{sec:PSFsub}

\subsection{Analysis of the observations}\label{sec:analysis}
The nuclear mid-IR emission of nearby active galaxies, as observed from
ground-based telescopes, is a combination of an
unresolved component, generally assumed to arise from dust heated
by the AGN, and extended emission.
The latter arises from dust in the NLR and/or dust heated by on-going star formation
activity, especially in local Seyferts that are also classified as luminous IR
galaxies \citep[see the review by][and references
therein]{PerezTorres2021},
or might be due to synchrotron emission, especially in radio-loud
low-luminosity AGN \citep{Mason2012}.

\begin{figure*}[!ht]
  \centering
  \includegraphics[width=7cm]{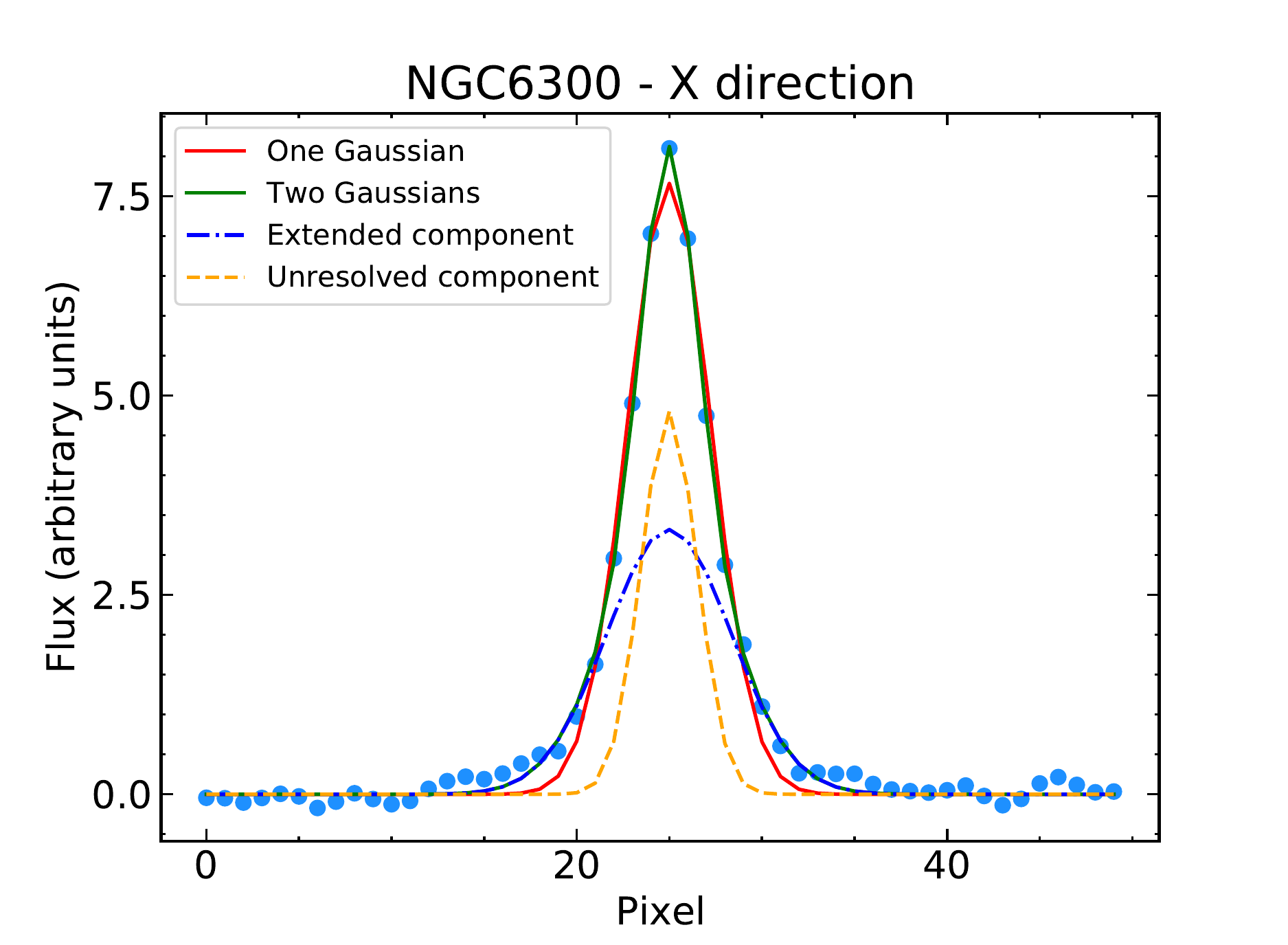}
  \includegraphics[width=7cm]{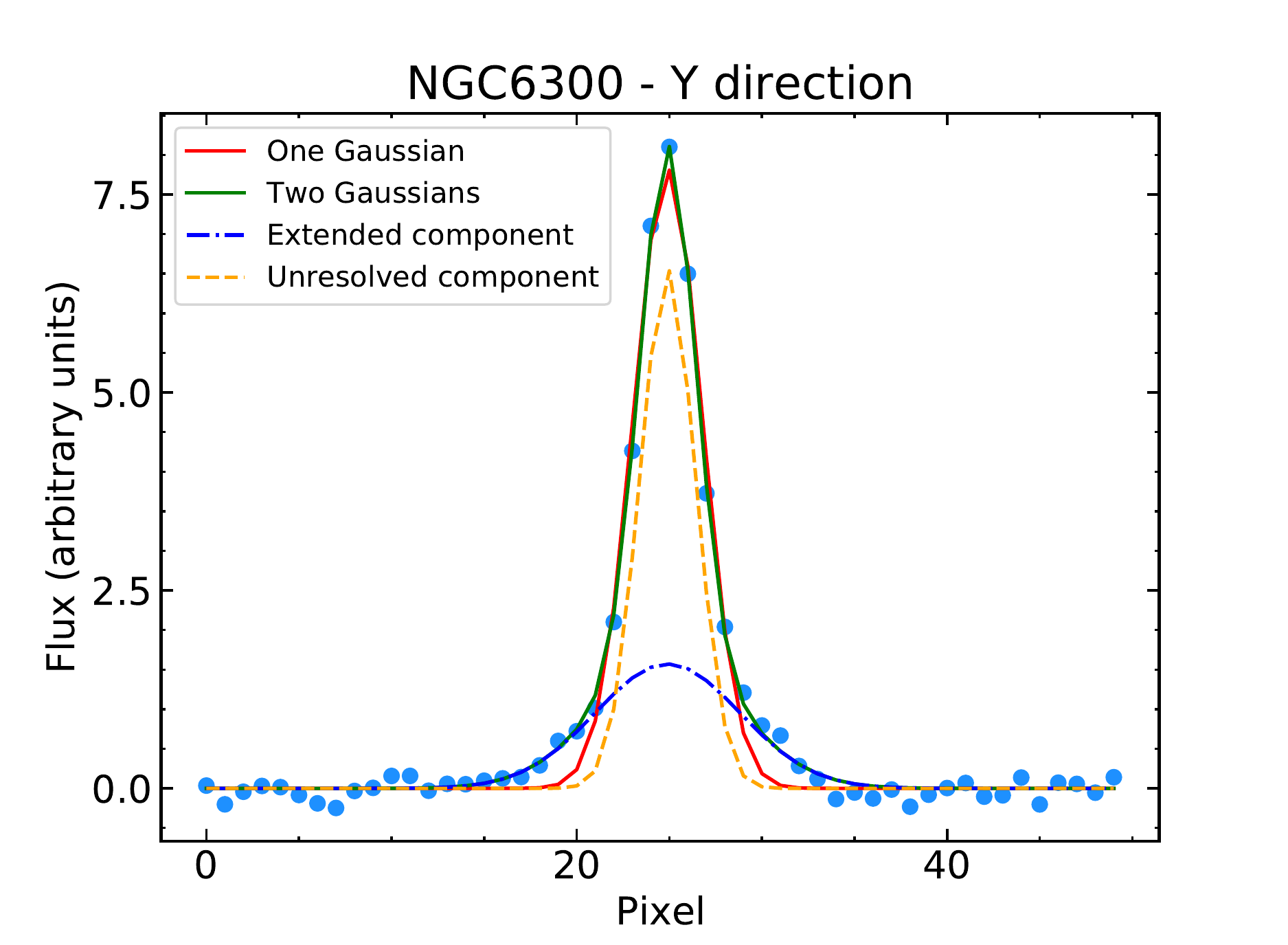}

  \includegraphics[width=7cm]{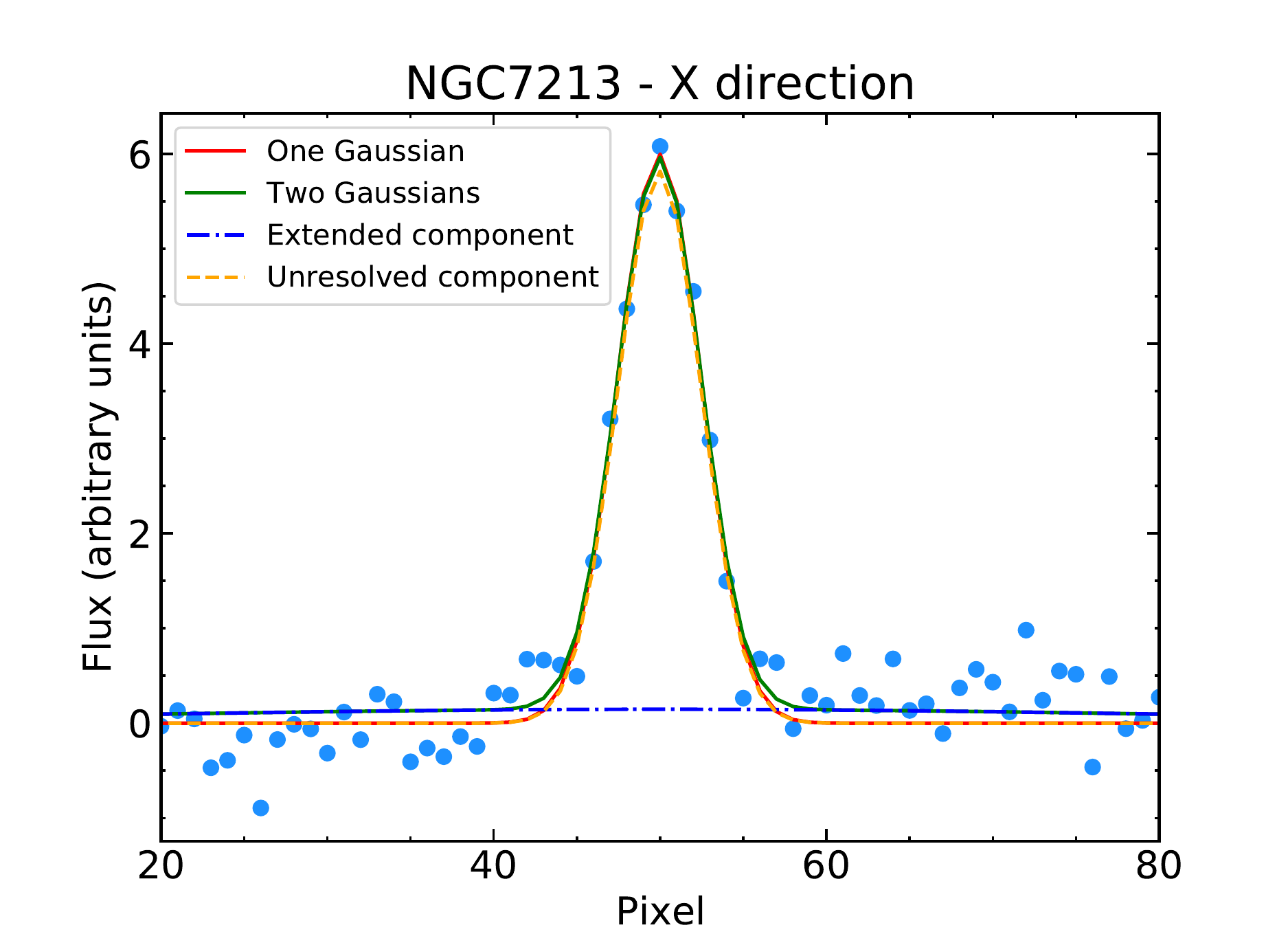}
  \includegraphics[width=7cm]{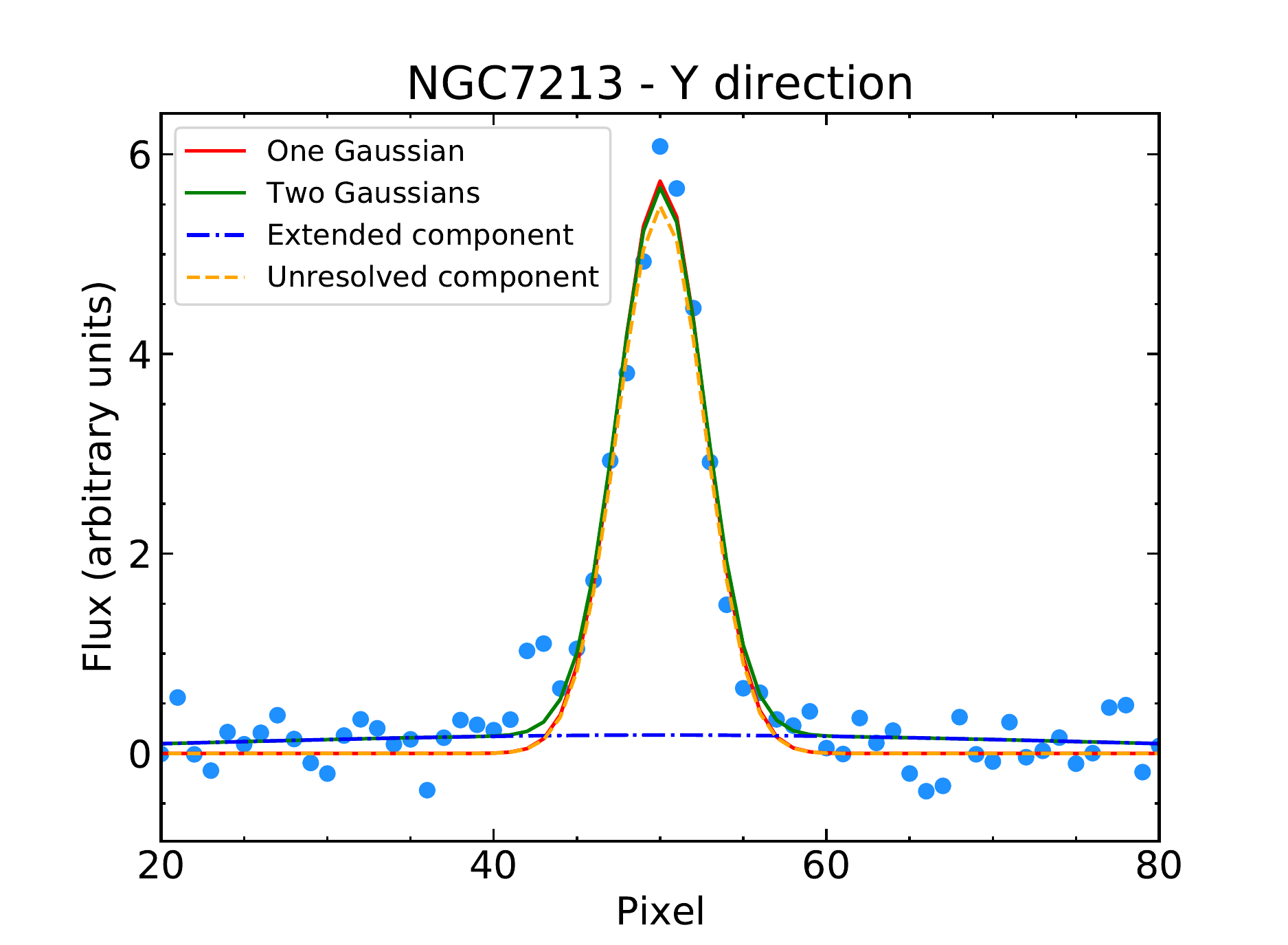}

  \caption{Examples of the fits of the 1D mid-IR profiles extracted
    along the X and Y directions for two galaxies in
    our sample. The top panels are for NGC~6300, which is a case with a large
    fraction of extended mid-IR emission, whereas the bottom panels
    are for NGC~7213, which appears unresolved at the angular
    resolution of the mid-IR observations. The cyan dots are the
    observations and the red lines are the
    result of fitting the galaxy profile with just one Gaussian
    function. The green lines are the result of fitting the galaxy
    profile with the sum of two Gaussian functions, one to represent
    the unresolved component (orange dashed lines) and the second the
    extended emission (blue dotted-dashed lines).}
              \label{fig:1Dprofiles}%
    \end{figure*}

To
separate  the unresolved and extended  emission in ground-based mid-IR
images, 
the majority of works  used the so-called PSF-scaling technique. The
main assumption is that an imaging
observation of a standard star taken close in time to the galaxy
observation represents the galaxy unresolved nuclear component. The
standard star image is then
scaled to the peak of the galaxy observation and subtracted
from the galaxy image. The scaling level (that
is, 100\%, 90\%, etc. at the peak of the galaxy emission) is visually
assessed from the PSF-subtracted
image and the azimuthally averaged
one-dimension (1D) emission profiles \citep[see, e.g., figures~1
in][]{Radomski2003, RamosAlmeida2011}.
Generally in Seyfert galaxies  a 100\% scaling
produces residual holes at the center of the galaxy and the scaling level is
determined when the residual image does not show a central hole and
the PSF-subtracted galaxy profile is flat in the central region.

In this work we used the same PSF-scaling technique but we 
first determined the scaling factor by fitting 1D emission
profiles. We extracted 1D profiles of one pixel width, 
centered at the peak of the emission along each of the X
and Y axes of the mid-IR galaxy and standard star images. We used the
original images before rotating to the usual orientation and smoothing
them. We started by  
fitting separately the X and Y
profiles of the corresponding standard stars with one Gaussian function
to derive the width $\sigma$ of the unresolved component. 
 We then used two Gaussians for the galaxy profiles along each of the X and Y
 directions. The first component models the unresolved component. We fixed
its  $\sigma$ to the value derived from the standard star (within $\pm
1\,$pixel), and allowed its
intensity and position to vary. For the second Gaussian we allowed all
the parameters  to vary to model the extended emission.
We derived the scaling factors 
for the unresolved component Gaussian along the X and
Y directions. We scaled the PSF images (that is, the standard star images) with the two factors and
subtracted them from the galaxy images. From the two PSF-subtracted
galaxy images, we chose the one that did not  over-subtract  the
unresolved emission.
We  note that there is not a strong dependence of the morphology of
the extended
mid-IR emission with  the scaling factor, except in
the very inner regions. \cite{Asmus2019} illustrated this in
their Fig.~4 for NGC~5643, which is included in our analysis.

\begin{figure*}
 \centering
  \includegraphics[width=8cm]{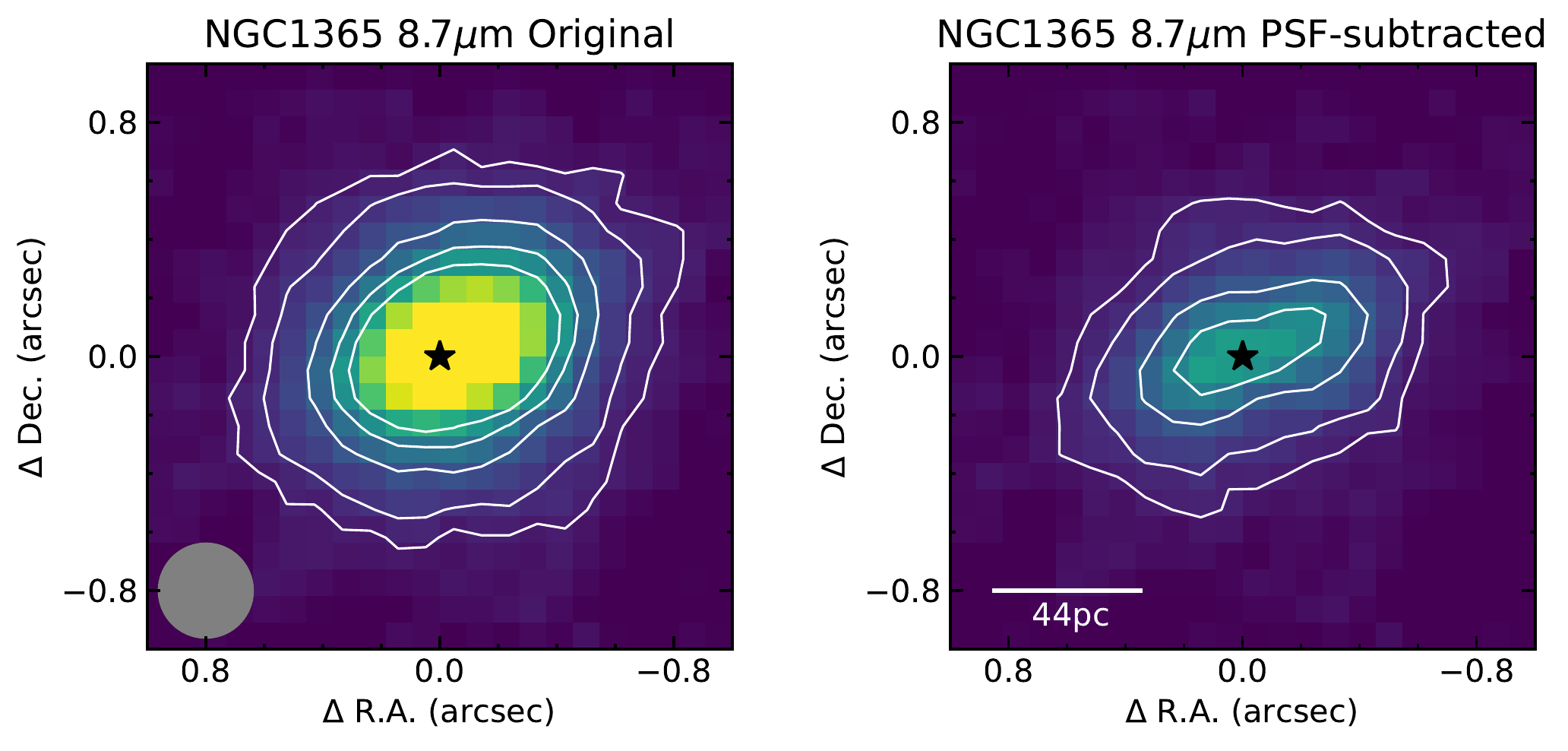}
\hspace{1cm}
  \includegraphics[width=8cm]{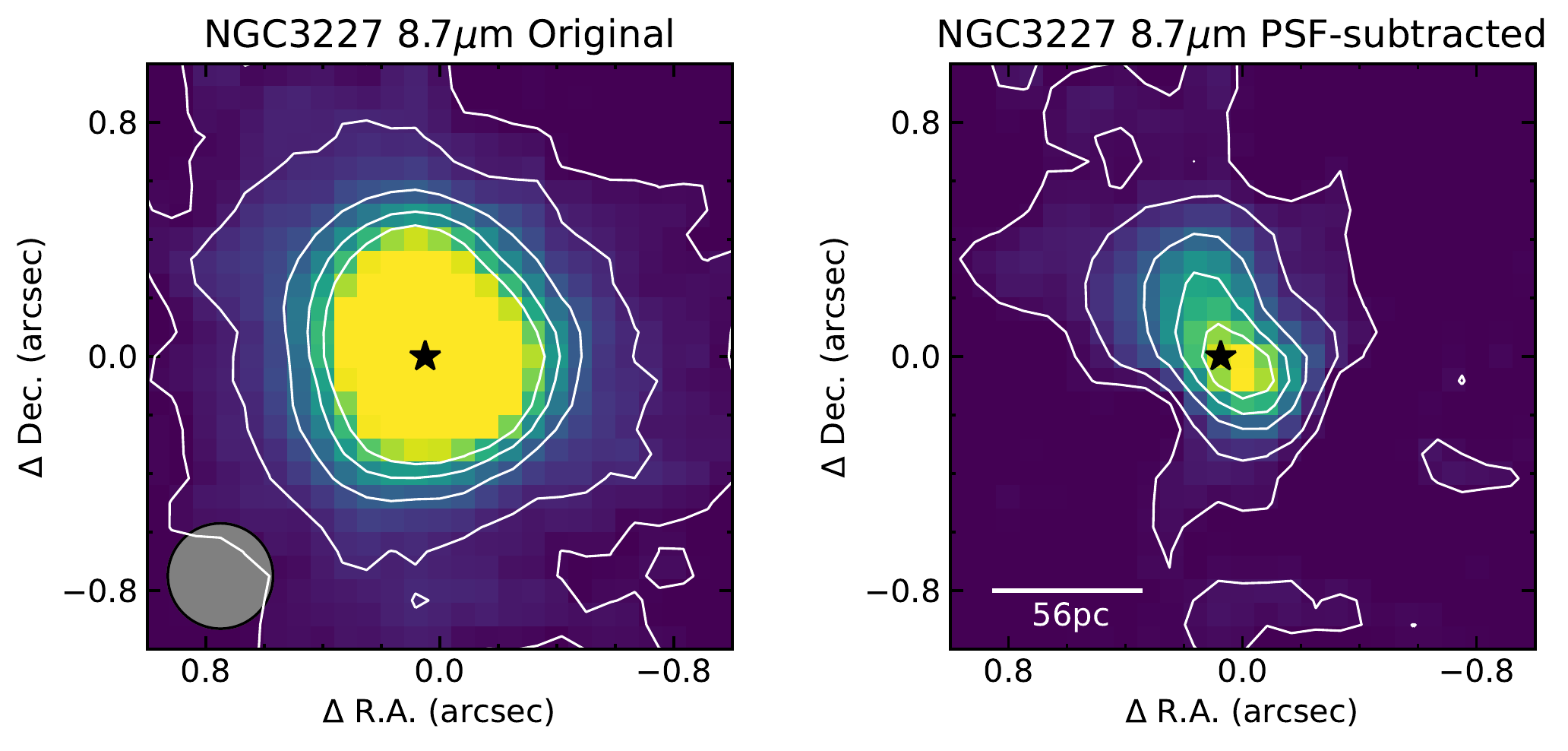}

 \includegraphics[width=8cm]{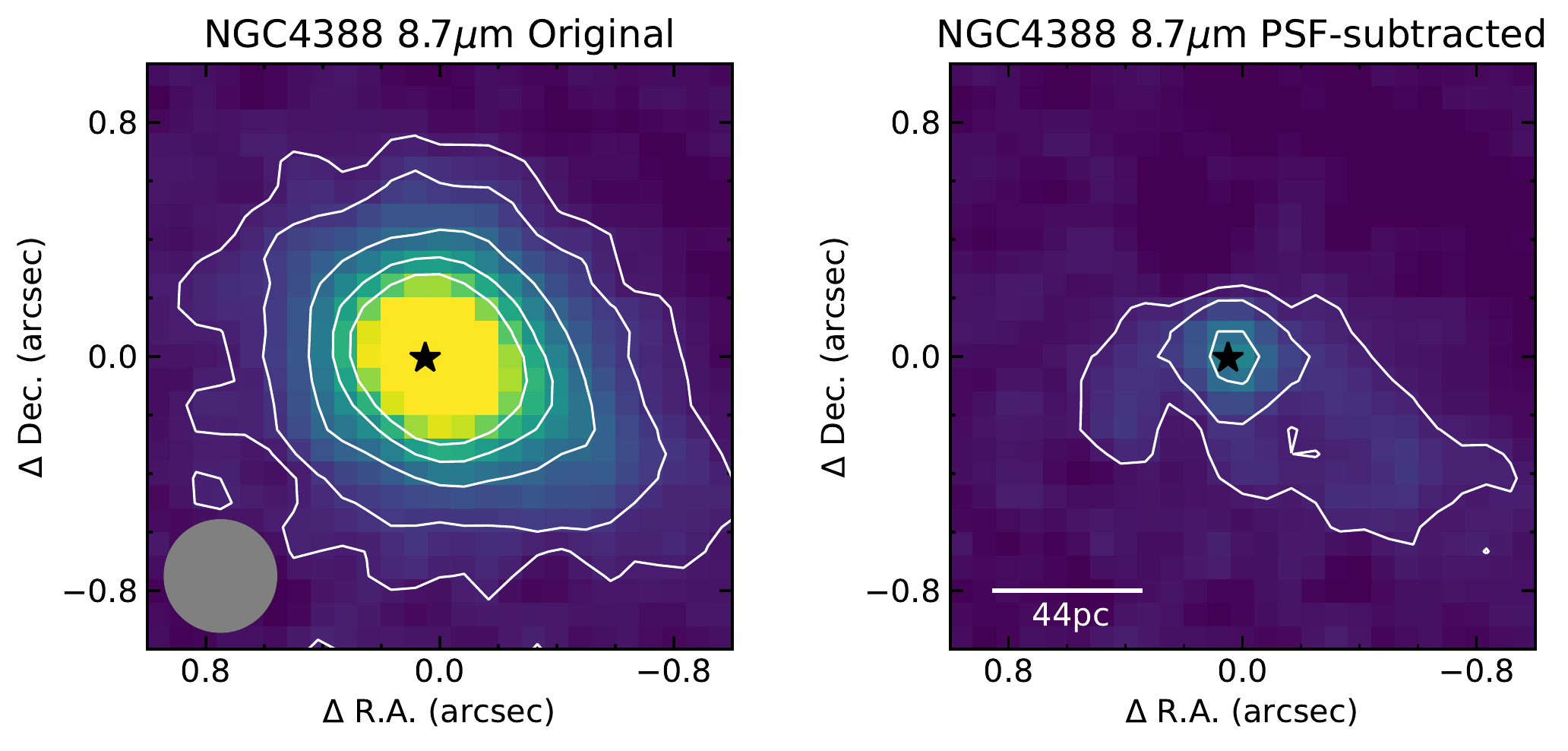}
\hspace{1cm}
  \includegraphics[width=8cm]{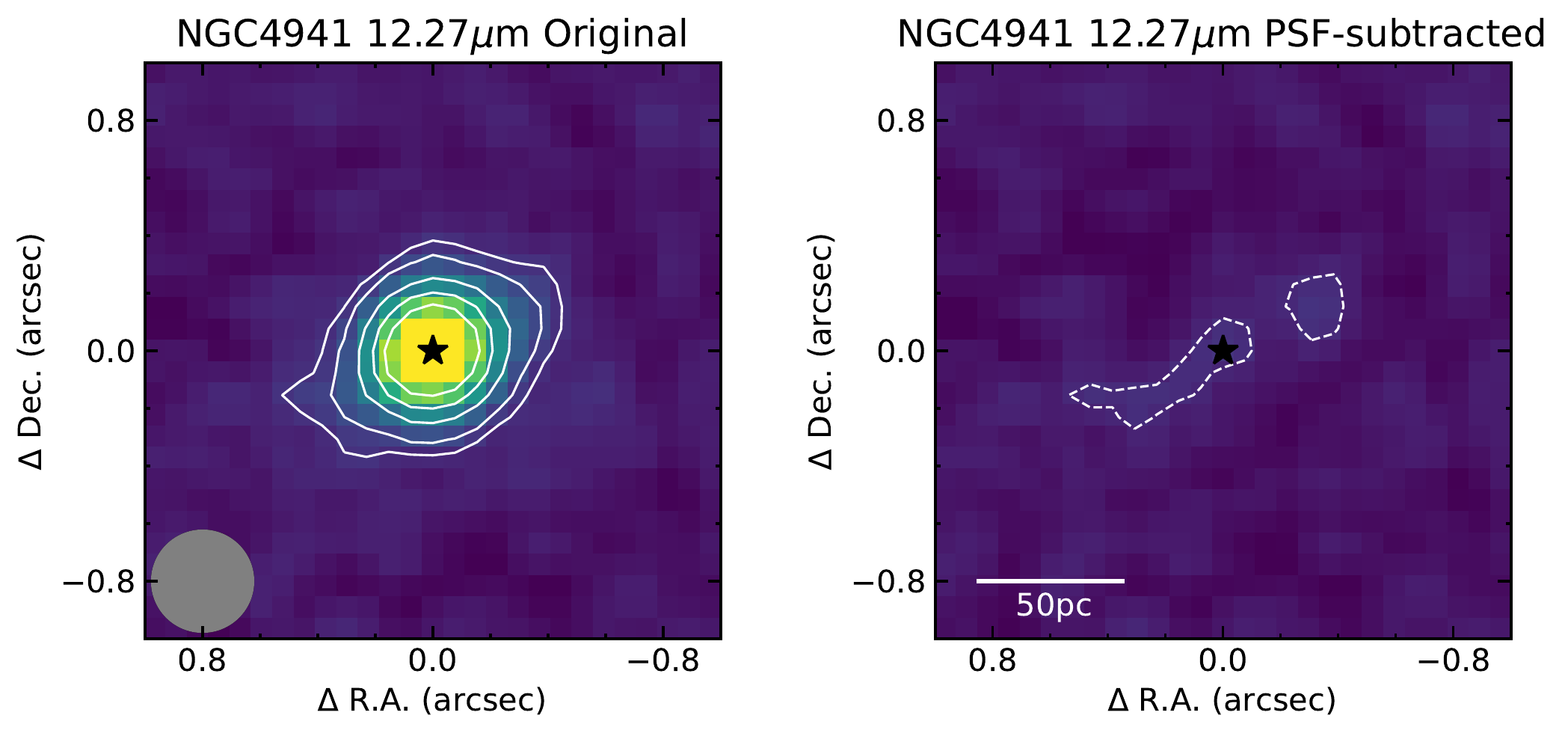}

    \includegraphics[width=8cm]{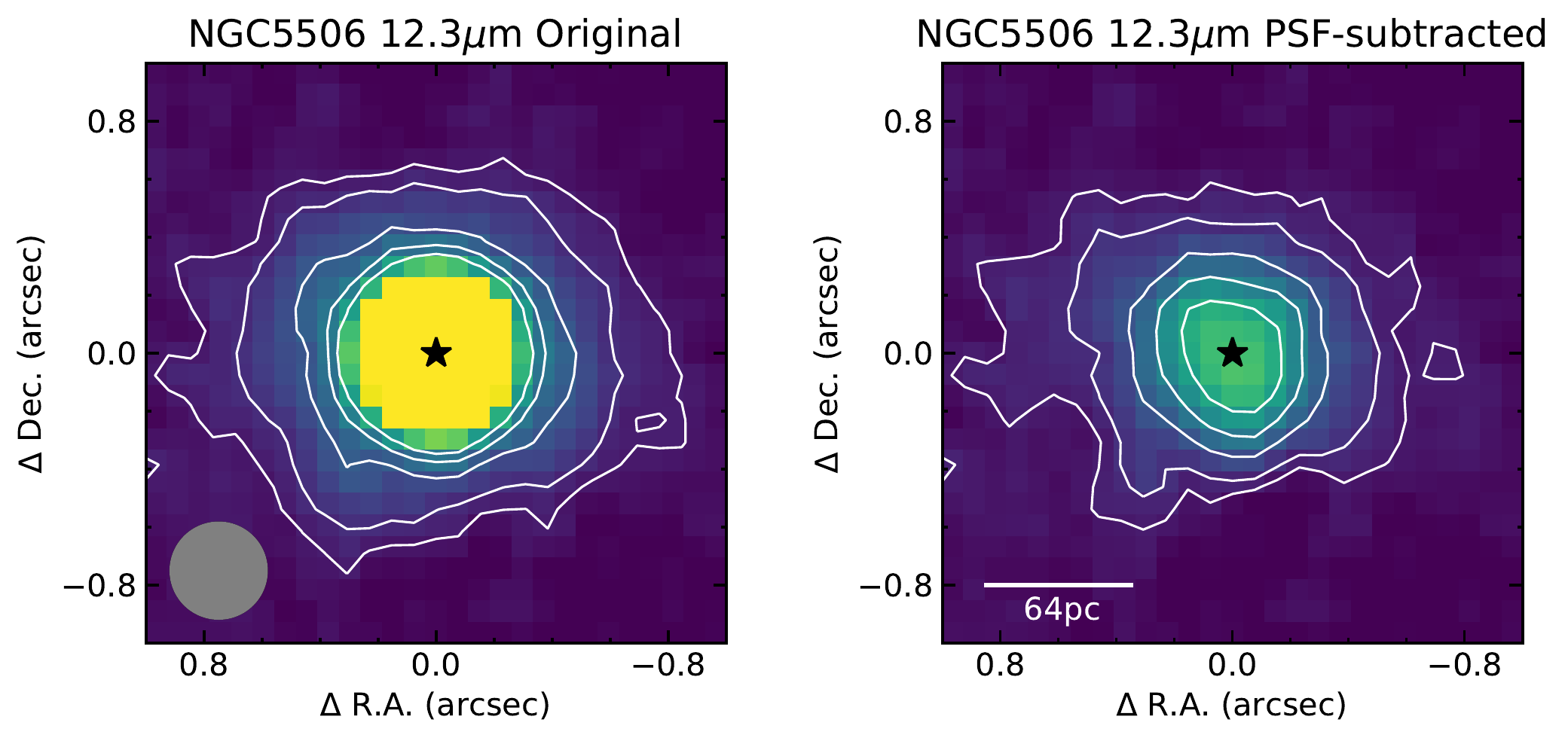}
\hspace{1cm}
\includegraphics[width=8cm]{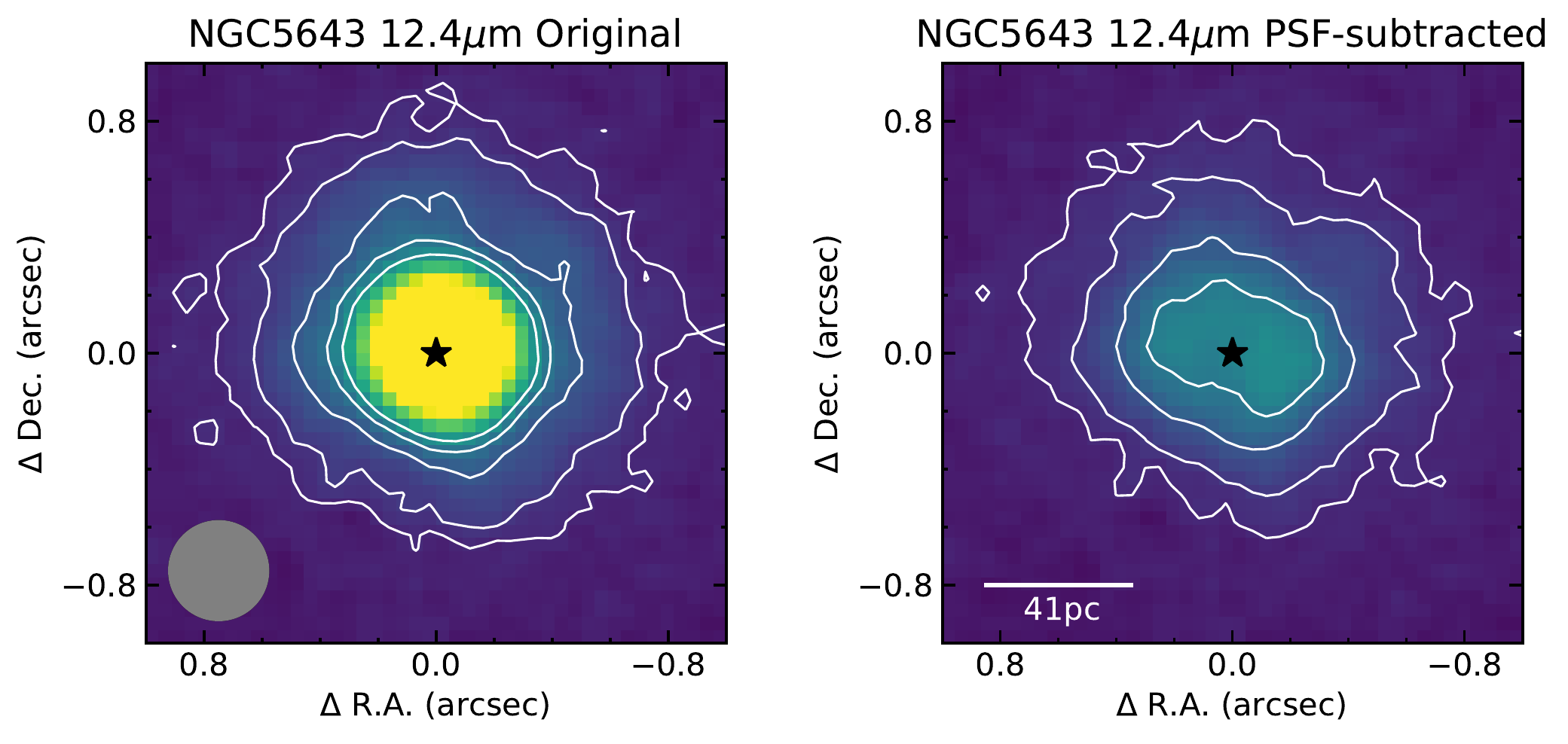}

 \includegraphics[width=8cm]{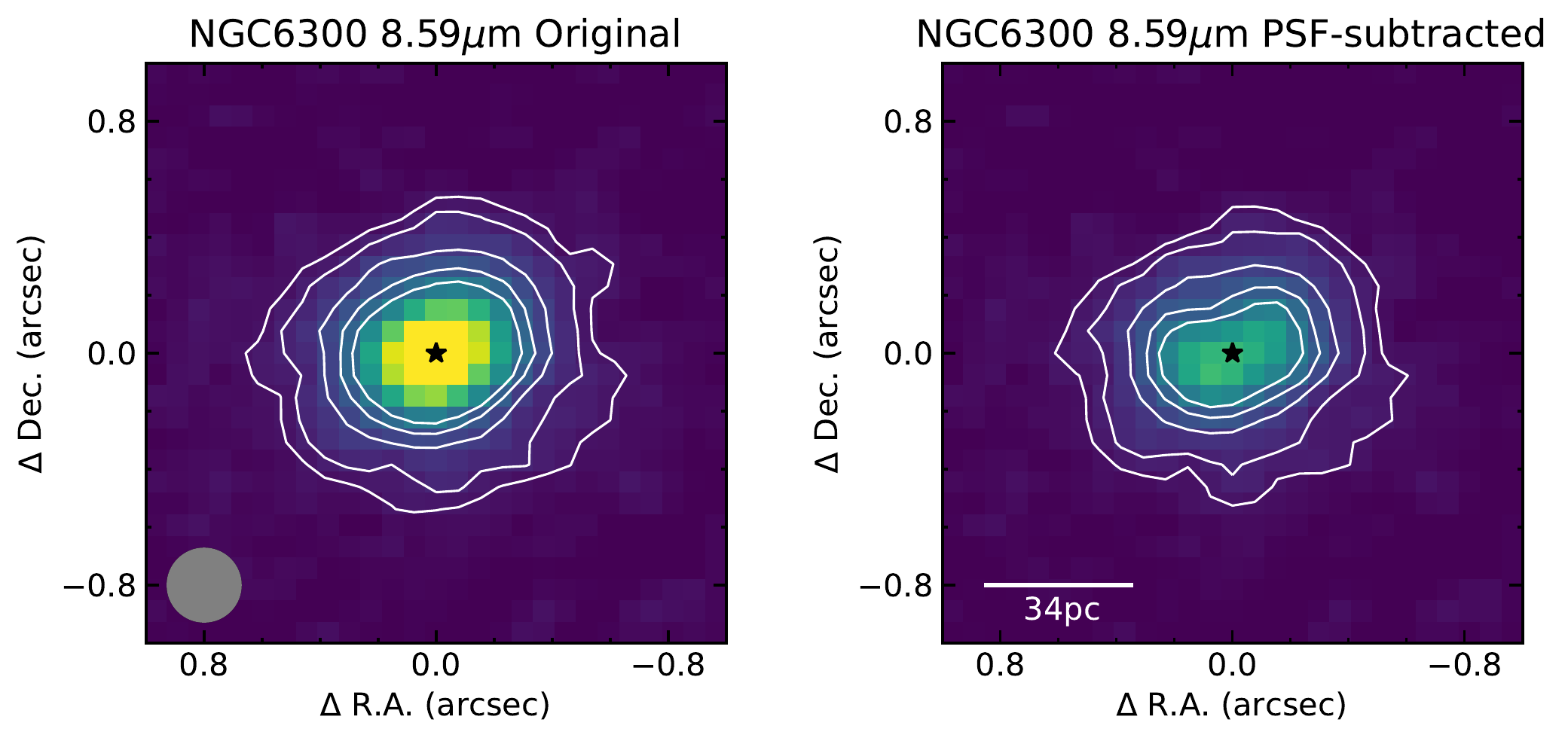}
\hspace{1cm}
  \includegraphics[width=8cm]{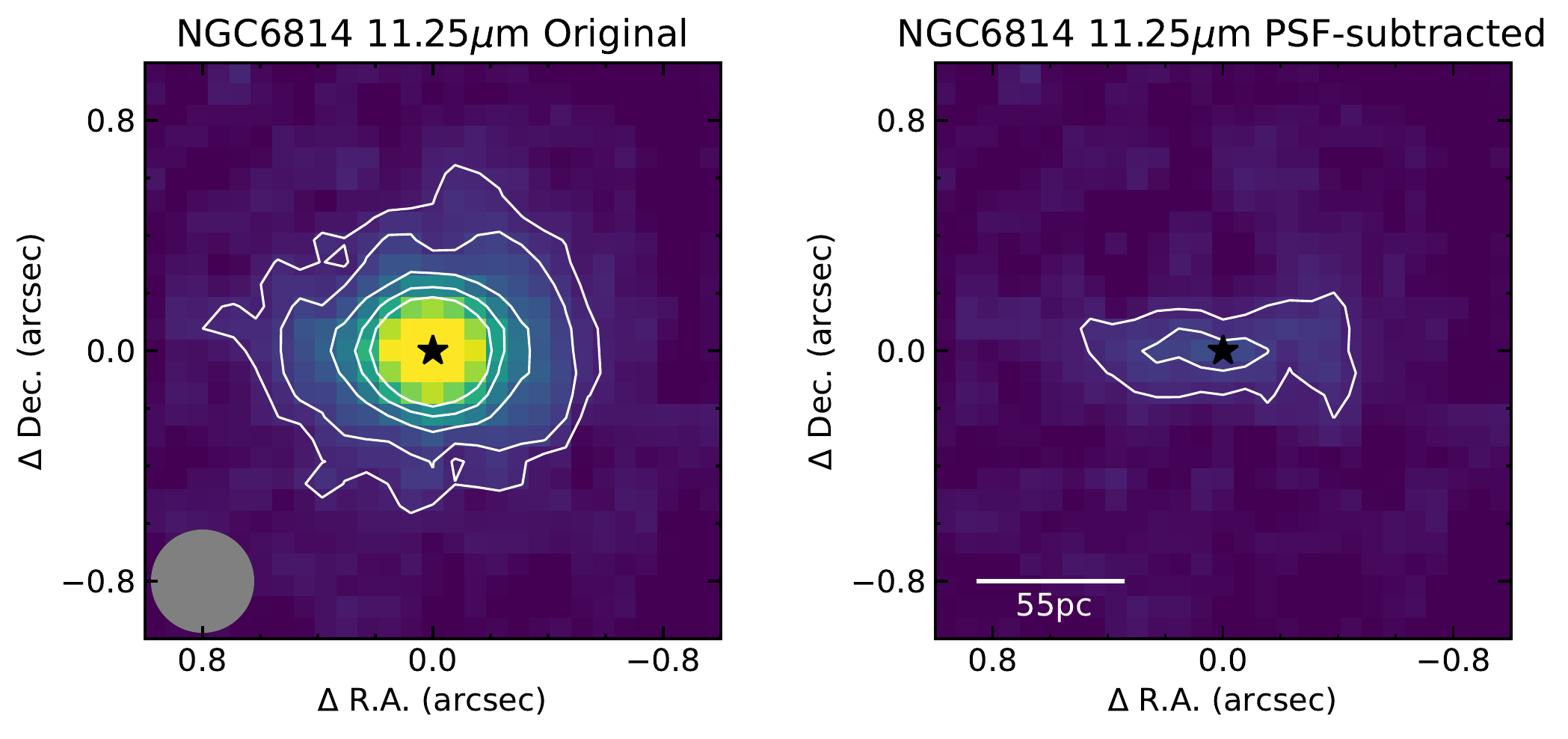}

    \includegraphics[width=8cm]{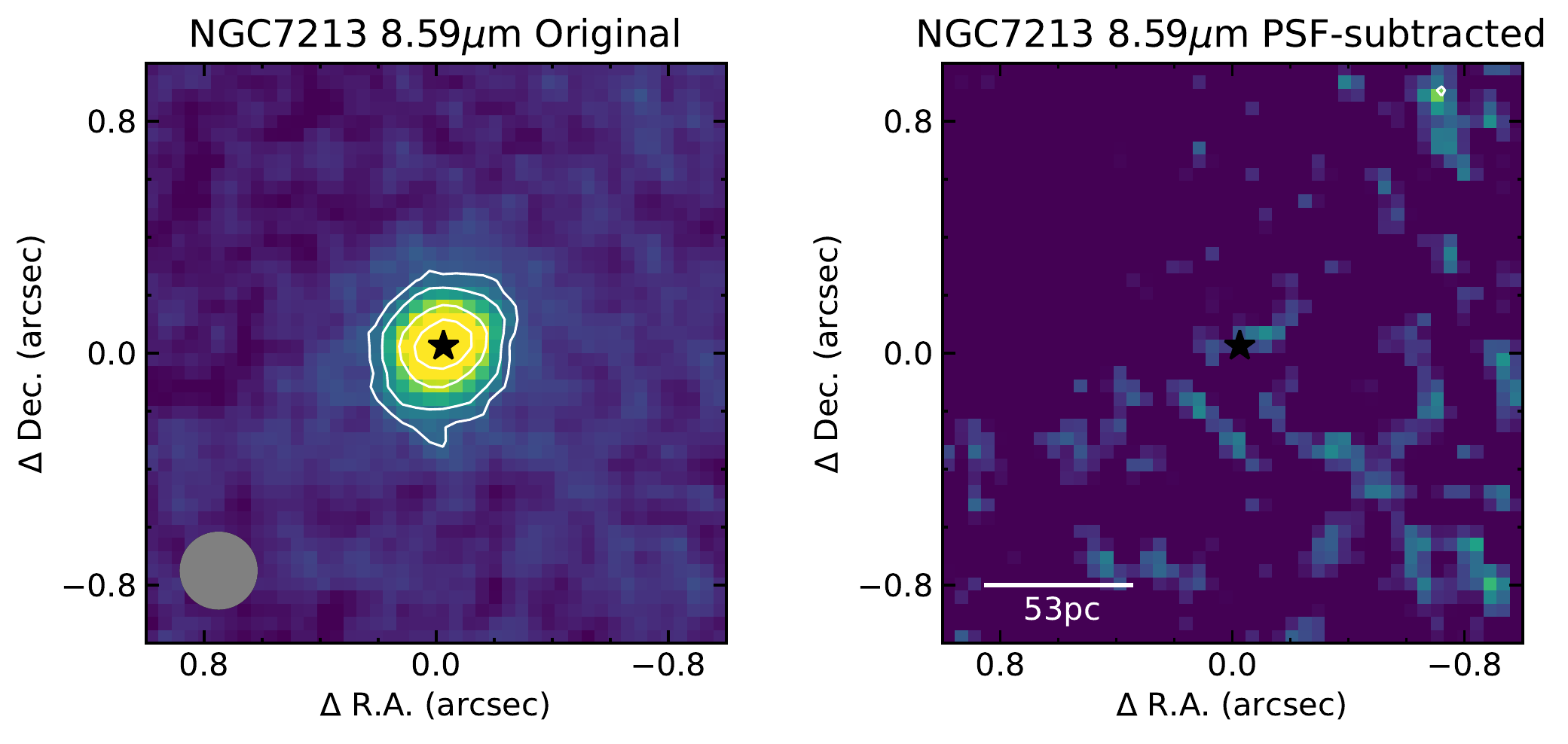}
\hspace{1cm}
\includegraphics[width=8cm]{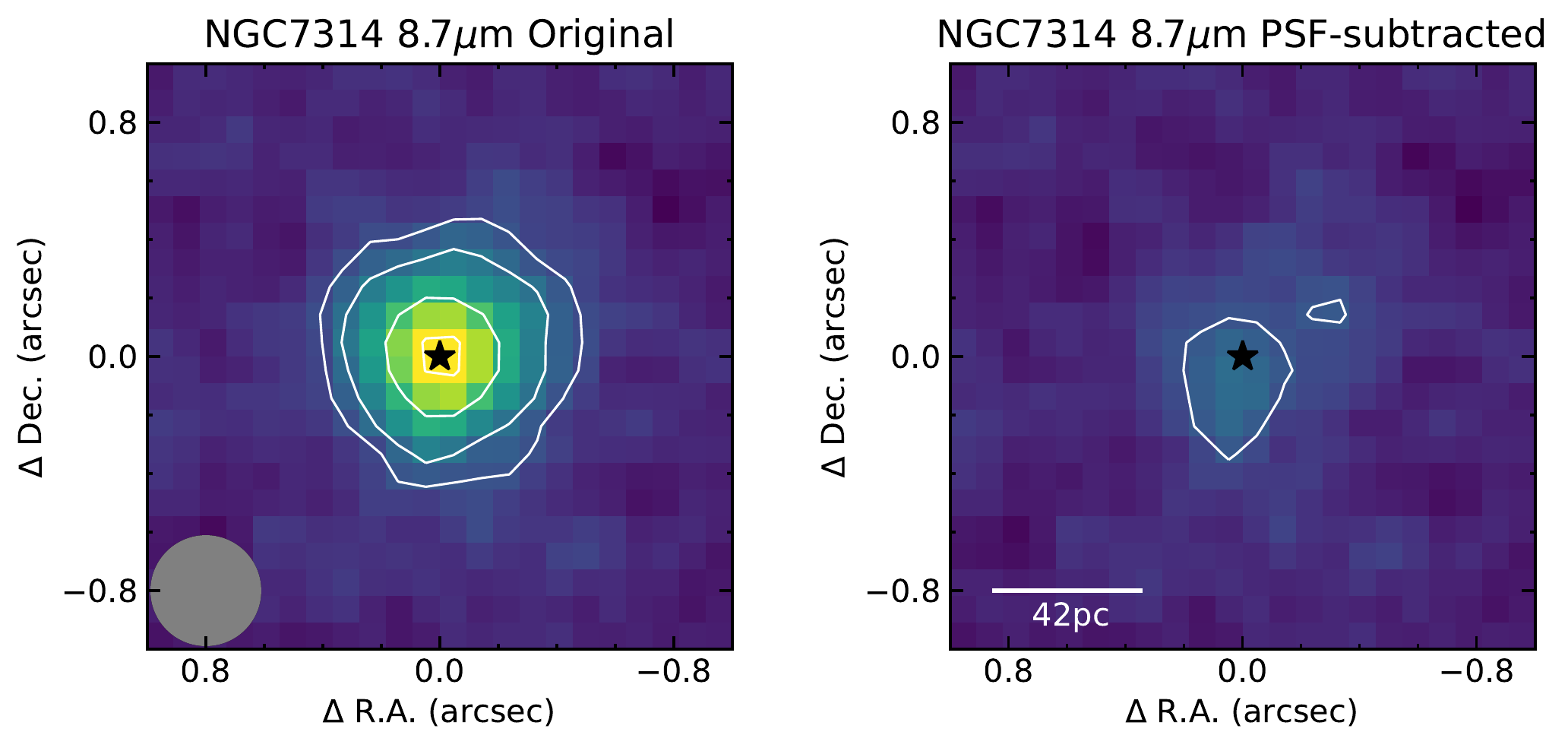}

  \includegraphics[width=8cm]{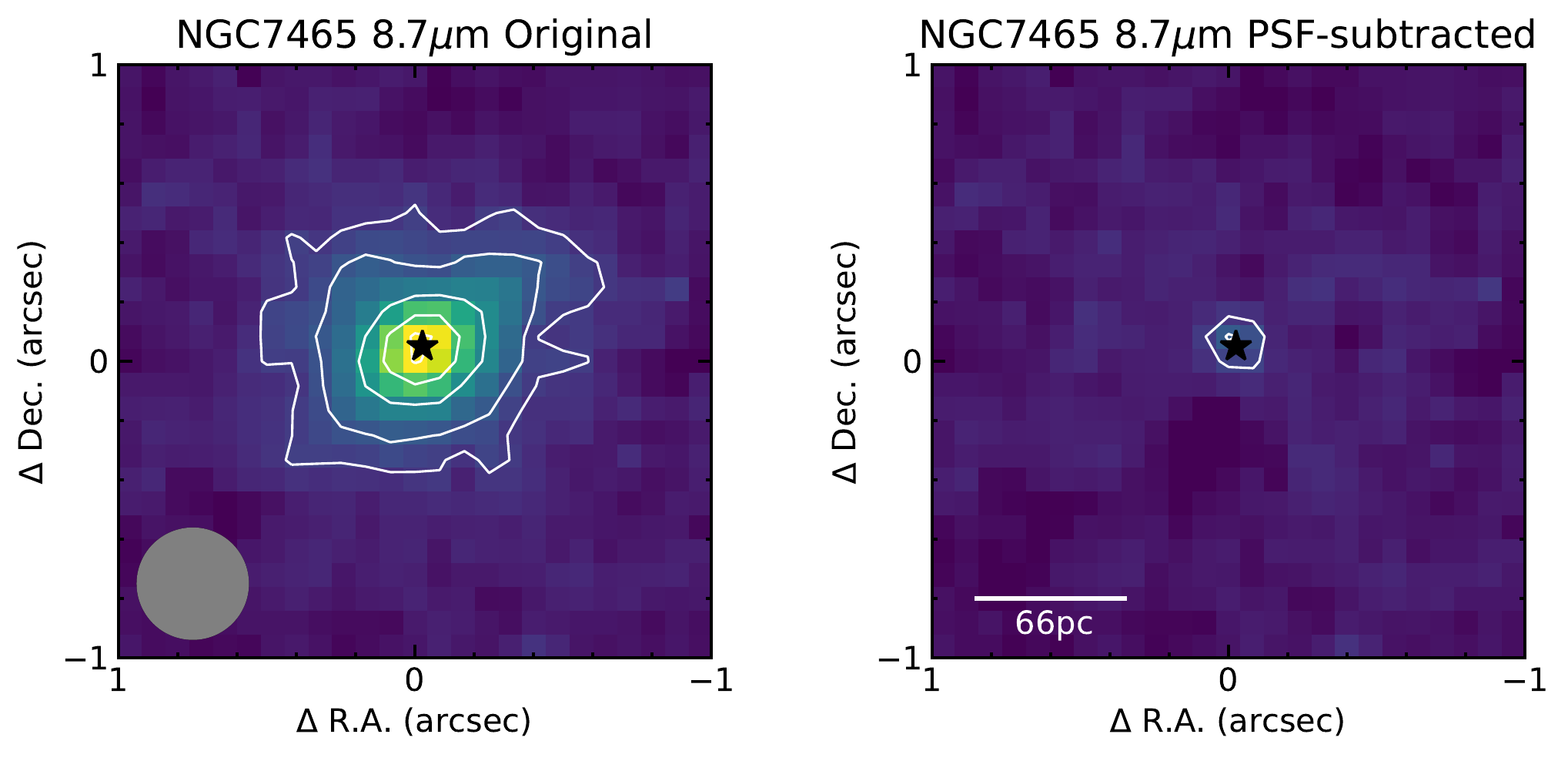}
\hspace{1cm}
  \includegraphics[width=8cm]{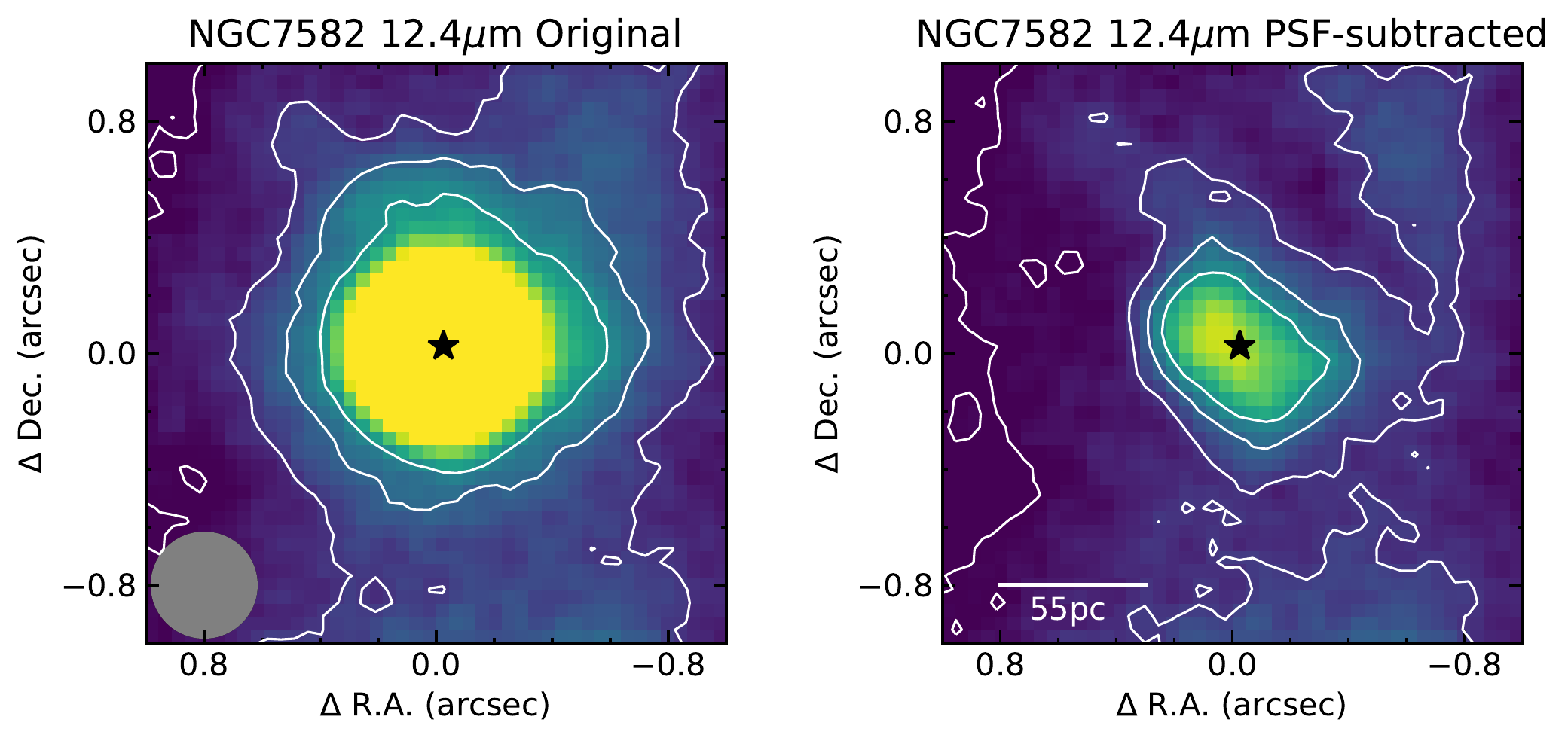}

  \vspace{-0.2cm}
  \caption{Mid-IR images of the central $2\arcsec \times
    2\arcsec$. For each galaxy,   the left and right panels are the
    original and the PSF-subtracted images, respectively. The
    grey circles indicate the FWHM of the standard star as fitted with
    a 2D Gaussian. The star
    symbol marks the peak of the mid-IR emission before PSF subtraction.
 The first contour with a solid line is at a $3 \times \sigma_{\rm bk}$ level and the
 next contours are at 5, 10, 15, $20 \times \sigma_{\rm bk}$ levels.
 For NGC~4941 the
 dashed lines in the PSF-subtracted image are  $2 \times \sigma_{\rm bk}$. All the
  images are oriented north up, east to the left. We smoothed the
original and PSF-subtracted  images with a Gaussian function (see text
for more details). For
each galaxy we display the two images with the same intensity and contour levels.}
              \label{fig:midIRimages}%
    \end{figure*}

\begin{figure}

  \includegraphics[width=6.cm]{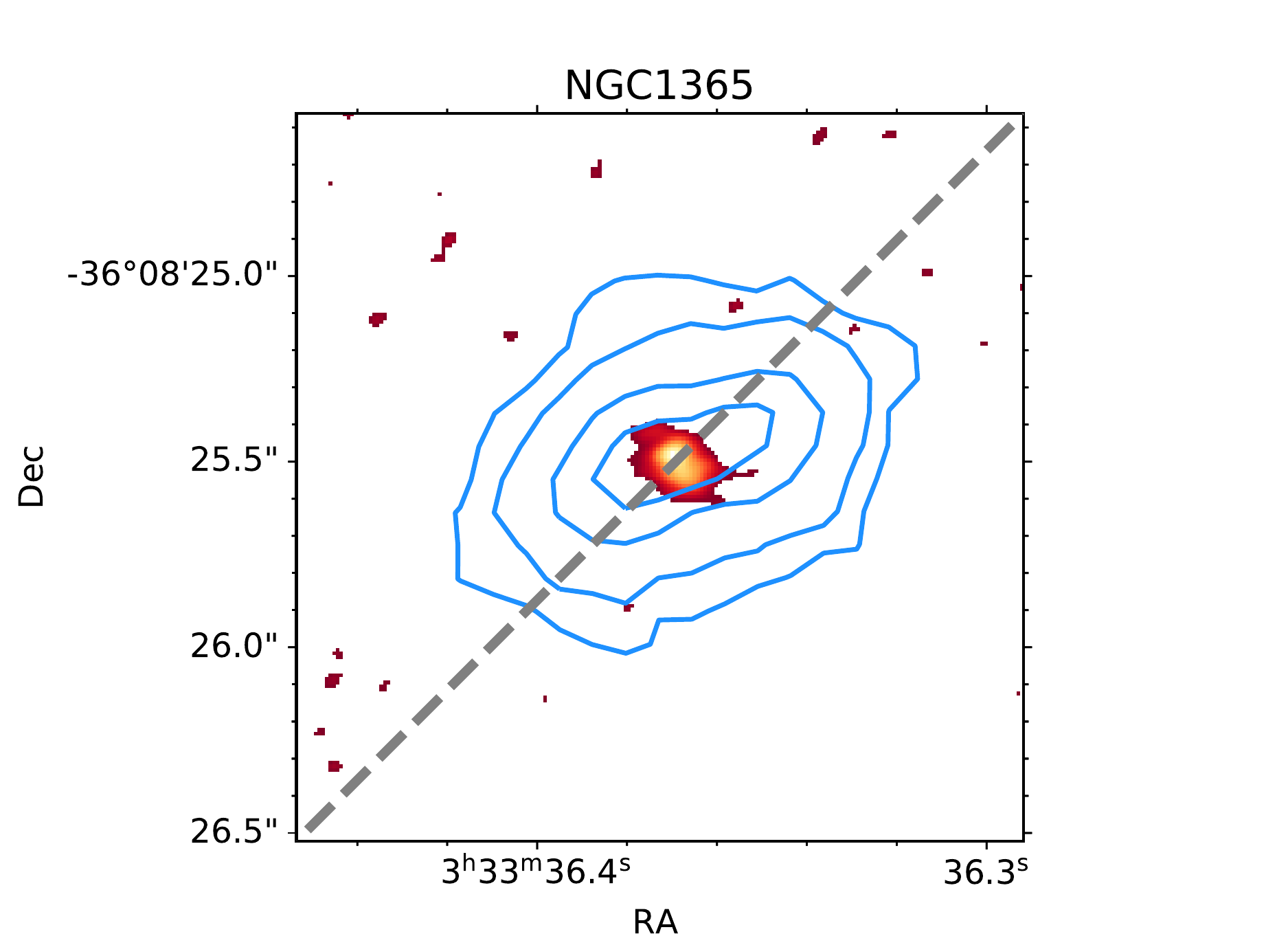}
  \hspace{-1.cm}
 \raisebox{0.55cm}{\includegraphics[width=3.4cm]{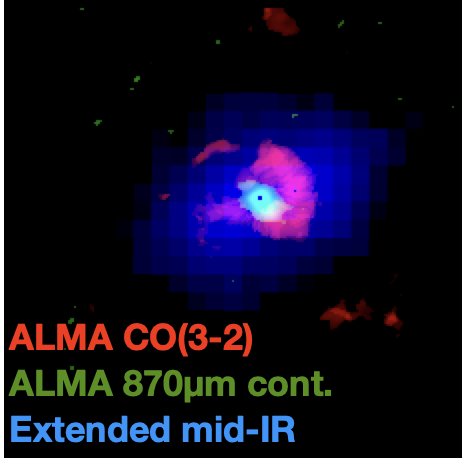}}

  \includegraphics[width=6.5cm]{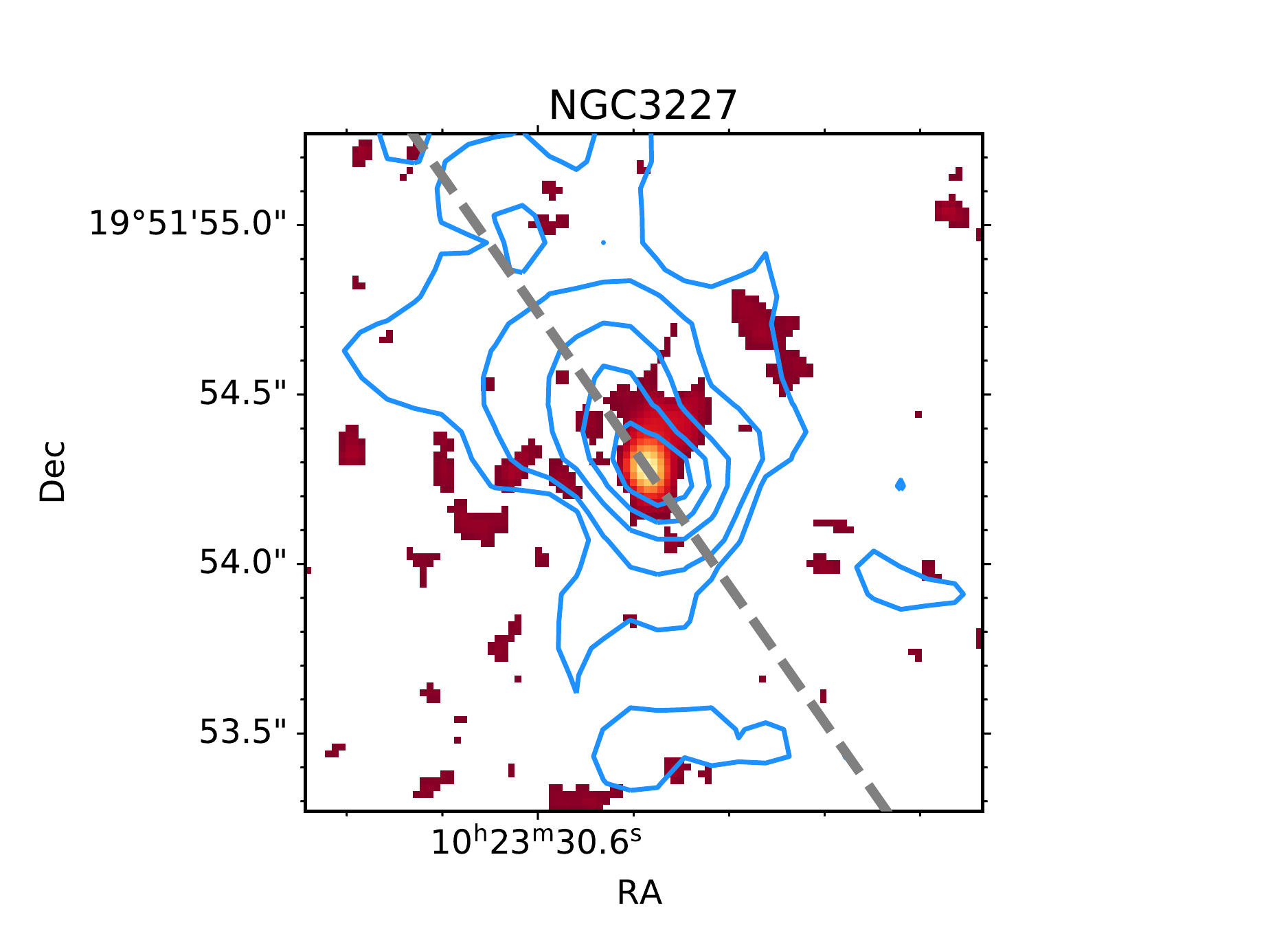}
  \hspace{-1.5cm}
 \raisebox{0.7cm}{\includegraphics[width=3.4cm]{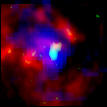}}

  \includegraphics[width=6.cm]{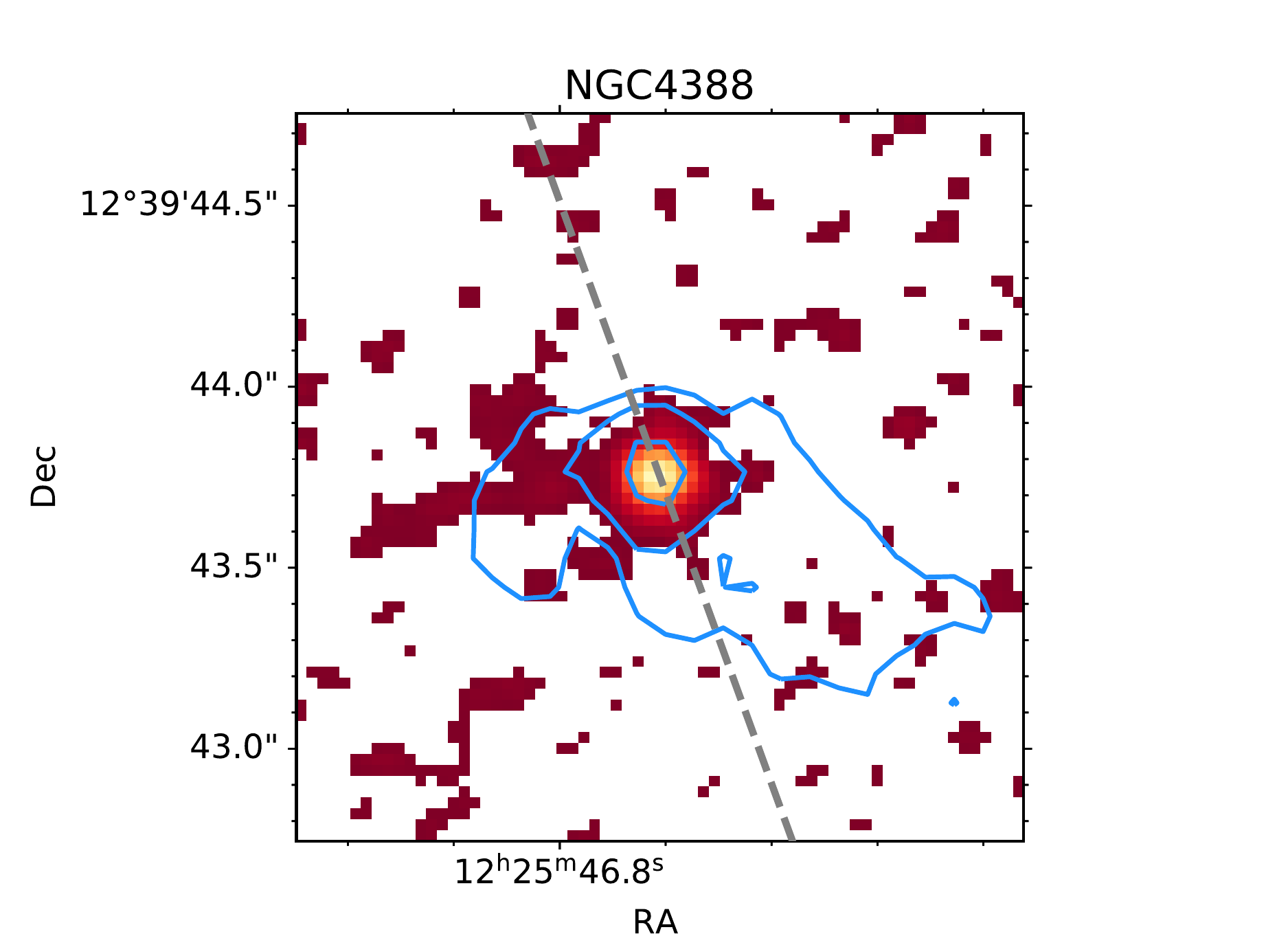}
  \hspace{-1.cm}
 \raisebox{0.55cm}{\includegraphics[width=3.4cm]{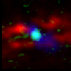}}

  \includegraphics[width=6.cm]{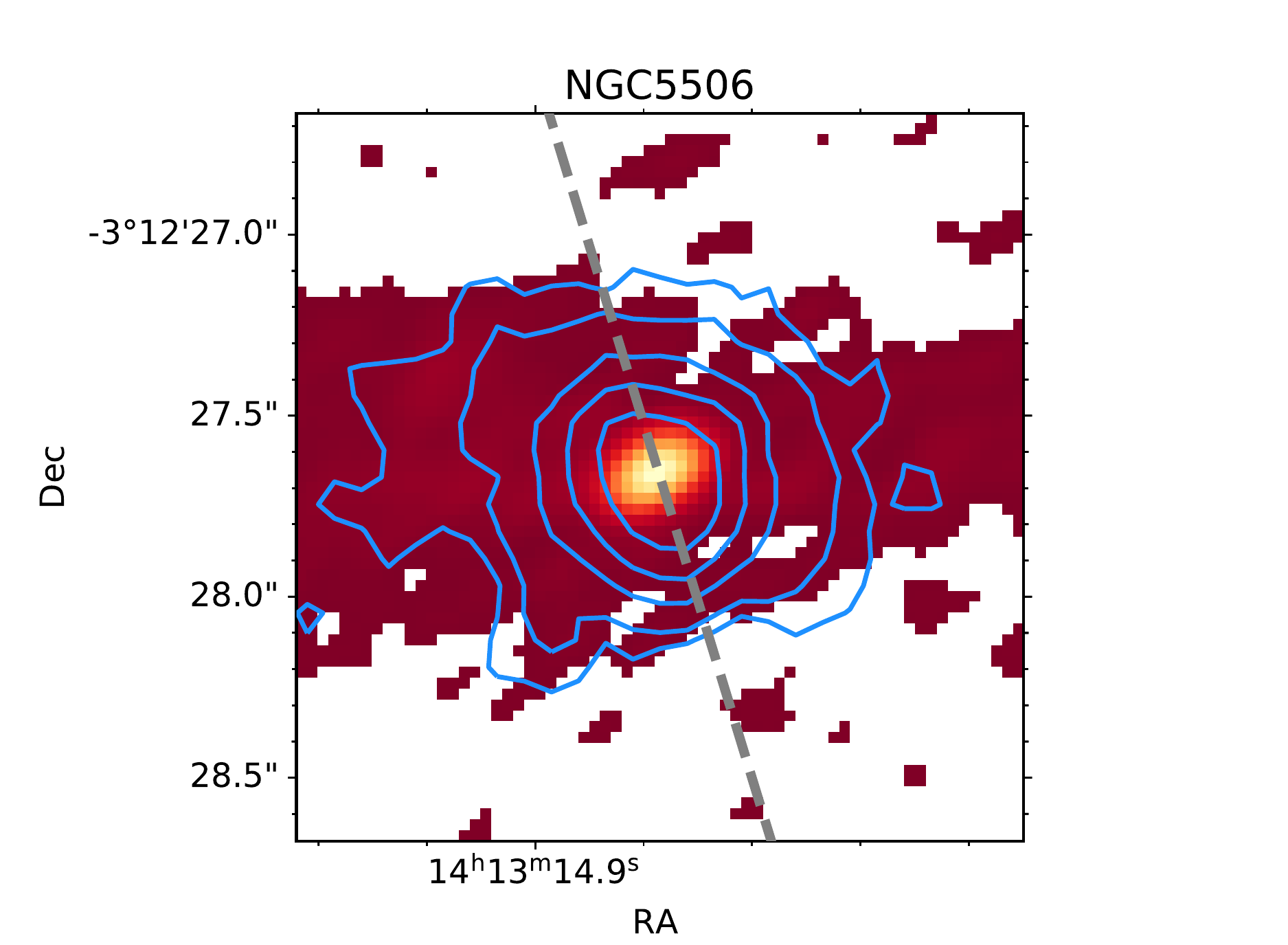}
  \hspace{-1.cm}
 \raisebox{0.55cm}{\includegraphics[width=3.4cm]{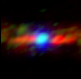}}

 \caption{Galaxies with extended mid-IR emission.  The left panels show in color
   the ALMA MSR $870\,\mu$m continuum images
    (with both the unresolved and the extended emission) from
    GB21. The blue contours are the mid-IR PSF-subtracted images with the
   same levels as in Fig.~\ref{fig:midIRimages}. The FoV
    is $2\arcsec \times 2\arcsec$.  We also plot with the dashed lines
   the
    approximate PA of the ionization cones, NLR and/or outflows (see
   Sect.~\ref{sec:NLR}). The right panels are RGB images constructed
   with the ALMA CO(3-2) image (red), ALMA $870\,\mu$m continuum (green), and 
extended mid-IR emission (blue). The FoV is approximately the same as
in the left panels.}             \label{fig:ALMAmidIRimages}%
\end{figure}

\begin{figure}
  \includegraphics[width=6.cm]{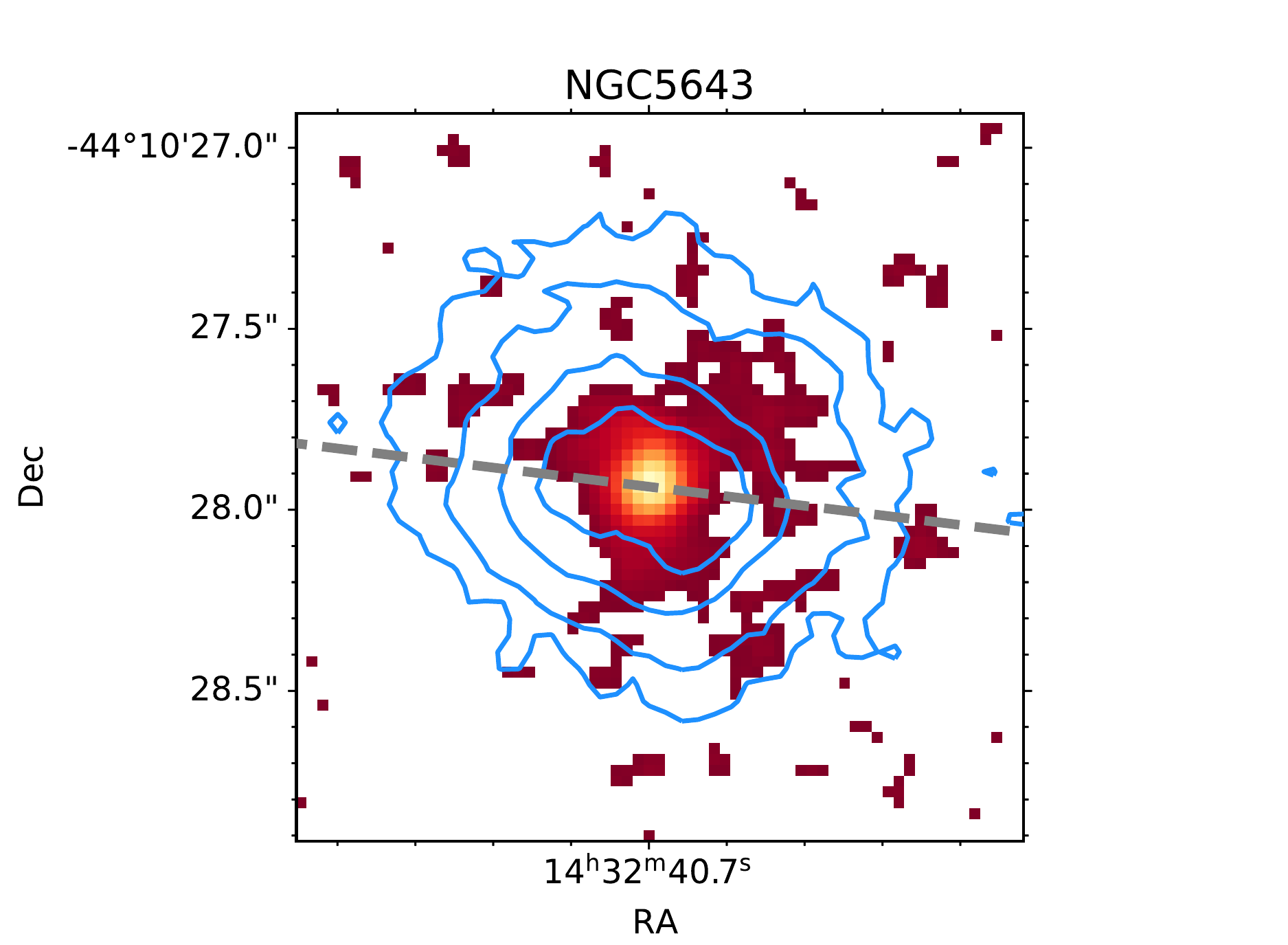}
  \hspace{-1.cm}
 \raisebox{0.55cm}{\includegraphics[width=3.4cm]{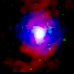}}

  \includegraphics[width=6.cm]{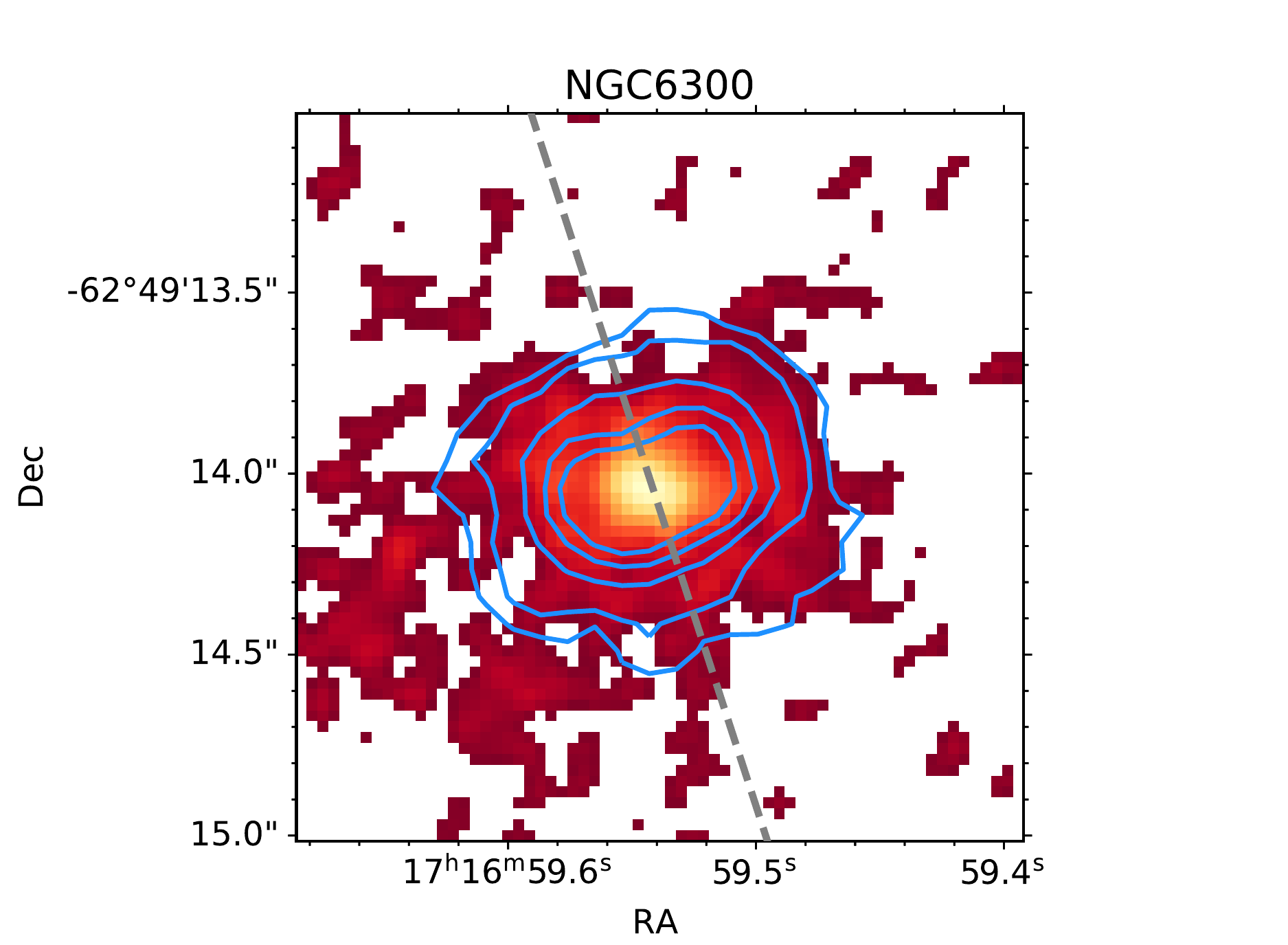}
  \hspace{-1.cm}
 \raisebox{0.55cm}{\includegraphics[width=3.4cm]{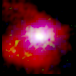}}

  \includegraphics[width=6.cm]{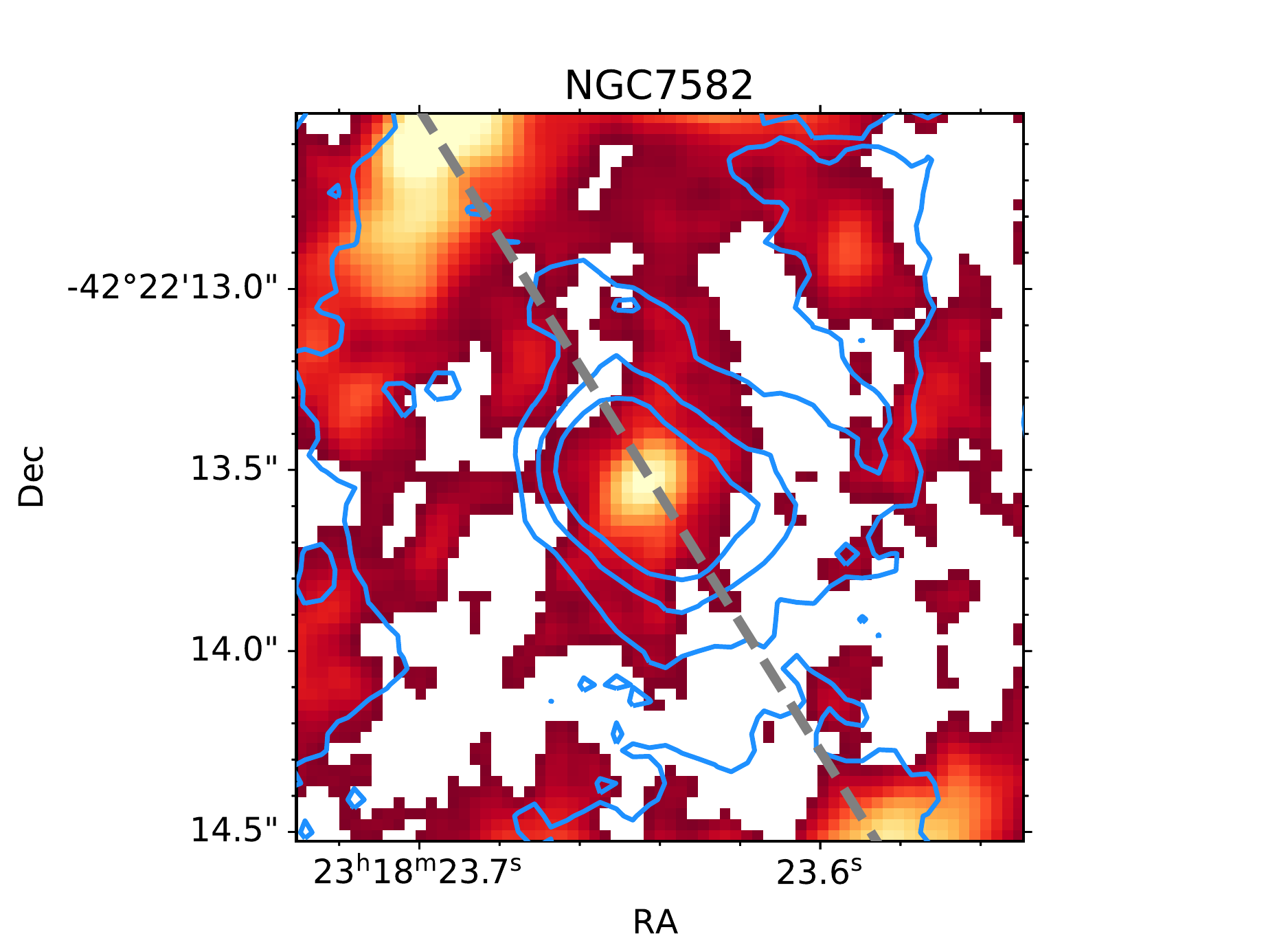}
  \hspace{-1.cm}
 \raisebox{0.55cm}{\includegraphics[width=3.4cm]{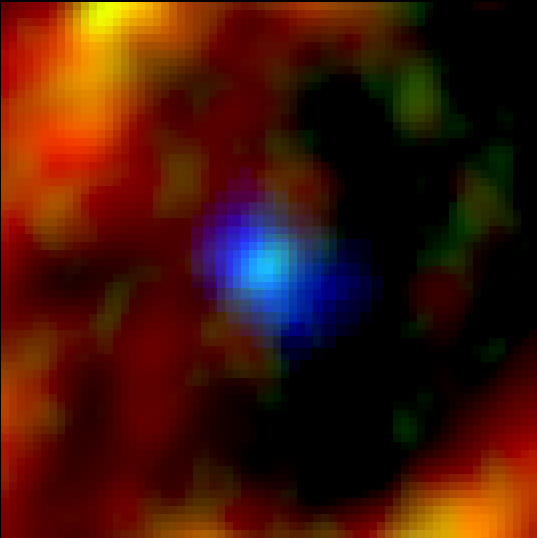}}

\setcounter{figure}{2}
  \caption{Continued.}
\end{figure}

Figure~\ref{fig:1Dprofiles} shows two examples of the fits to the X and
Y direction profiles. The top panels are for NGC~6300,  a galaxy in our sample with a large
mid-IR extended component and the bottom panels NGC~7213,  which appears
unresolved in our mid-IR image. In the case of
NGC~6300, the 1D profiles show that the extended emission is more
prominent along the X-direction (east-west direction in this case) than along the
Y-direction. The derived scaling factors were 63\% and 84\% for the X
and Y directions, respectively. The PSF-subtracted galaxy image using the
scaling factor along the Y direction had negative residuals and
thus we used  the scaling factor along the X direction to produce the
PSF-subtracted galaxy image. For NGC~7213,  
both 1D profiles extracted along the X and Y directions were well modelled
with a dominant unresolved component, which resulted in a scaling factor
of the PSF image of 97-98\%.

\subsection{Results}

In Fig.~\ref{fig:midIRimages} we display for each galaxy in our sample
the original mid-IR image and the resulting PSF-subtracted images for
a field of view (FoV) of $2\arcsec \times 2\arcsec$. This FoV includes most
of the mid-IR extended emission detected for  our sample  with
the instruments listed in Table~\ref{tab:midIR}. The only exceptions are NGC~3227 and
NGC~7582 that show extended emission over larger scales due to the
presence of circumnuclear rings of star formation
\citep{AlonsoHerrero2016, Wold2006, GarciaBernete2016}.
We also smoothed the images using Gaussian functions with FWHM of
between 1.1  and 1.7  pixels, corresponding to 0.07 to  0.13\arcsec. In
all the images, the first contour
shown with a  solid line is at a $3 \times \sigma_{\rm bk}$ level, where $\sigma_{\rm
  bk}$ is the standard deviation of the background
measured on the original images, that is, before we smoothed the
images and when needed, rotated them. The other contours are shown 
at 5, 10, 15, and $20 \times \sigma_{\rm bk}$ levels.

The PSF-subtracted images of six out of the twelve galaxies  present bright extended mid-IR
emission (Fig.~\ref{fig:midIRimages}). These are 
NGC~1365, NGC~3227, NGC~5506,
NGC~5643, NGC~6300, and NGC~7582. The contours show that the extended emission is
detected up to at least a $20 \times \sigma_{\rm bk}$
level. For NGC~4388 we also detected extended mid-IR emission at a
lower significance but in a similar orientation as that derived by
\cite{Asmus2016} using a different dataset. The sizes of the extended  mid-IR emission range from
approximately 50\,pc in NGC~6300 to 190\,pc in NGC~5506.
We used the {\it ellipse} task within {\sc
   iraf} on the PSF-subtracted images to  derive the position angle
 (PA) of the extended mid-IR emission (PA$_{\rm   MIR-ext}$). The
 typical uncertainties in these measurements are $\pm 5^{\circ}$.
We report these values in Table~\ref{tab:midIR} for the inner $\sim
1\arcsec$, which is the location of most of the extended mid-IR emission.  PA$_{\rm MIR-ext}$
 is relatively constant with radius in NGC~1365, NGC~3227, NGC~4388,
NGC~6300, and NGC~7582.  
For NGC~5643 we found that the values are between approximately
$48^\circ$ 
in the inner  regions ($\sim 0.45\arcsec$) changing progressively to approximately
$60-70 ^\circ$ in the outer
contours.  In the most
extreme case in our sample NGC~5506, the PA$_{\rm MIR-ext}$ values vary
from $30^\circ$ in the inner region (central $\sim 0.5\arcsec$) to nearly
$90 ^\circ$ in the outer regions.

Three galaxies, NGC~6814, NGC~7314,
and NGC~7465, only show evidence of faint extended emission 
at  $3-5 \times \sigma_{\rm bk}$ levels. 
Finally, the
PSF-subtracted images of NGC~4941 and NGC~7213 do not show evidence of
extended emission.

The fraction of unresolved  flux over total flux measured within 
aperture diameters of $1.5-1.6\arcsec$ (see Table~\ref{tab:midIR}) are between 40 and
72\% for the seven galaxies with extended mid-IR emission. These
apertures cover the extent of the mid-IR nuclear emission and are
large enough to include the PSF wings. 
 For the five galaxies with faint or no detected extended mid-IR
 emission, the unresolved emission contributes between 80 and 100\%,
 within similar apertures.
This is in good agreement with previous works
\citep{Asmus2014, GarciaBernete2016, Asmus2019}.
In summary, at our angular resolutions the
unresolved component accounts for between 60 and 100\% of the mid-IR
emission in the central
1.5\arcsec$\sim$150\,pc for the majority of the galaxies in our sample.

\begin{table*}
\caption{Observational and modelled torus and wind properties.}             
\label{tab:Torus}      
\centering                          
\begin{tabular}{c cccccccc ccc}        
  \hline
  \hline
  \rule{0pt}{2.6ex}
  \smallskip 
      & \multicolumn{7}{c}{Torus/disk} &
                                                \multicolumn{3}{c}{Wind}\\
  \cmidrule(lr){2-8} \cmidrule(lr){9-11}
\rule{0pt}{2.6ex}
  \smallskip 

&         \multicolumn{2}{c}{ALMA} &       
                                                 \multicolumn{5}{c}{$i_{\rm torus}$                                                      ($^\circ$)}
  & $i_{\rm cone}$ ($^\circ$) & PA$_{\rm cone}$($^\circ$) & $f_{\rm
                                                            wd}$\\
\cmidrule(lr){2-3} \cmidrule(lr){4-8}
\rule{0pt}{2.6ex}
  \smallskip 

  Galaxy            & $d$ (pc)&PA$_{\rm A}(^\circ)$ &  {\sc clumpy}  &
                                                              {\sc cat3d-wind}
                                          & {\sc xclumpy}&  ALMA &
                                                                   Kinem.
                                                & Model &
                                                                    Opt/IR
                 & {\sc cat3d-wind}\\ 
\hline

NGC~1365    & 28&   50 & [8-40]  & \dots & 53  & $>$48 & \dots & \dots & $-45$ &
                                                                       \dots\\
NGC~3227    & 41&  166 & [9-35] &[28-32] & 20 (f)  &  $>$50  & 30 & 75 & 35 & $>0.5$\\
NGC~4388    & 32&  $-$43$^*$ & [13-20] & [50-52]  & 70 & \dots & &
                                                                    \dots
               & $-165$        & $>0.7$\\
NGC~4941 & 78 & $-$29$^*$ &  \dots& [35-41] & \dots & \dots & \dots &
                                                                      \dots
                 & 40
                 & [0.2-0.3]\\
NGC~5506    & 129 & 87 & [5-40] & low & 33 & $>$55 & & 10& 18 & $>0.7$ \\
NGC~5643    & 43 & 4  &  [65-81] & \dots & 74 (f)  & $>$39 & 60 & 25 & 82 & \dots \\
NGC~6300    & 64 & 85 & [14-21] & low & 53 & $>$44 & 57 &\dots &  18 & [0.2-0.3]\\
NGC~6814    & 33  & 57$^*$ & [7-34] & 30 & 45 (f) & \dots & \dots &
                                                                    \dots
                 & 33 & [0.2-0.3]\\
NGC~7213    & 40 & 71 & [0-30] & low & 45 (f) & $>$30 &  \dots & \dots &
                                                                     no &
                                                                       $<$0.2\\
NGC~7314 &  60 & 21 & [5-37] & high & 45 (f) & $>$58 & \dots & \dots&
                                                                       110 & $>$0.7\\
NGC~7465 & 67 & 4 & \dots & \dots & \dots & $>$53 & \dots& \dots& $-72$
                 & \dots\\
NGC~7582    & 91 & $-$18 & [53-82] & \dots & 41 & $>$59 & 59 &  \dots & $-122$&  \dots\\
\hline
\end{tabular}
\tablefoot{Columns for the torus/disk components are as follows. $d$ is the ALMA 
 MSR $870\,\mu$m torus size (diameter of the extended component at the
 $3\sigma$ level) and PA$_{\rm A}$ the position angle
  PA and inclination, from GB21. $^*$The $870\,\mu$m PA is
  not along the  equatorial direction (i.e., torus).
The torus/disk inclination ($i_{\rm torus}$) constraints from
modelling of the IR emission with
the CLUMPY torus models are from \cite{AlonsoHerrero2011,
  AlonsoHerrero2012, RuschelDutra2014, Ichikawa2015,
  GarciaBernete2019, MartinezParedes2020}, the CAT3D-WIND
models from \cite{GonzalezMartin2019}, and from modelling of
X-ray observations are from \cite{Walton2010, Tanimoto2020,
Ogawa2021}. The torus inclinations from the ALMA  MSR $870\,\mu$m images
are from GB21 and from the ALMA
CO(3-2) and CO(2-1) kinematics are from \cite{AlonsoHerrero2018,
  AlonsoHerrero2019} and GB21. Columns for the wind components are as
follows. The ionization cone inclination
($i_{\rm cone}$) constraints are
from the modelling of the NLR by
\cite{Fischer2013}. PA$_{\rm cone}$ is the position angle of the
ionization cones and the references are 
given in the text in Sect.~\ref{sec:NLR}.
The wind-to-disk ratio $f_{\rm wd}$
constraints are from modelling with the CAT3D-WIND
models from \cite{GonzalezMartin2019}.}

\end{table*}


\subsection{Comparison with ALMA $870\,\mu$m and CO(3-2)
  observations}\label{sec:comparisonALMA}

To compare the mid- and far-IR morphologies, we first adjusted the
astrometry of the mid-IR images of the seven galaxies with extended
emission. We assumed that the peak of the
mid-IR emission coincides with that of the ALMA $870\,\mu$m
continuum emission and adjusted the mid-IR coordinates
accordingly. Figure~\ref{fig:ALMAmidIRimages} (left panels for 
each galaxy) shows in color the
original ALMA continuum images, that is, including both the unresolved
and the extended emission, with the contours showing
the extended mid-IR emission from the PSF-subtracted images.
In the six out of the seven galaxies with extended
mid-IR emission, GB21 concluded that the ALMA $870\,\mu$m extended
emission traces cold dust mostly along the  equatorial material in the
torus, with sizes between
28 and 129\,pc (median of 42\,pc, Table~\ref{tab:Torus}).


In NGC~1365, NGC~3227, and NGC~7582 as well as the inner regions of
NGC~5506, the extended mid-IR emission appears to
be approximately perpendicular (in projection) to the extended ALMA $870\,\mu$m
continuum emission. The latter for these four galaxies is identified
as the torus. In NGC~5643, the orientation of the extended mid-IR emission is
mostly along $PA_{\rm MIR} = 60-70 ^\circ$, which is nearly
perpendicular  (in projection) to 
the nearly north-south ALMA $870\,\mu$m
continuum emission.  Finally, in NGC~6300 the extended mid- and far-IR
emission show similar orientations, and thus the extended mid-IR
emission is along the equatorial direction of the torus.


GB21 showed that the nuclear CO(3-2) morphologies of the GATOS Seyferts are in
most cases rather different from the $870\,\mu$m emission. 
We constructed red-green-blue
(RGB) images  using the ALMA CO(3-2) images in red and $870\,\mu$m
continuum in green for the seven Seyferts with extended mid-IR
emission.
In blue we show only the extended mid-IR emission to
emphasize the low surface brightness emission at these wavelengths.
In all  galaxies but NGC~6300 (see right panels of
Fig.~\ref{fig:ALMAmidIRimages}),  the extended mid-IR
emission appears to {\it fill} the nuclear regions where the CO(3-2)
emission is fainter. 
On  larger scales the CO(3-2) emission is more extended than both 
the mid- and far-IR continuum images.
In NGC~5643 the extended
mid-IR emission is not only along the east-west direction, where there
is faint CO(3-2) emission, but also along the nuclear  molecular gas mini-spiral.

\subsection{Comparison with ionization cones, NLR, and nuclear outflows}\label{sec:NLR}

The seven galaxies with extended mid-IR emission have bright ionization cones/NLR that
are identified from optical and
near-IR narrow-band imaging and/or integral field unit (IFU)
spectroscopy. We derived the PA of the cones by visual inspection of the
[O\,{\sc iii}]$\lambda$5007 images from
\cite{Venturi2018} for NGC~1365, from \cite{Schmitt2003} for NGC~4388,
from \cite{Fischer2013} for NGC~5506, from  \cite{GarciaBernete2020}
for NGC~5643, and from \cite{Thomas2017} for NGC~7582. For NGC~3227 we
used {\it Hubble Space Telescope}
(HST)
narrow-band images from
\cite{AlonsoHerrero2019}. For NGC~6300, \cite{Davies2014} detected a
nuclear molecular gas outflow using observations of the H$_2$
$2.12\,\mu$m line. We list the values of
PA$_{\rm cone}$ in Table~\ref{tab:Torus} and plot them in
Fig.~\ref{fig:ALMAmidIRimages}. All seven
galaxies have large projected opening angles of the cones/outflows,
ranging from 85 to $115^\circ$, approximately.

Six of the seven GATOS Seyferts  with extended mid-IR emission
show nearly the same orientations for the extended
mid-IR emission  and the  ionization
cones/outflows
 (Fig.~\ref{fig:ALMAmidIRimages}, left panels). The exception is
 NGC~6300 (see below).
The differences in (projected) orientations for these six Seyferts are less than
$10-20^\circ$. These are in good agreement with the typical values found
by \cite{Asmus2016, Asmus2019} for a larger sample of nearby AGN with prominent
extended mid-IR emission. The extended mid-IR emission in
NGC~3227, NGC~4388, and NGC~7582 is more prominent on the optical bright side of the ionization
cone. In NGC~1365, NGC~5506, and NGC~5643 the mid-IR extended emission
is more symmetric around the AGN position and approximately along the direction of
the ionization cone. In the innermost region of NGC~5643, the extended
mid-IR emission is at PA$_{\rm MIR}\sim 48^\circ$, which could be due to emission in
the northeast inner walls of the ionization cone \citep[see figure~22
of][]{Fischer2013}, becoming progressively more aligned with the
ionization cone in the outer regions. In the inner regions of
NGC~5506, PA$_{\rm MIR}\sim 30^\circ$, which is nearly aligned with
the orientation of the cone, whereas at larger radial distances the
extended mid-IR emission becomes more equatorial.

The extended mid-IR emission of NGC~6300 appears to be perpendicular to the orientation of the
ionization cone. The 
morphology and orientation of the mid-IR extended emission are, on the
other hand, similar to that seen in the {\it HST} $V-H$ color map
\citep{Martini2003, Davies2014}, which likely traces dust in the host
galaxy. The
outer regions of NGC~5506  also show a large difference between the 
projected orientations
of the cones and mid-IR extended emission. This indicates that the
extended mid-IR emission along the approximate east-west direction
also traces  dust emission in the highly inclined host 
galaxy, as seen again from the $V-H$ color map \citep{Martini2003}.
In NGC~7582, the extended mid-IR emission additionally probes
dust in regions in 
the circumnuclear ring of star formation \citep{Wold2006}.

\begin{figure*}
  \hspace{1.5cm}
  \includegraphics[width=15.cm]{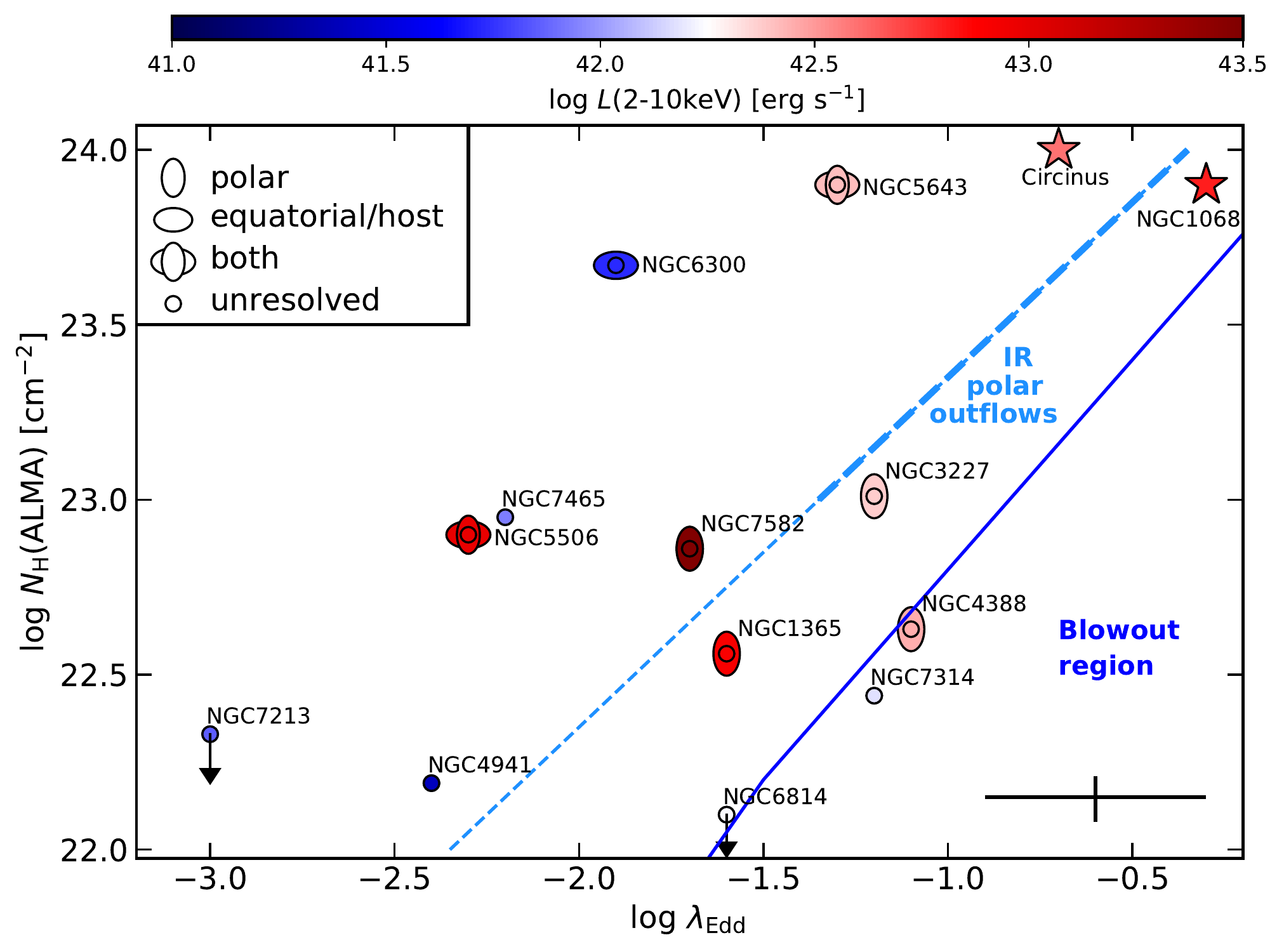}
  \caption{Diagram showing the nuclear
    hydrogen column
    densities against the Eddington ratios  for the GATOS Seyferts.
    The hydrogen column densities
    are  from  ALMA CO(3-2) based H$_2$ estimates  at
    the AGN position from   GB21. 
    The symbols are color-coded in terms of the intrinsic $2-10\,$keV
    luminosities. The different symbols indicate the mid-IR
    morphologies (see Table~\ref{tab:midIR}). Circinus and NGC~1068,
    which 
    are marked with star symbols,    also show polar mid-IR emission
    but are not included in our sample. The error bars represent
     typical uncertainties in the observations, namely,  0.3dex for the Eddington ratios and a
     15\% uncertainty in the absolute flux calibration of the ALMA
     data. 
    Below the solid curve from \cite{Fabian2008} is the blowout region where 
    outflows are likely to clear material in the nuclear regions \citep[see also][]{Ricci2017Nature}.
    The dashed line from \cite{Venanzi2020} indicates the limit
    where the AGN radiation acceleration balances gravity and the IR
    radiation pressure dominates giving rise to polar dusty
    outflows.   Their radiation-dynamical  simulations
    were for  a column density range
    $N_H\simeq 10^{23}-  10^{24}\,{\rm cm}^{-2}$, which we
    mark with the thicker
    dashed line. 
    }\label{fig:dustywinds}
\end{figure*}


NGC~4941, NGC~7465 and
NGC~7314 have extended optical NLRs, which are 
detected with optical IFU observations and/or HST imaging
\citep[see][and Sousa et al. in preparation,
respectively]{ErrozFerrer2019, Ferruit2000}. In NGC~6814, the NLR and
the coronal line region are rather compact and there is evidence for ionized
outflows on these scales \citep[$\sim
1\arcsec$, see][]{MuellerSanchez2011}.
In these four galaxies we do
not detect extended mid-IR emission.

\section{Dependence of warm dust morphologies on the Eddington ratio and
  nuclear column density}\label{sec:columnEddington}

In the previous section, we showed that there is extended mid-IR
emission approximately along
the polar direction in six
out of the 12 GATOS Seyferts analyzed in this work. It is yet unknown
whether this polar dust emission is simply due to dust near the edges
of the  NLR being illuminated
by the AGN, part of a nuclear dusty outflow or  both. On the other hand,
another four galaxies in our
sample show NLR emission but no evidence for bright extended mid-IR
emission.
In this section we investigate if the nuclear column densities and
Eddington ratios in the galaxies in our
sample are likely to launch dusty outflows.

The gas and dust in the immediate surroundings of an active nucleus are subject to
the AGN radiation pressure. The idea of infrared radiation pressure
was first put forward by \cite{PierKrolik1992_IRradpressure} and developed further by
\cite{Krolik2007}, as an explanation to support the vertical height of the
torus. In the classical Eddington limit, only the
electron scattering is taken into account. Because the opacity of the
dust is much greater than the Thomson opacity, there is an 
effective Eddington ratio, which \cite{Krolik2007} estimated to be of the order
of 10\% the classical Eddington ratio.  In other words, in the
presence of dust, the effects
of the radiation pressure are {\it boosted}. Thus, at high Eddington
ratios, although not necessarily $\lambda_{\rm Edd} \ge 1$, and
moderate column densities, nuclear outflows may be present 
\citep[see][and references therein]{Fabian2008}. These outflows might even clear out the
gas from the nuclear region \citep[see also][]{Ricci2017Nature}.

\cite{Venanzi2020} demonstrated semi-analytically that IR-dominated winds can be
launched more efficiently when the acceleration due to the AGN
radiation pressure ($a_{\rm AGN}$) and the acceleration due to the 
gravitational force ($a_{\rm g}$) from the central black hole are
balanced \citep[see also][]{Tazaki2020}.
Their radiation dynamical simulations  for the typical Eddington ratios of Seyfert
galaxies  showed that IR-dominated
outflows would take place at moderate column densities of a 
few $N_H \sim
10^{23}\,{\rm cm}^{-2}$. At higher column densities of approximately  $N_H\ge
10^{24}\,{\rm cm}^{-2}$, they predicted that the lifting of the
dusty clumps is suppressed and this material settles in the disk
plane.

In Fig.~\ref{fig:dustywinds},  we show the hydrogen column densities or
upper limits 
derived toward the AGN position of the GATOS Seyferts against the
Eddington ratios.  We used the ALMA estimates of $N_{\rm H2}$
from GB21 (see also Table~\ref{tab:Sample}), which were derived using the 
CO(3-2) data and the 
canonical CO-to-H$_2$ conversion factor of the Milky Way. The  ALMA $N_{\rm
  H2}$  values correspond to physical
resolutions of $7-10\,$pc. To derive
$N_{\rm H}$ we assumed that the molecular gas phase is dominant 
on these scales \citep[see for instance the simulations
of][]{Wada2016}. We note that these values are not the measured column
densities of the individual clumps. However, they likely give an estimate of
the average properties of the clumps modulo the (unknown) filling
factor. In X-rays, the derived  $N_{\rm H}$ are only along the line of
sight.  We therefore assume that $N_{\rm H}$(ALMA) for each galaxy is  representative of
those clouds located in the dusty wind launching region.  
We also included in this figure NGC~1068 and Circinus, which
show clear polar mid-IR emission
\citep[see][respectively]{Cameron1993, Packham2005}.
 For the latter, the ALMA hydrogen column densities are from 
    \cite{GarciaBurillo2019} and \cite{Izumi2018}, respectively.

We also plot in this figure the predicted {\it blowout region}  (the area
below the solid line) derived by 
\cite{Fabian2008}, which is not populated by AGN. 
The area in this figure near the dashed line \citep[computed when  $a_{\rm AGN}/a_g\equiv 1$,
see][]{Venanzi2020}, shows the region where we would
expect to find AGN with  IR-dominated outflows.  As discussed by these authors, below
approximately $N_{\rm H}=10^{22}\,{\rm cm}^{-2}$ which is equivalent to an
optical depth in the near-IR below one, the driving of infrared
radiation is not effective.

We find that the GATOS Seyferts
with extended mid-IR emission in the  (projected) polar direction
(NGC~1365, NGC~3227, NGC~4388, and NGC~7582)
as well as  NGC~1068 and Circinus are located close to the region where polar dusty
outflows are more likely to be launched. Among
these galaxies, NGC~3227 and NGC~1068 present  evidence of 
molecular outflows  in CO transitions on the {\it torus} scales
\citep[$\sim 20-40\,$pc, see][]{AlonsoHerrero2019,
  GarciaBurillo2019}. The ALMA observations of Circinus show, on the
other hand, a nuclear outflow with a modest velocity
\citep{Izumi2018}. These authors predicted that a significant mass in
the wind will fall back to the disk. 
NGC~3227, NGC~4388  and  NGC~7582 as well as
NGC~1068  show evidence of molecular outflows  on 
 physical scales {\it larger than} the  torus \citep[up to a few
 hundred parsecs, see][and
 GB21]{Davies2014, GarciaBurillo2019, DominguezFernandez2020}, which are more
 likely due to the interaction between the AGN wind and/or a radio jet
 if present and molecular gas
 in the disk of the host galaxy.
 
NGC~5643 and NGC~5506 appear to
have extended both polar and equatorial mid-IR emission. In
NGC~6300 the mid-IR emission might be due to dust in the equatorial direction
of the torus as well as in the host galaxy. According to the
\cite{Venanzi2020} simulations, galaxies with relative high Eddington ratios
that lie to the left of the dashed line might be more conducive to having
equatorial outflows. This is the case 
for NGC~5643 \citep[see][]{AlonsoHerrero2018}. Moreover, these
equatorial outflows are not likely to be efficient (maybe not just yet) 
at clearing the nuclear regions of NGC~5643 and NGC~6300 (and others),
as shown observationally by GB21.

NGC~6814  and NGC~7314 are close to the
{\it blowout}  region, where polar dusty outflows should not be
prominent.   The CO(3-2) map of NGC~6814 of GB21 (their
Fig.~10) shows that there is little molecular gas in the nuclear and
circumnuclear regions of this galaxy, in agreement with being in
the {\it blowout} region and an unresolved mid-IR morphology. NGC~7314,
on the other hand, contains more molecular gas with two  CO peaks located
symmetrically around the AGN in the inner 40\,pc (Fig.~13 of
GB21). This might indicate that some
clearing of molecular gas at the AGN location already took place in the
nuclear regions of this galaxy. For this
galaxy, the mid-IR emission is mostly unresolved.
Finally the low nuclear column density and Eddington ratio of
NGC~ 4941 amd NGC~7213
place these galaxies in a region in Fig.~\ref{fig:dustywinds} where dusty
winds are not launched. This is in agreement with the unresolved
mid-IR morphology.

\begin{table*}
\caption{CAT3D-WIND model parameters used in this work.}             
\label{tab:modelparameters}      
\centering                          
\begin{tabular}{l c c c c c c c c c c c c}        
\hline\hline                 
  Name      &  Symbol & Range & NGC~7213, NGC~6814\\
\hline
Disk radial index & $a$ & [$-0.5$, $-2$] & $-1$\\ 
  Wind radial index & $a_w$ &  [$-0.5$, $-2$] & $-0.5$\\
  Wind-to-disk cloud ratio & $f_{\rm wd}$ & [0.3, 0.6, 0.9, 1.2] & 0.15\\
  Number of clouds along disk equator& $N_0$    & 10 & 5\\                 
  Inclination ($^\circ$) & $i$ & [30, 45, 60, 75] & [0, 15, 30]\\
Disk height  & $h$ & [0.1, 0.2, 0.3] & 0.1\\
Wind angular width ($^\circ$) & $\sigma_\theta$ & [5, 10, 15] & 10\\
  \hline
\end{tabular}
\tablefoot{The common parameters are a cone opening angle of
  $\theta_w=45^\circ$ and the optical depth of the clouds $\tau_v=50$.}

\end{table*}

In summary,  seven galaxies in our sample show both  Eddington
ratios and nuclear column densities favorable to launching polar and equatorial
dusty winds. The remaining have either low nuclear hydrogen column
densities and/or low Eddington ratios, and thus dusty winds might not
be likely.

\section{CAT3D-WIND mid- and far-IR model images}\label{sec:comparisonmodelCAT3D-WIND}
In Sect.~\ref{sec:comparisonALMA} we showed that there is a
diversity of nuclear (1-1.5\arcsec) mid-IR morphologies in
the GATOS Seyferts with the unresolved component dominating the
emission. At $870\,\mu$m, on the other hand, GB21 found that the extended
component is dominant on these scales.
In this section we use the radiative transfer model of a clumpy disk
and wind dubbed
CAT3D-WIND and presented in
\cite{Hoenig2017} to generate model images in the mid- and far-IR. The
goal is to explore the
 morphologies at these wavelengths for several disk and wind
 configurations and geometries as informed by fits to their IR
 emission. In Sect.~\ref{sec:comparisondatamodel} we then simulate 
 CAT3D-WIND model images with angular resolutions corresponding to
 those of our mid-IR and 
 ALMA observations. 

\subsection{Brief description of the models}\label{sec:modeldescription}

\begin{figure}
  \hspace{0.5cm}
  \includegraphics[width=8.cm]{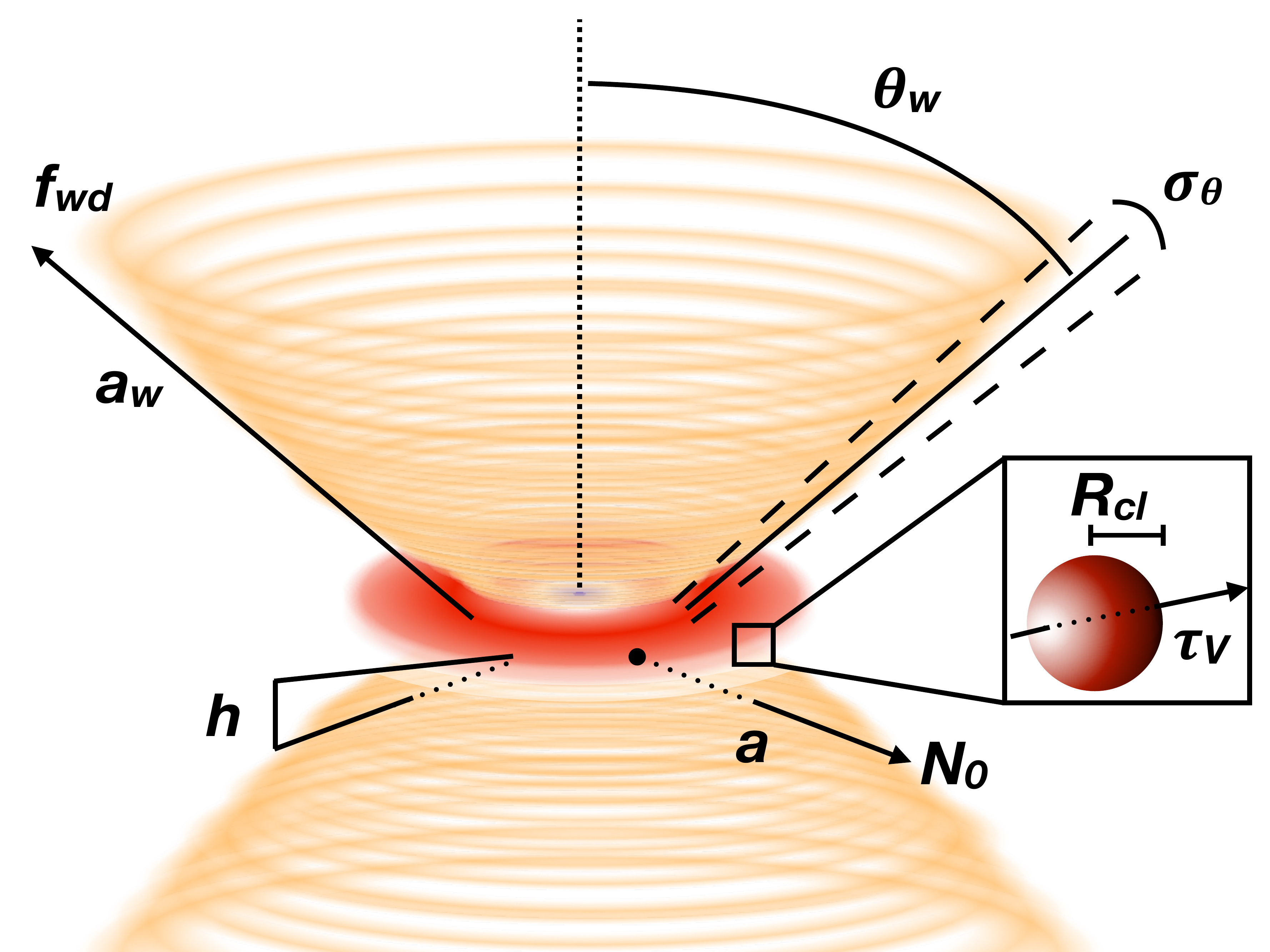}
  \caption{Cartoon showing the CAT3D-WIND geometry. In red is the
    clumpy disk
    component and in orange the dusty clumpy wind, which is represented as a hollow cone. The
    model parameters are also marked (see
    Sect.~\ref{sec:modeldescription} and Table~\ref{tab:modelparameters}).}\label{fig:modelcartoon}
\end{figure}

The CAT3D-WIND model  includes both a traditional dusty clumpy
disk (or geometrically thin torus), which is  based on that of 
\cite{Hoenig2010model}, and a dusty clumpy wind perpendicular to the
torus.  The latter is included to account  for the polar dust
emission detected in local AGN \citep[see, e.g.,][]{Asmus2016,
  Asmus2019}. The clouds have an optical depth of $\tau_V=50$  and
$N_0$ describes the number of clouds along the 
equatorial direction of the torus. The clouds are distributed
following a radial power law ($\propto r^a$)  with
indices $a$ for the disk and $a_w$ for the wind. The wind-to-disk
ratio $f_{\rm wd}$ is the number of clouds along the cone with respect to
$N_0$.  The vertical distribution of the disk follows a Gaussian
distribution $\propto
\exp^{-z^2/(2(hr)^2)}$ with a dimensionless scale height $h$ and the vertical distance from the
mid-plane $z$ in units of  the sublimation radius ($r_{\rm sub}$).
The wind is
modelled as a hollow cone with a half-opening angle of $\theta_w$ and
walls with an angular width of $\sigma_\theta$. Figure~\ref{fig:modelcartoon} shows
a sketch of the model and its parameters.

The inner radius is determined by the dust
$r_{\rm sub}$, which the model takes as $r_{\rm mod}=0.3\,$pc  at
$L_{\rm bol} = 10^{46}\,{\rm erg\,s}^{-1}$ and the scaling is then
$r_{\rm sub} = r_{\rm mod} \times (L_{\rm bol}/10^{46}\,{\rm
  erg\,s}^{-1})^{1/2}$ \citep[see][for more details]{Hoenig2010model}. $L_{\rm bol}$ is the accretion disk bolometric
luminosity computed as $L_{\rm bol} = 8 \times \nu L_\nu(V)$.

\subsection{Selection of model parameters and generation of model images}\label{sec:parameters}

The CAT3D-WIND model has a large  number of parameters to describe the
disk and the wind components \citep[see table~1 of ][]{Hoenig2017}. Since we are interested in comparing the
observed mid- and far-IR observations of the GATOS Seyferts with
CAT3D-WIND model images, in this work we 
concentrate on some of the main parameters that define the geometry
and thus are more likely to control the morphologies.
These are the
indices of the radial distribution of
the clouds in the disk and the wind ($a$ and $a_w$), the wind-to-disk ratio of
clouds ($f_{\rm wd}$), as well as the inclination. As we shall see, the
disk height and angular width of the cone walls  play an important
role on the self-obscuration of the disk and wind
in the mid-IR. 

 \begin{figure}
 \includegraphics[width=9cm]{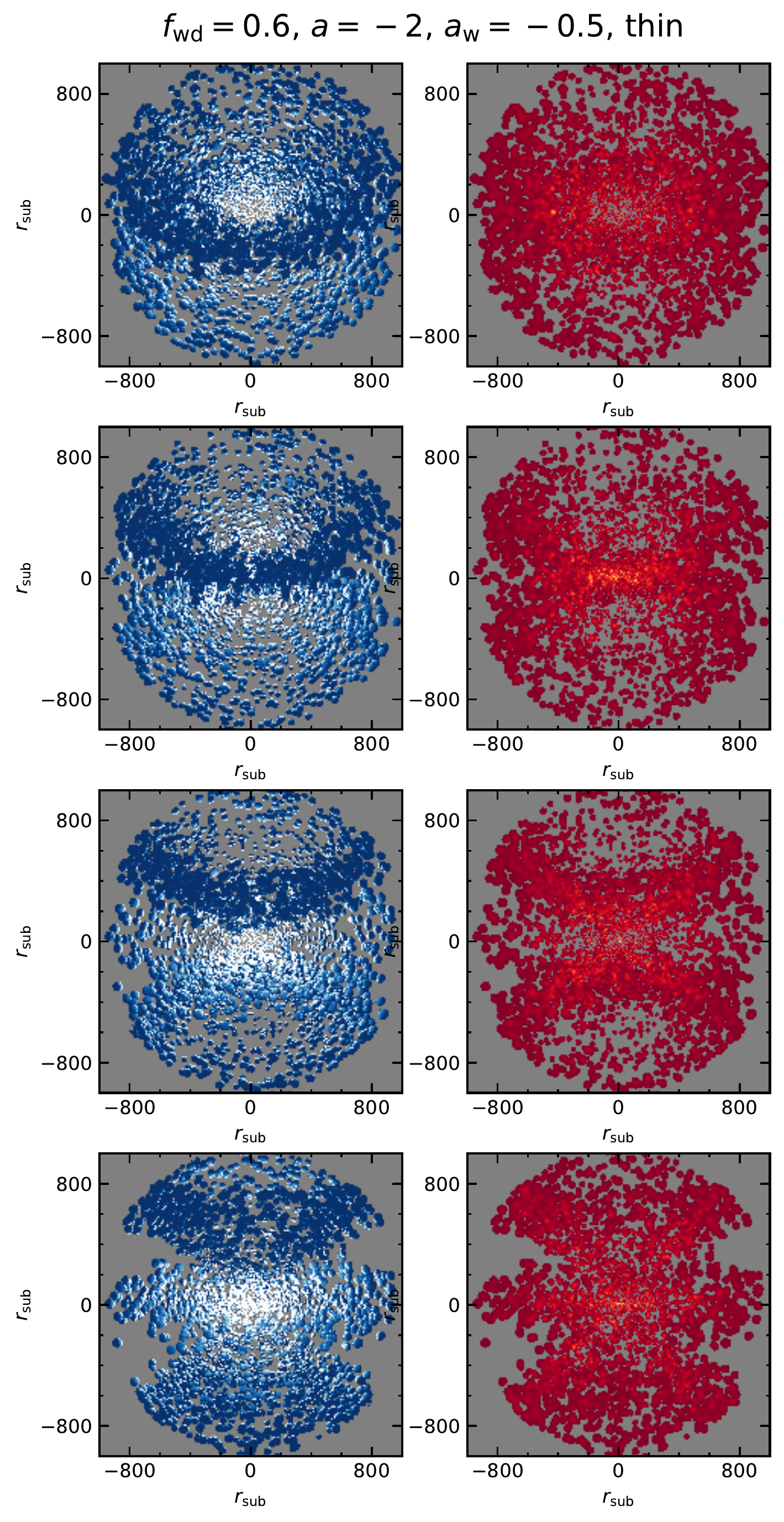}
  \caption{CAT3D-WIND model images with a ``thin'' ($h=0.1$  and
    $\sigma_\theta =5^\circ$) geometry and $f_{\rm wd}=0.6$. We show
    the $8\,\mu$m (in blue colors, left panels) and
    $700\,\mu$m (in orange colors, right panels) emission for a compact disk ($a=-2$) and
    an extended wind ($a_w=-0.5$). The resolution of the model images is
    $10\,r_{\rm sub}$ per pixel. From top to bottom the inclinations
are   $i=30 ^\circ$ (nearly face-on disk), 45$^\circ$, 60$^\circ$, and $75^\circ$ (nearly edge-on disk). 
The rest of parameters are listed in
Table~\ref{tab:modelparameters}. The color scales are such that dark
blue and dark orange  indicate fainter
emission, white/yellow is brighter emission and grey means no emission. }
              \label{fig:CAT3D-WINDcompactdiskextendedwind_thin}%
    \end{figure}

    \begin{figure}[!ht]
    \includegraphics[width=9cm]{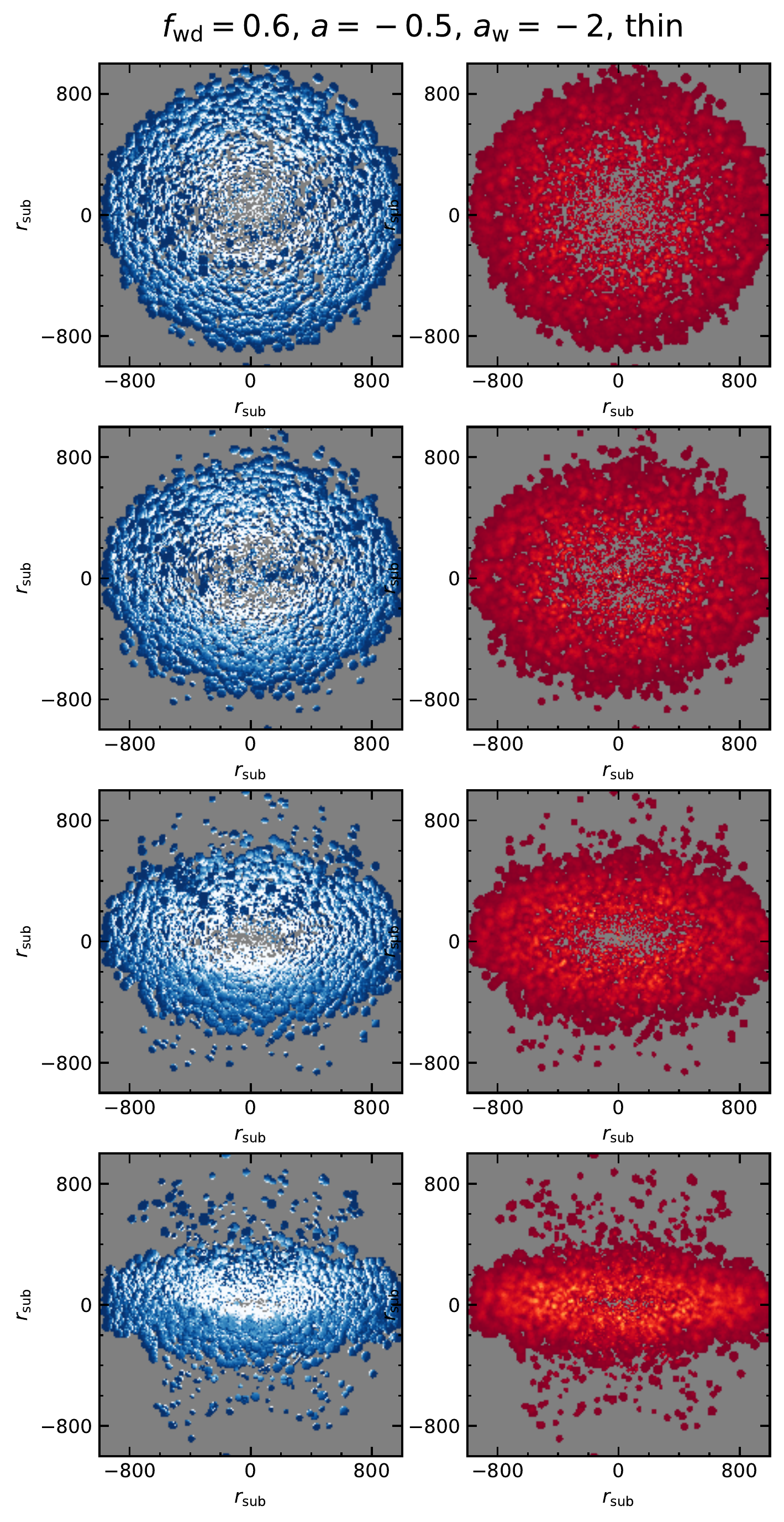}
  \caption{CAT3D-WIND model images for a ``thin'' geometry and $f_{\rm wd}=0.6$ at $8\,\mu$m (in blue colors, left panels) and
    $700\,\mu$m (in orange colors, right panels) for an extended disk ($a=-0.5$) and
    a compact wind ($a_w=-2$). From top to bottom the inclinations
are   $30 ^\circ$ (nearly face-on disk), $45 ^\circ$, $60 ^\circ$, and $75^\circ$ (nearly edge-on disk). 
 }
              \label{fig:CAT3D-WINDextendeddiskcompactwind_thin}%
    \end{figure}

\cite{GonzalezMartin2019} fit  {\it Spitzer}/IRS spectroscopy
of a large sample of local AGN using a variety of torus models, including
CAT3D-WIND. They found that CAT3D-WIND provided a good fit for
approximately half of their sample. In particular, they reproduced
better the mid-IR spectra of type 1 AGN than type 2 as well as the
spectra of the more luminous AGN in their sample. Eight of the GATOS Seyferts
are in their sample. Rather than focusing on the individual
fits, we based our
choice of the CAT3D-WIND parameters  on the  derived ranges (see
Table~\ref{tab:modelparameters} and below) for these
Seyferts, regardless of whether we detected a 
bright extended mid-IR component or not. 
We included the following
combinations of cloud radial distributions:
\begin{itemize}
\item compact disk - extended wind, $a=-2$ and $a_w = -0.5$
\item extended disk - compact wind, $a=-0.5$ and $a_w = -2$
\item  compact disk - compact wind, $a=-2$ and $a_w =-2$
\item extended disk - extended  wind, $a=-0.5$ and $a_w = -0.5$
\end{itemize}

We used four values of the wind-to-disk ratio of clouds: $f_{\rm wd} =0.3, 0.6,
0.9, 1.2$ and four values of the inclination: $i=30^\circ$ (nearly face-on
view of the disk), $45, 60$ and $75^\circ$ (nearly edge-on view of the disk). 
We fixed the following parameters, $N_0=10$ and the half-opening angle of the
wind $\theta_w=45^\circ$. We took ranges of the  
angular width of the wind $\sigma_\theta=5, 10, 15^\circ$ and  height of
the disk $h=0.1, 0.2, 0.3$. We distributed the dust over a region of  $2000\,r_{\rm sub} \times 2000\,r_{\rm
  sub}$ in size. This corresponds to typical physical sizes of tens of parsecs
for our sample (see Sect.~\ref{sec:results}), which are needed to
reproduce the relatively large ALMA tori.

We additionally generated CAT3D-WIND model images for the specific cases
of NGC~7213 and NGC~6814,  for which \cite{GonzalezMartin2019}
derived some
model parameters outside the range found for the rest of the galaxies
in our sample. In particular, they fitted
relatively low values of the number
of clouds along the disk equator, the fraction of wind-to-disk
cloud ratio, and 
inclination. The low
$N_0$ fitted
agree with the relatively little amounts of nuclear cold molecular gas \citep[see][and
GB21, see also Table~\ref{tab:Sample}]{AlonsoHerrero2020} detected in
these two galaxies. Table~\ref{tab:modelparameters} summarizes the CAT3D-WIND parameters
selected to generate model images  in this work.



\begin{figure}
  
  \includegraphics[width=9cm]{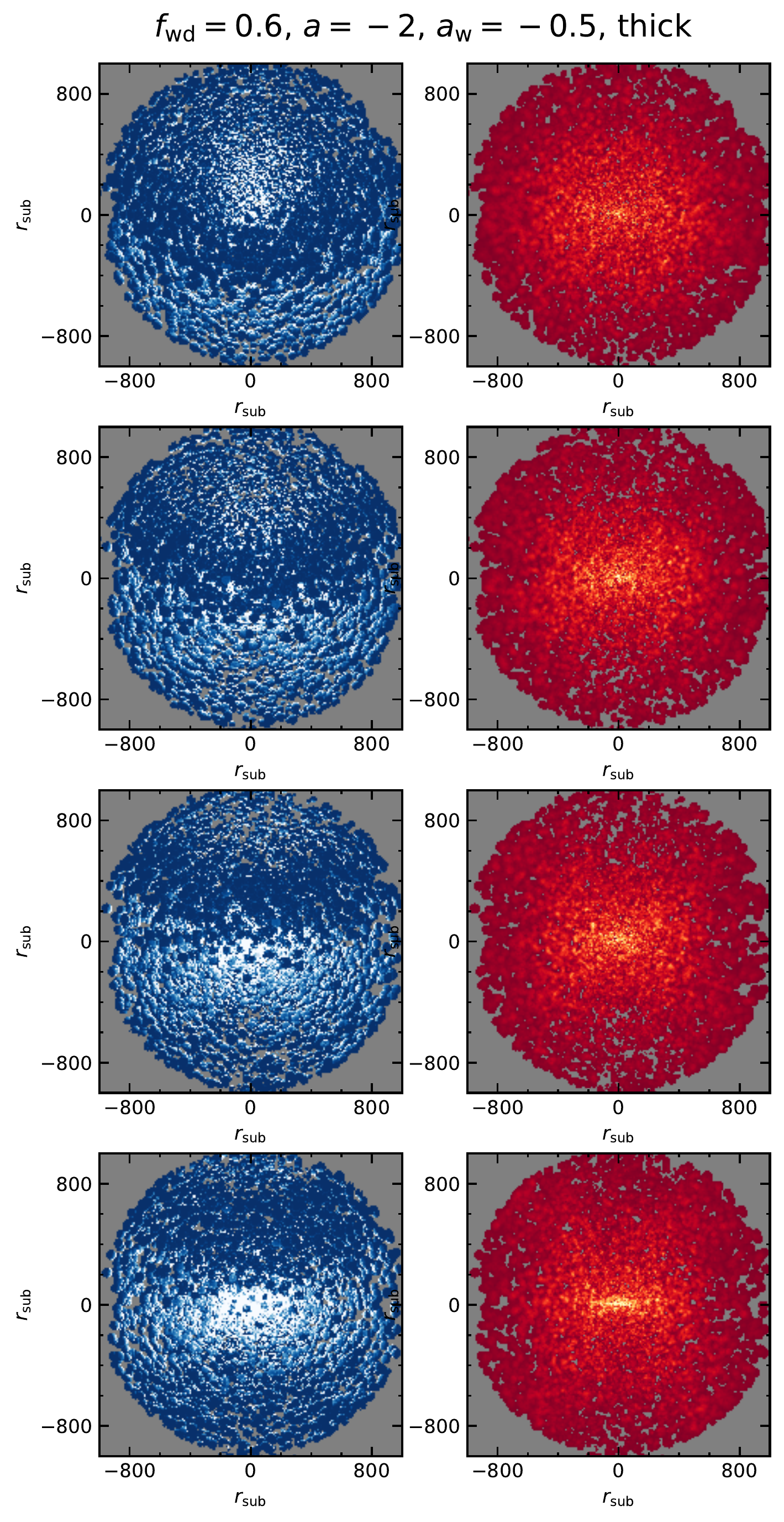}
  \caption{CAT3D-WIND model images with a ``thick'' geometry  ($h=0.3$
    and $\sigma_\theta=15^\circ$) and $f_{\rm wd}=0.6$. We show the $8\,\mu$m (in
    blue colors, left panels) and 
    $700\,\mu$m emission  (in orange colors, right panels) for a compact disk ($a=-2$) and
    an extended wind ($a_w=-0.5$). From top to bottom the inclinations
are   $i=30^\circ$ (nearly face-on disk), $45 ^\circ$, $60 ^\circ$, and $75^\circ$ (nearly edge-on disk). 
The rest of parameters are listed in the notes of
Table~\ref{tab:modelparameters}. The color scales are as in
Fig.~\ref{fig:CAT3D-WINDcompactdiskextendedwind_thin}. }
              \label{fig:CAT3D-WINDcompactdiskextendedwind}%
    \end{figure}

    \begin{figure}[!ht]
    \includegraphics[width=9cm]{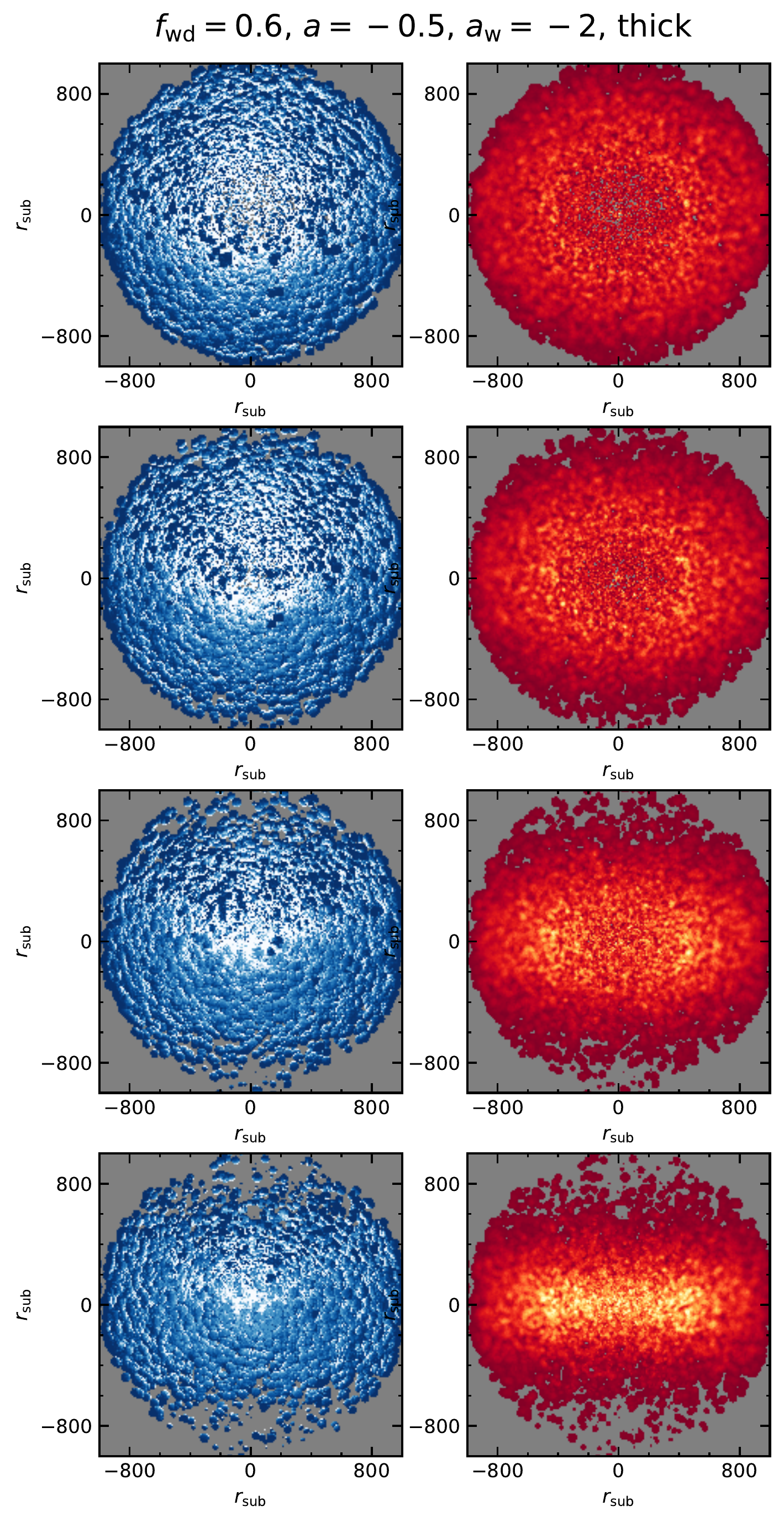}
  \caption{CAT3D-WIND model images  for a ``thick'' geometry and
    $f_{\rm wd}=0.6$ at $8\,\mu$m (in blue colors, left panels) and
    $700\,\mu$m (in orange colors, right panels) for an extended disk ($a=-0.5$) and
    a compact wind ($a_w=-2$). From top to bottom the inclinations
are   $30 ^\circ$ (nearly face-on disk), $45 ^\circ$, $60 ^\circ$, and $75^\circ$ (nearly edge-on disk). 
 }
              \label{fig:CAT3D-WINDextendeddiskcompactwind}%
    \end{figure}

\begin{figure}[!ht]
    \includegraphics[width=9cm]{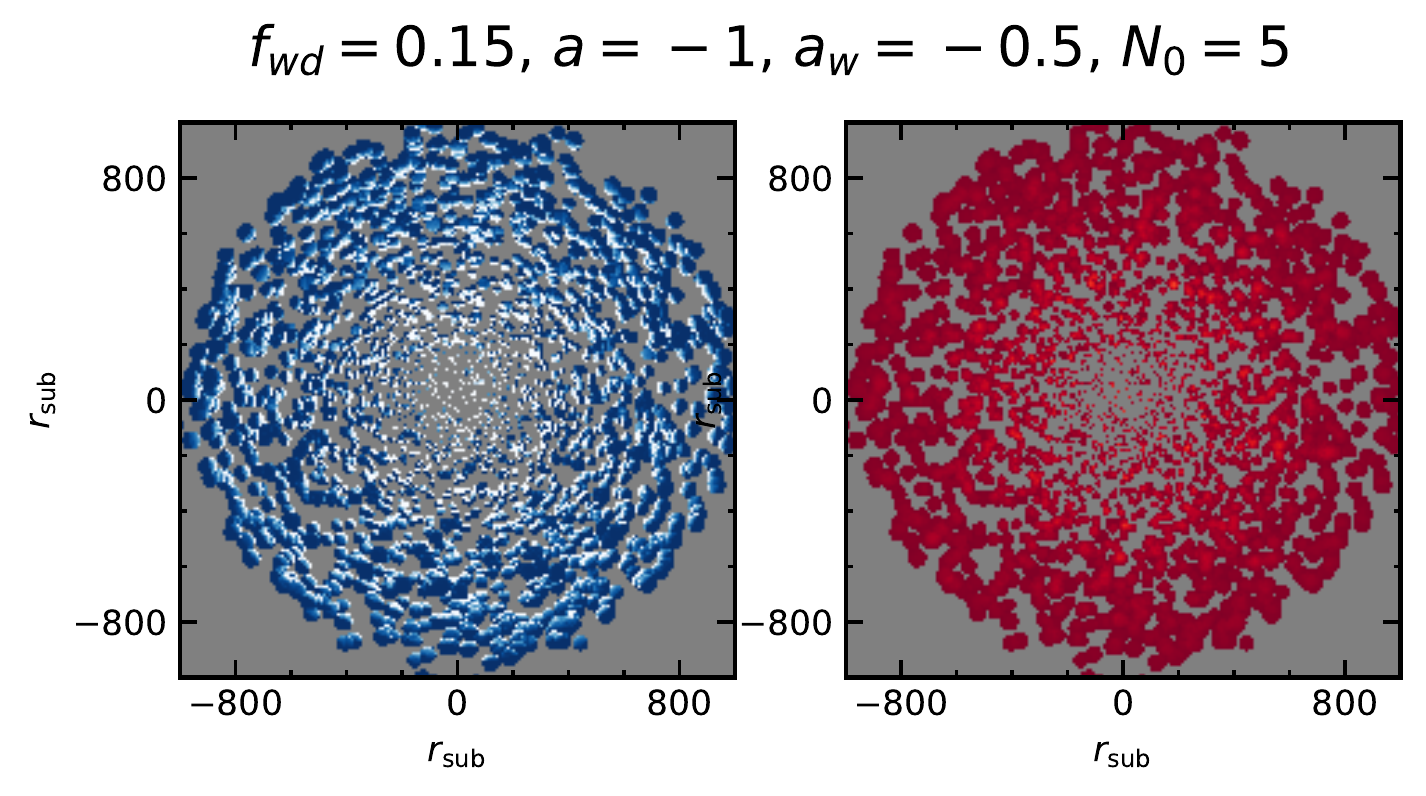}
  \caption{CAT3D-WIND model images at $8\,\mu$m (in blue colors, left panel) and
    $700\,\mu$m (in orange colors, right panel) for the NGC~7213 and
    NGC~6814 parameters listed
    in the fourth column of Table~\ref{tab:modelparameters} and an inclination of $i=15 ^\circ$.  }
              \label{fig:CAT3D-WINDNGC7213}
    \end{figure}

  We generated model monochromatic images at two mid-IR wavelengths, namely, $\lambda =8$ and
$\lambda =12\,\mu$m to encompass approximately those of the central
wavelengths of the 
observation filters. We used as a far-IR wavelength  $\lambda =
700\,\mu$m, which is  one  of the wavelengths of the model standard grid
(the next one is $1000\,\mu$m). 
Taking into account the model parameters (see Table~\ref{tab:modelparameters})  and the three wavelengths, 
we generated a total of 1728 + 9 model images\footnote{The
    CAT3D-WIND model
  images and corresponding SEDs can be
  found at http://cat3d.sungrazer.org.}.
We resampled the images
from the model  natural resolution  of $2\,r_{\rm sub}$ to 
 a pixel size of $10\,r_{\rm sub}$. As we shall see in
 Sect.~\ref{sec:results}, this resolution is higher than that
 achieved with the current ground-based
 mid-IR imaging and ALMA angular resolutions.
We also note
that the $8$ and $12\,\mu$m model images are similar and thus, we only
show in the figures an discuss in the following sections
the former.

\subsection{``Thin'' geometries}\label{sec:CAT3DWINDmodelimages_THIN}
We start by presenting the model images for a ``thin'' geometry, that
is, a disk height $h=0.1$ and  cone walls with 
$\sigma_\theta=5^\circ$. This geometry suffers less from
self-obscuration
effects. Figures~\ref{fig:CAT3D-WINDcompactdiskextendedwind_thin} and \ref{fig:CAT3D-WINDextendeddiskcompactwind_thin}
show the mid- and far-IR images for the compact disk- extended wind
and extended disk - compact wind combinations, at the four selected inclinations and for an
intermediate value of the wind-to-disk ratio, $f_{\rm wd}=0.6$ (see
Appendix for images with $f_{\rm wd}=1.2)$. In 
Figs.~\ref{fig:CAT3D-WINDcompactdiskcompactwind_thin} and 
\ref{fig:CAT3D-WINDextendeddiskextendedwind_thin} in the Appendix we
show the compact disk - compact wind and extended wind - extended disk
configurations. 
In all
the figures, we
plotted the images in a linear 
scale and with the same scaling values for all geometries and inclinations. 

The wind component is clearly observed at intermediate-to-high inclinations and is
more prominent, as expected, when the wind is in an extended
configuration (Figs.~\ref{fig:CAT3D-WINDcompactdiskextendedwind_thin}
and \ref{fig:CAT3D-WINDextendeddiskextendedwind_thin}).
Moreover, the characteristic
``X''-shape of the wind is most notable in the far-IR. At this wavelength the
dust is optically thin. The disk
and wind can have comparable far-IR 
brightnesses  in the compact disk - extended wind
configuration at relatively high inclinations ($i=60^\circ$ and
$75^\circ$, two bottom panels of
Fig.~\ref{fig:CAT3D-WINDcompactdiskextendedwind_thin}).
The disk component, on the other hand, is brighter 
than the wind in
the far-IR for the compact wind configurations
(see Figs.~\ref{fig:CAT3D-WINDextendeddiskcompactwind_thin} and
\ref{fig:CAT3D-WINDcompactdiskcompactwind_thin}). 
In the extended disk configurations
(Figs.~\ref{fig:CAT3D-WINDextendeddiskcompactwind_thin} and \ref{fig:CAT3D-WINDextendeddiskextendedwind_thin}),
the far-IR emission of the disk shows a ring-like morphology at all inclinations
whereas in  the compact disk configurations
(Figs. ~\ref{fig:CAT3D-WINDcompactdiskextendedwind_thin} and
\ref{fig:CAT3D-WINDcompactdiskcompactwind_thin})  the far-IR emission
peaks  at the central position.

In the mid-IR,
two-sided polar wind emission is also present at $i=45^\circ$
for the compact disk and extended wind configuration (see second panel
from the top in
Fig.~\ref{fig:CAT3D-WINDcompactdiskextendedwind_thin}). However, even
for this ``thin'' geometry, it suffers moderately from 
self-obscuration due to the cone walls. Clearer two-sided polar 
morphologies in the mid-IR are seen in other geometries with the
CAT3D-WIND models \citep[see figure~1 of][]{Hoenig2017}  and other
models \citep[see for instance,][]{Schartmann2014, Gallagher2015, Stalevski2017}. We note that even at $i=75^\circ$
the disk component is quite prominent in the mid-IR in our ``thin''
geometry.
For the wind-to-disk ratio considered in this section ($f_{\rm wd}=0.6$) and
the majority of  
the disk-wind configurations and inclinations,
the mid-IR emission comes  mostly
from  the inner part of the disk/cone. For higher values of $f_{\rm
  wd}$,
bi-conical and one-sided polar emission
morphologies in the mid-IR become more apparent (see
Figs.~\ref{fig:CAT3D-WINDcompactdiskextendedwind_fwd1.2} and
\ref{fig:CAT3D-WINDextendeddiskextendedwind_fwd1.2} in
the Appendix).
At lower inclinations, part of the
mid-IR emission arises from the far side of the cone at radial
distances close to the heating source. It is also possible to observe
both bright mid- and far-IR emission along the equatorial direction of the disk
in the extended disk - compact wind configuration at high inclination
(see the bottom panels of
Fig.~\ref{fig:CAT3D-WINDextendeddiskcompactwind_thin}).

\subsection{``Thick'' geometries}\label{sec:CAT3DWINDmodelimages_THICK}

Figures~\ref{fig:CAT3D-WINDcompactdiskextendedwind} and
\ref{fig:CAT3D-WINDextendeddiskcompactwind}, and
Figs.~\ref{fig:CAT3D-WINDcompactdiskcompactwind} and 
\ref{fig:CAT3D-WINDextendeddiskextendedwind} in the Appendix
present the images for  ``thick'' geometries, that is, 
a disk height of
 $h=0.3$ and cone walls with $\sigma_\theta=15^\circ$, and $f_{\rm wd}=0.6$. Note that this
 ``thick'' geometry does not mean a geometrically thick torus. None of the generated model images 
produces clear {\it bi-conical} dust emission morphologies in the
mid-IR and  at high inclinations ($i=75^\circ$). Our value of $\sigma_\theta=15^\circ$ produces
relatively  {\it thick} cone edges that obscure the near side of the cone. A
similar effect also takes place for large values of $f_{\rm wd}$ when the
number of clouds in the cone line-of-sight becomes too large.

At low and intermediate inclinations ($i=30, 45^\circ$) there is one-sided polar dust emission in the
mid-IR in all geometries (Figs.~\ref{fig:CAT3D-WINDcompactdiskextendedwind},
\ref{fig:CAT3D-WINDextendeddiskcompactwind}  and
\ref{fig:CAT3D-WINDextendeddiskextendedwind}) except for the compact disk -
wind configuration. This mid-IR 
polar dust emission is only seen projected on the far side of the
disk and is due to self-obscuration of the near side of
the cone and disk. This morphology appears in all the values of the wind-to-disk
ratio  explored in this work. Torus models
without the wind component can also produce this one-sided polar
emission at intermediate inclinations and relatively large torus angular widths
\citep[see, e.g.,][]{Schartmann2008,
  Siebenmorgen2015, LopezRodriguez2018, Nikutta2021}.
The details of the mid-IR morphology in the model images
also depend on how obscuring the cone
edges are (that is, the thickness of the cone walls).

In the two extended disk configurations (Figs.~\ref{fig:CAT3D-WINDextendeddiskcompactwind}  and
\ref{fig:CAT3D-WINDextendeddiskextendedwind}), the mid- and far-IR
emissions have complementary morphologies. At low and intermediate
inclinations  ($i=30-60^\circ$) most of
the mid-IR emission comes from the inner regions of the disk and the
edges of the cone, whereas the far-IR emission is produced in
the outer regions with an apparent  ring-like morphology. At
high inclinations most of the far-IR emission 
comes from the disk that shows  a {\it puffed-up} morphology due not
only to the thicker disk but also to the additional
contribution from cold dust in the wind.

There is also polar dust emission in the far-IR with the
characteristic ``X''-shape  at $i=60$ and $i=75^\circ$
for the extended wind configurations
(Figs. ~\ref{fig:CAT3D-WINDcompactdiskextendedwind} and
~\ref{fig:CAT3D-WINDextendeddiskextendedwind}).  The constrast of the
``X''-shape is however lower than in the ``thin'' geometry.
Nevertheless, in the model with the steep cloud distribution
profile in the disk (compact
disk), this shape is seen more clearly. This is because in this
configuration there are not too many clouds left at larger distances,
so relatively more far-IR emission is coming from the cone
edges.

The compact disk - wind configuration produces relatively similar
morphologies in the mid- and far-IR (see
Fig.~\ref{fig:CAT3D-WINDcompactdiskcompactwind}), with a more roundish
shape at lower inclinations and more elongated at higher
inclinations, as expected. The mid-IR morphologies of this configuration are
similar to those seen in clumpy torus models without a wind component
but with similar steep radial dust
distributions \citep[see][]{Schartmann2008}. 

\subsection{The NGC~6814 and NGC~7213 geometry}\label{sec:CAT3DWINDmodelimages_NGC7213}
Figure~\ref{fig:CAT3D-WINDNGC7213} shows the specific model mid- and
far-IR images for NGC~6814 and
NGC~7213 at $i=15^\circ$.  In the mid-IR a large fraction of the emission comes from
the inner regions of the disk since the radial distribution of the clouds in the disk
is moderately steep ($a=-1$). Moreover, because in this model the wind component
is not prominent the effects of self-obscuration from the cone walls
appear to be small. In the far-IR, the emission is relatively faint
and diffuse due to the low number of clouds along the equatorial
direction ($N_0 = 5$), and it comes from the outer part of the disk
and shows a nearly ring-like morphology. 


\section{Comparison between CAT3D-WIND model images and
  observations}\label{sec:comparisondatamodel}

The good angular resolutions achieved with ground-based mid-IR
instruments on 8-10m class telescopes and ALMA  resolve extended
components in the mid and far-IR in nearby Seyferts (see
Sect.~\ref{sec:comparisonALMA} and references cited there). In this
section we adapt the resolutions of the CAT3D-WIND model images to
those of  our data,  and make a qualitative comparison between models and observations. We refer the reader to \cite{LopezRodriguez2018} and
\cite{Nikutta2021} for a
similar comparison for NGC~1068 using the \cite{Nenkova2008II} model
images produced with HyperCAT.

\subsection{Summary of torus and wind properties of the sample}
The CAT3D-WIND models produce a diversity of mid and far-IR morphologies
(Sects.~\ref{sec:CAT3DWINDmodelimages_THIN},
\ref{sec:CAT3DWINDmodelimages_THICK} and
\ref{sec:CAT3DWINDmodelimages_NGC7213}) depending on  
our viewing angle (inclination) as well as the radial
distributions assumed for the disk and wind components and the
thickness of the disk and cone walls.
In this section we compile values
of the torus inclination and sizes as well as the wind properties derived from
different types of modelling to aid the comparison between models and
observations in Sect.~\ref{sec:results}.

For the torus/disk component, we list in Table~\ref{tab:Torus} the
estimated inclinations using fits to the  high angular resolution  IR
SEDs and mid-IR spectroscopy using the {\sc
  clumpy}  models \citep{Nenkova2008I, Nenkova2008II} and  fits to the
{\it Spitzer}/IRS spectroscopy with
the CAT3D-WIND torus models \citep{Hoenig2017}. 
We also included the inclination values derived from fits to the X-ray emission
using the {\sc
  xclumpy} model \citep{Tanimoto2019, Tanimoto2020,
  Ogawa2021}. Additionally, there are 
lower limits to the torus 
inclination based on the ellipticity of the ALMA extended $870\,\mu$m
continuum images from GB21. 
We also provided in this table  estimates of the nuclear
disk/torus inclination based on the modelling of  the ALMA CO(3-2) and CO(2-1)
kinematics, and the sizes and PA of the  extended $870\,\mu$m sizes
from GB21.

We additionally summarized in Table~\ref{tab:Torus}
some information about the
wind/NLR/cones components. In the  simplest scenario these
components are perpendicular to
the nuclear disk/torus. We included the PA of the NLR (see Sect.~\ref{sec:NLR}
for references), as well as the
inclinations and PA of the cones  derived
from the modeling of the NLR kinematics  by \cite{Fischer2013} for the
galaxies in common with our work. The constraints on $f_{\rm wd}$ are
from \cite{GonzalezMartin2019}.

\subsection{Fiducial model images for GATOS Seyferts}\label{sec:results}

To match the angular resolutions of the models
to those of the observations, we need estimates of $r_{\rm sub}$.
For NGC~1365, recent GRAVITY  $K$-band
interferometry derived  
a radius of 0.035\,pc (scaled to our distance) for the hot dust
emission \citep{Dexter2020}.
Using the relationship between the absorption-corrected $2-10\,$keV luminosity
and $r_{\rm sub}$ 
\citep{Jensen2017},  we obtained values ranging from  $r_{\rm sub}=0.01\,$pc
for NGC~4941 to $r_{\rm sub}=0.07\,$pc for NGC~7582. We can derive
another estimate using the radius-luminosity relation seen from
GRAVITY observations  \citep{Dexter2020}, which for the typical bolometric
luminosities of our sample provides $r_{\rm sub} = 0.04\,$pc. We take
this as a representative value for our sample. For NGC~7213 we use
the value of $r_{\rm sub} = 0.029\,$pc (scaled to our distance) from 
\cite{Hoenig2010data}
based on $K$-band reverberation mapping from \cite{Kishimoto2007}.

    \begin{figure}[!ht]
    \includegraphics[width=9cm]{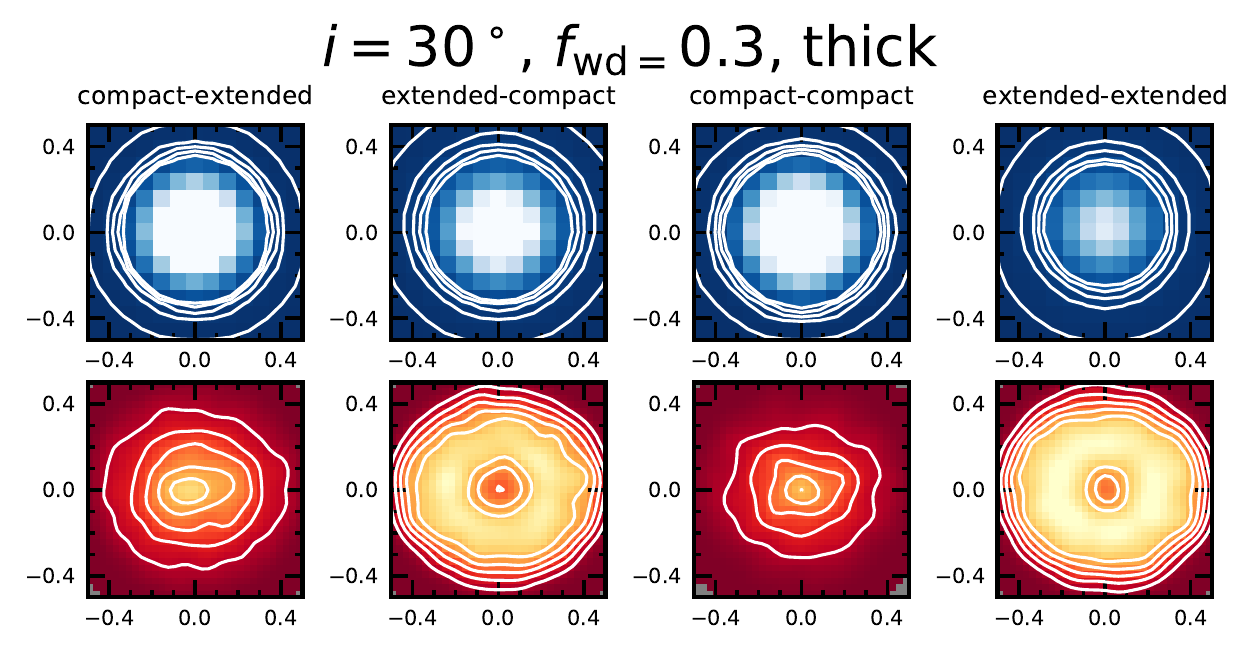}
    \includegraphics[width=9cm]{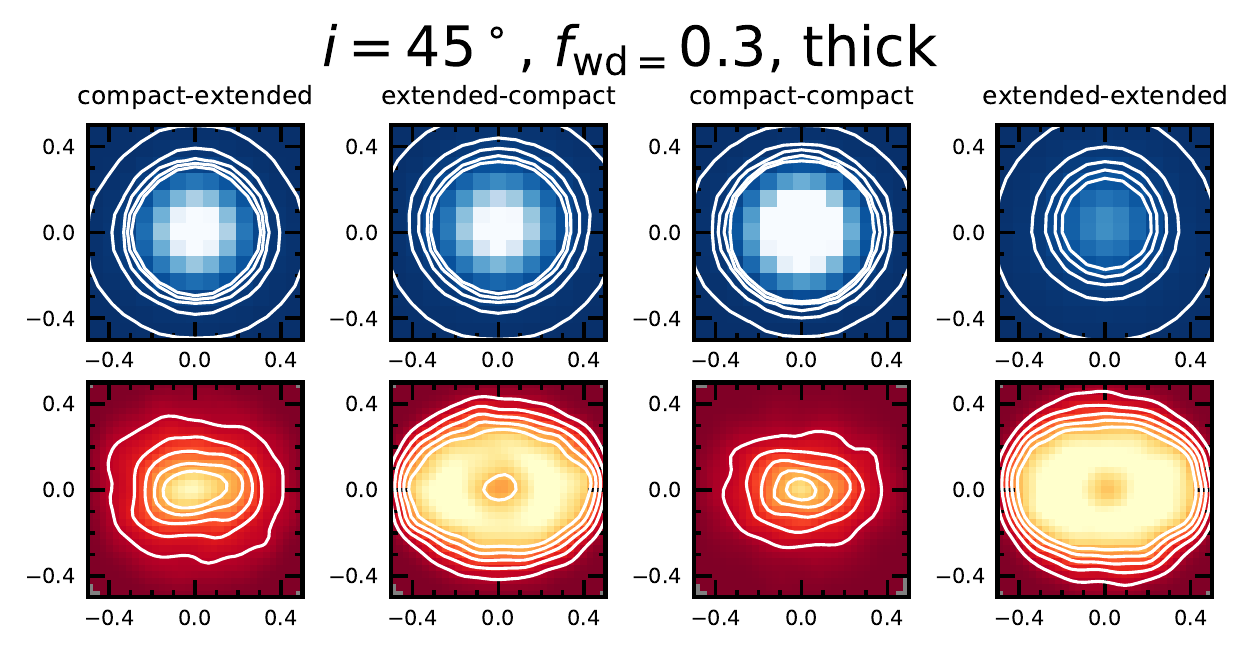}
    \includegraphics[width=9cm]{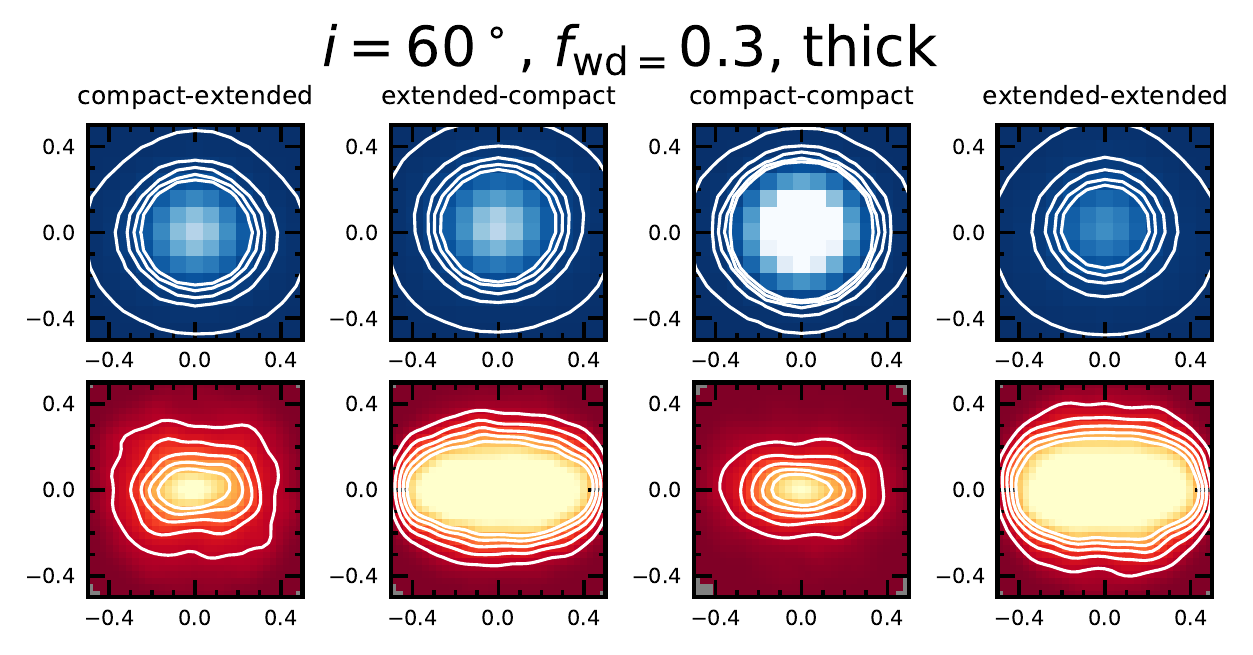}
    \includegraphics[width=9cm]{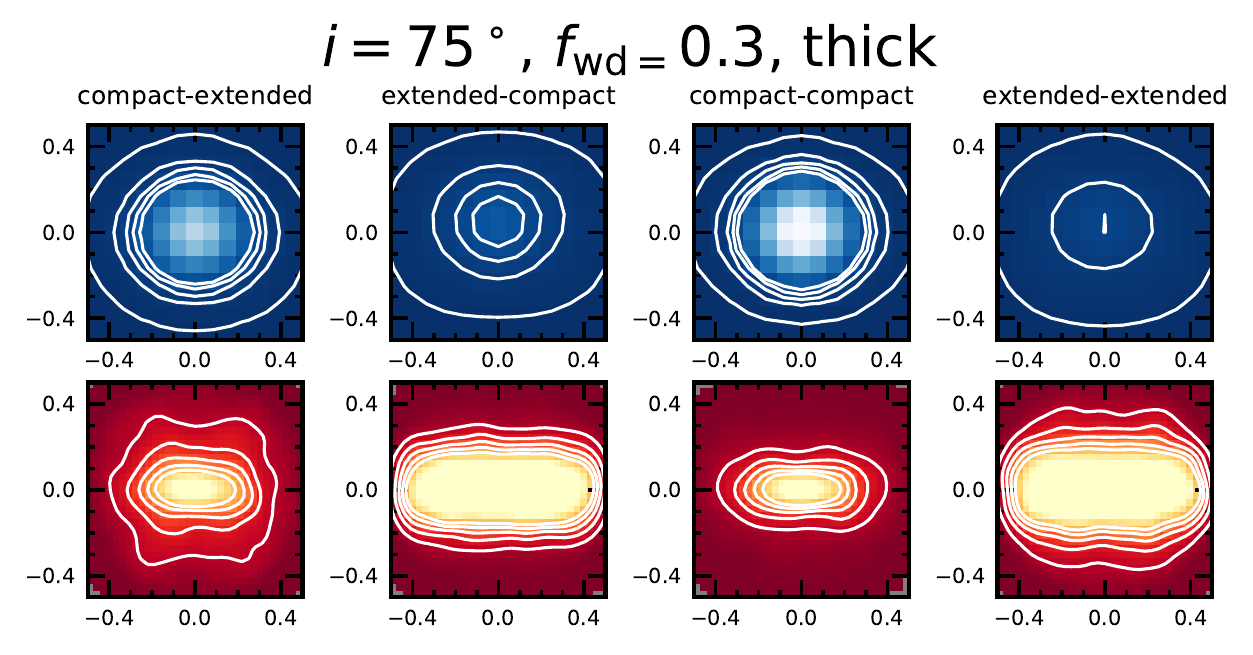}
    \caption{Simulated CAT3D-WIND images  matching approximately the
      angular resolutions of the observations of the GATOS Seyferts, namely $0.3\arcsec$
      in the mid-IR and $0.1\arcsec$ in the far-IR.  The images cover
      $1\arcsec \times 1\arcsec$ or $80\,{\rm pc} \times 80\,{\rm pc}$
      for the fiducial simulations. The models have a
      ``thick'' geometry and a mild wind with a wind-to-disk
    ratio of $f_{\rm wd}=0.3$. For each of the adopted inclinations, the top panels are
    the simulated  model images at $8\,\mu$m (in blue colors) and
the bottom panels at    $700\,\mu$m (in orange colors). The vertical
panels are the different disk-wind configurations discussed in
Sects.~\ref{sec:CAT3DWINDmodelimages_THIN} and
\ref{sec:CAT3DWINDmodelimages_THICK}. At a given wavelength, the image
brightness scale 
and the contour levels are the same in all the panels.  }
              \label{fig:simulated_fwd0.3_thick}
    \end{figure}

    \begin{figure}[!ht]
    \includegraphics[width=9cm]{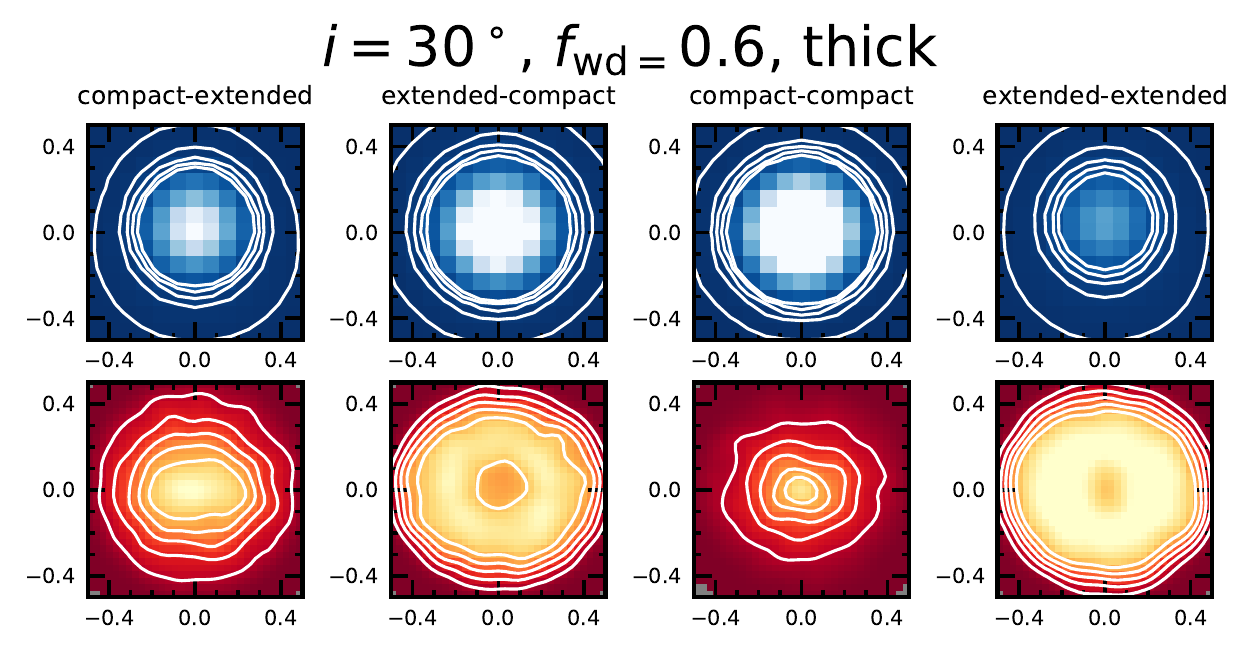}
    \includegraphics[width=9cm]{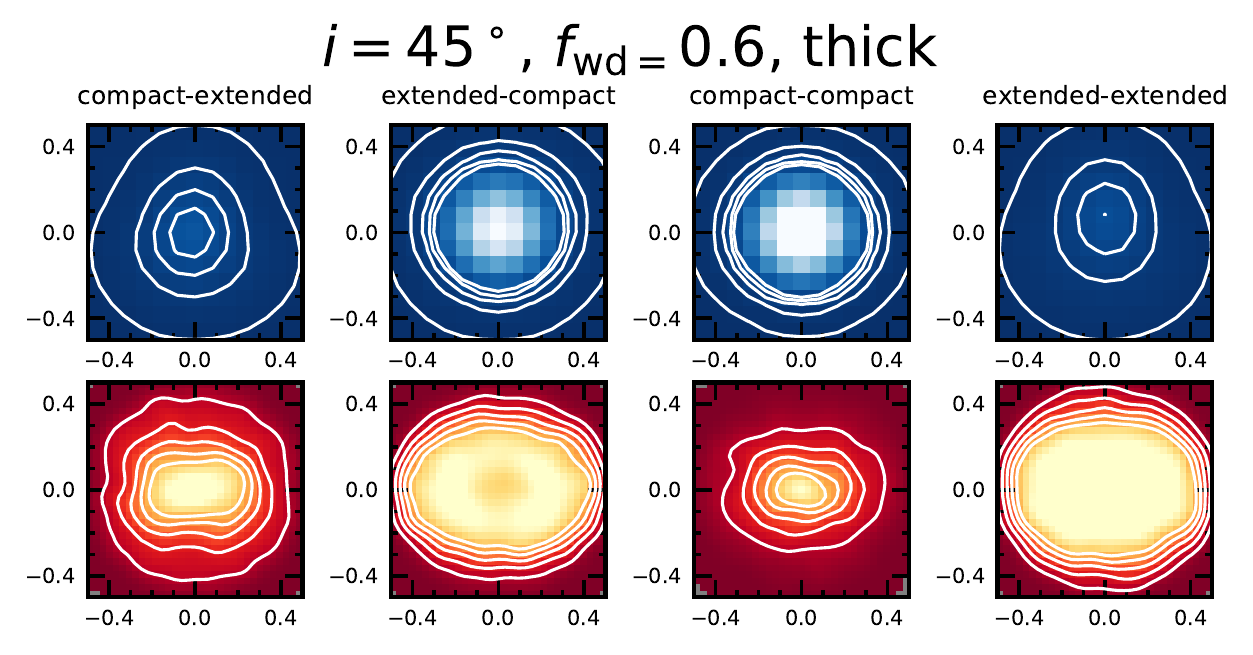}
    \includegraphics[width=9cm]{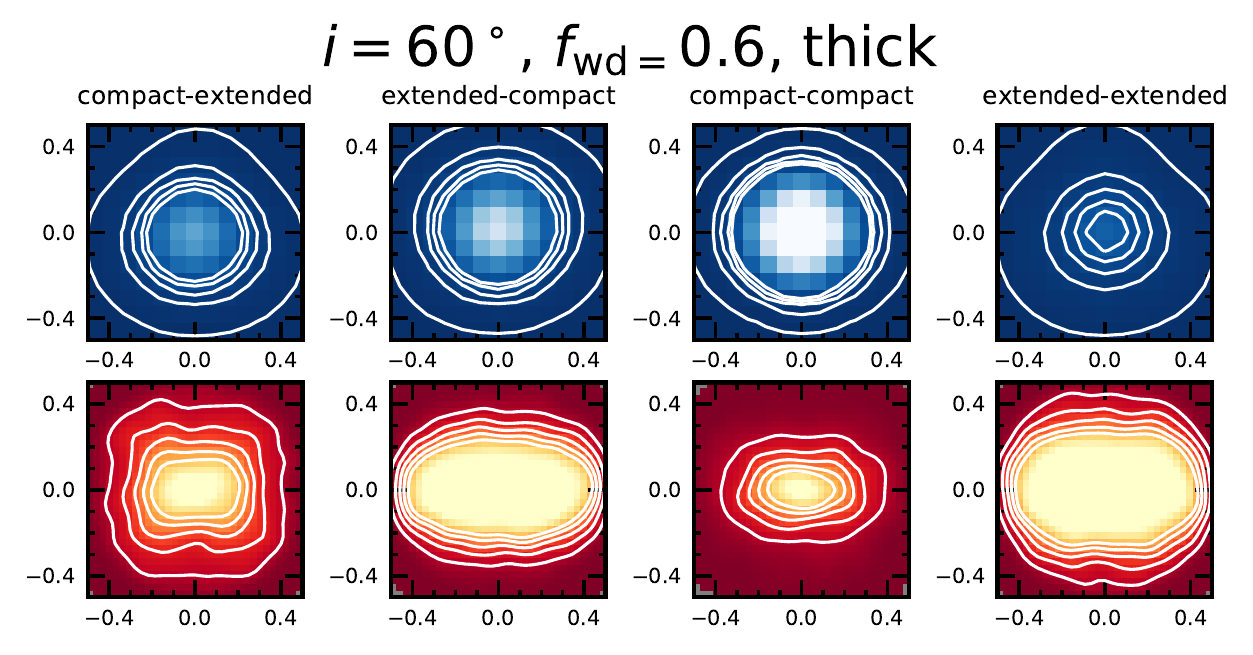}
    \includegraphics[width=9cm]{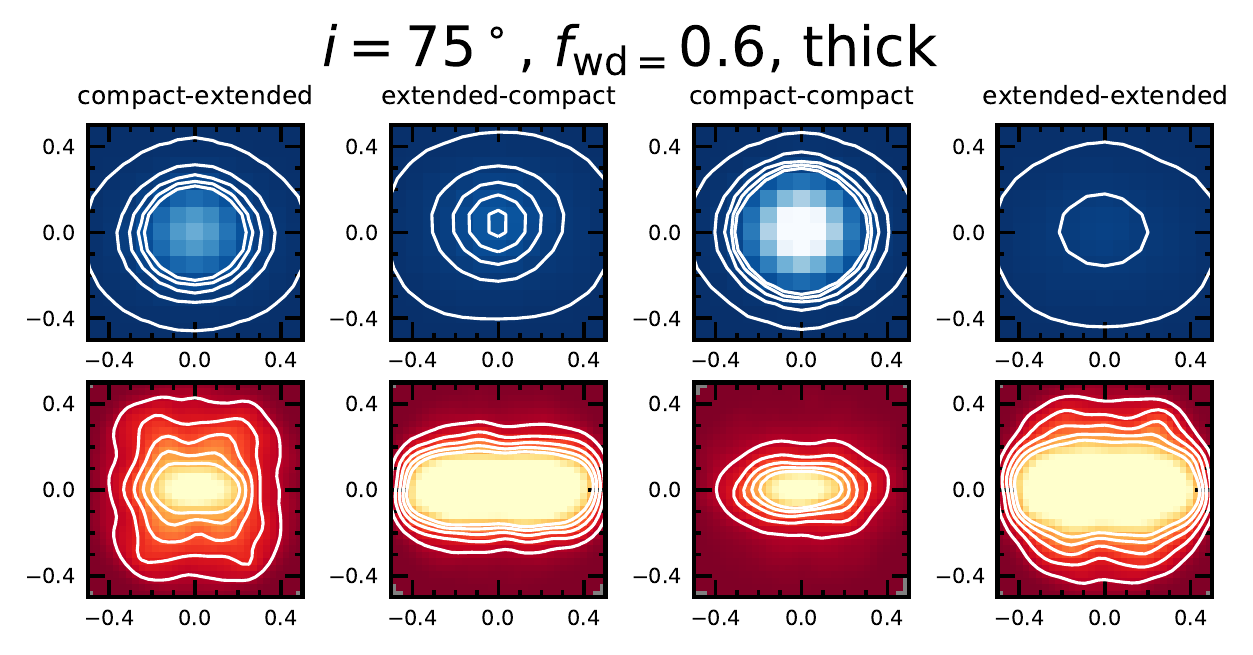}
    \caption{Simulated CAT3D-WIND images for the GATOS Seyferts. Parameters are as in
      Fig.~\ref{fig:simulated_fwd0.3_thick} but for a wind-to-disk
      ratio of $f_{\rm wd}=0.6$. The image brightness scales and
      contour levels are as in Fig.~\ref{fig:simulated_fwd0.3_thick}.}
              \label{fig:simulated_fwd0.6_thick}
    \end{figure}

\begin{figure}[!ht]
    \includegraphics[width=9cm]{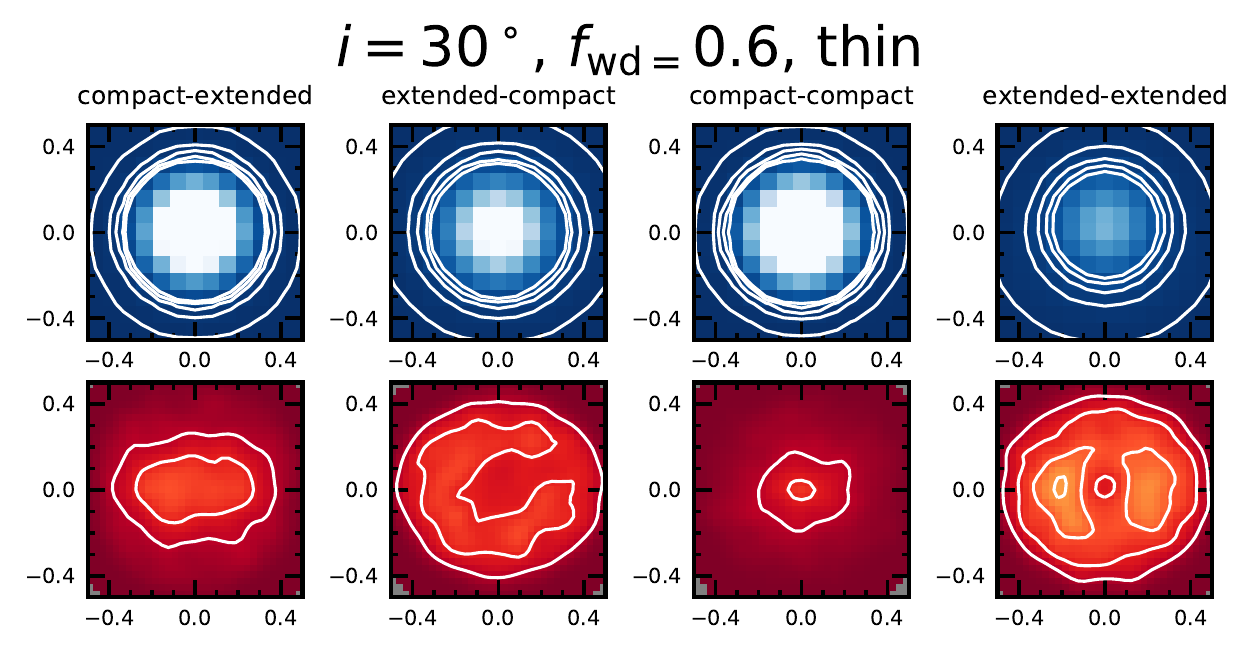}
    \includegraphics[width=9cm]{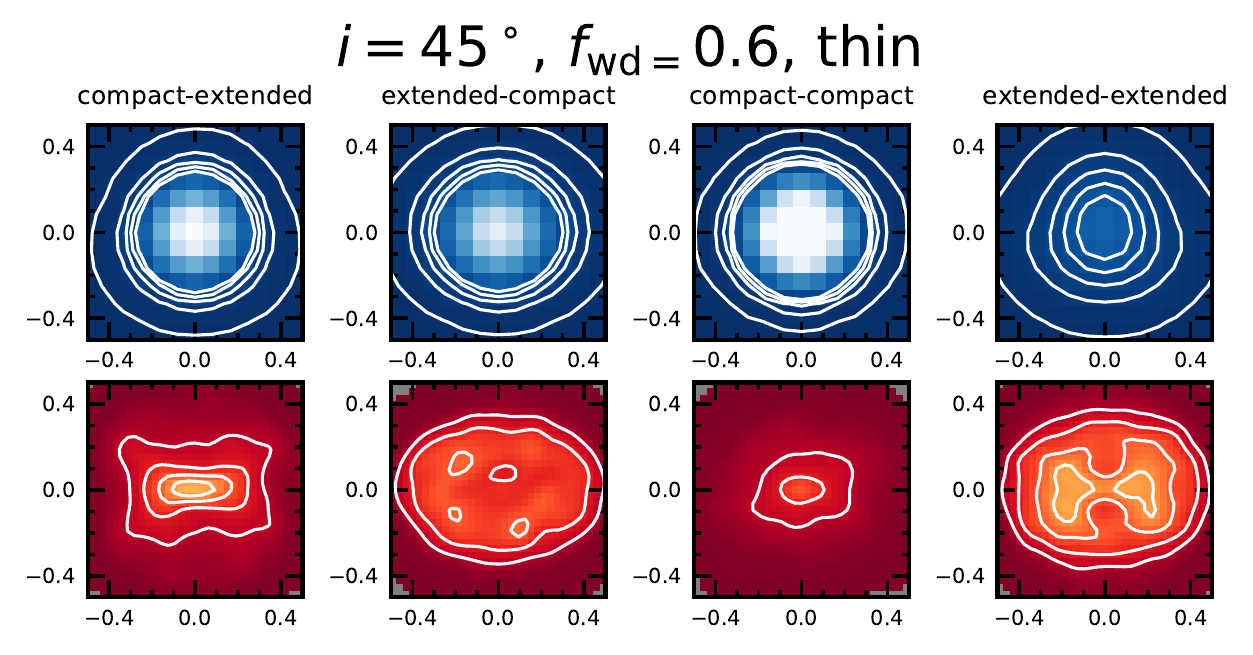}
    \includegraphics[width=9cm]{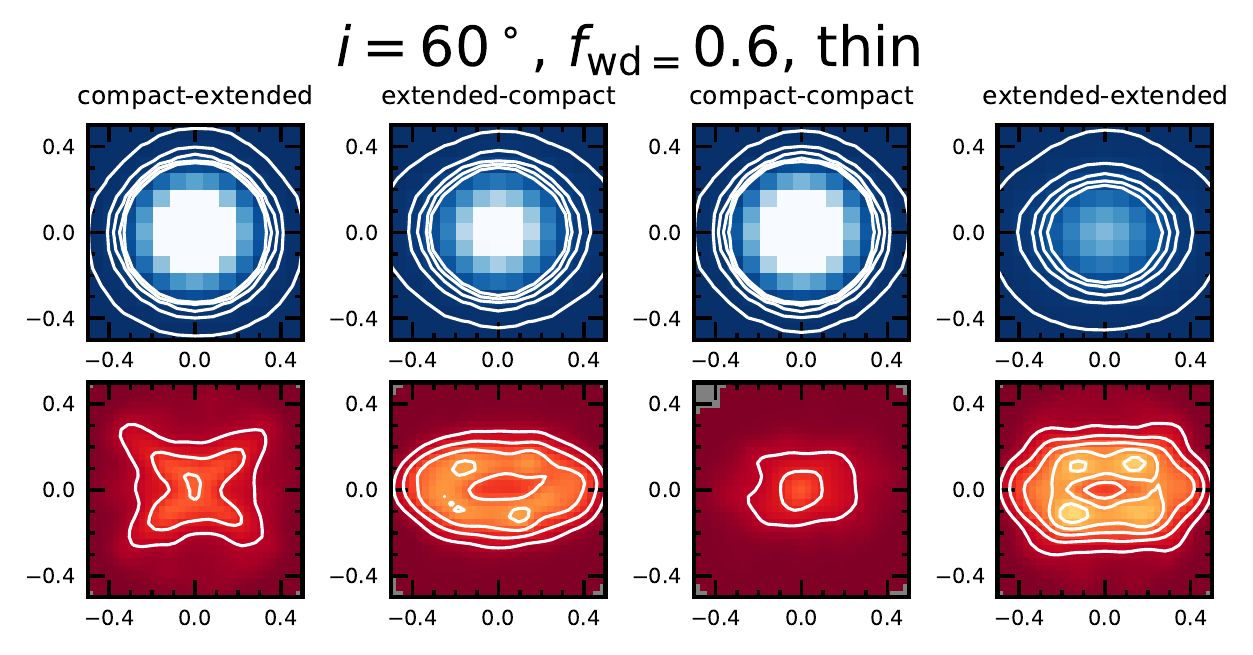}
    \includegraphics[width=9cm]{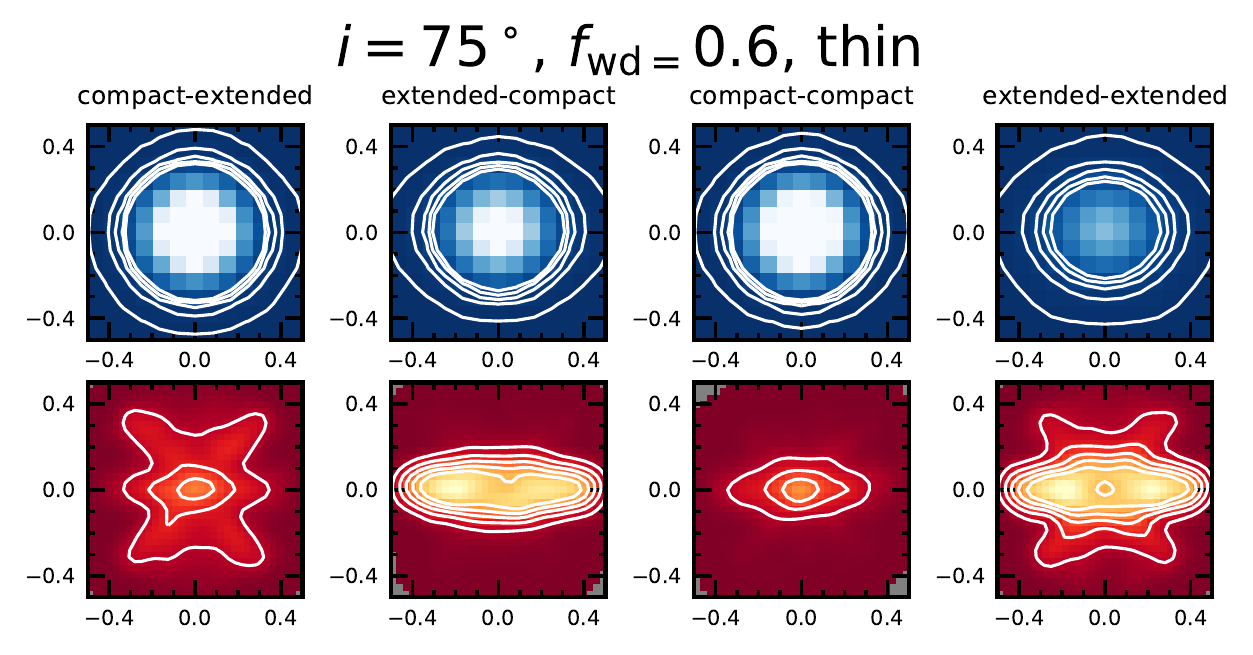}
  \caption{Simulated CAT3D-WIND model images for angular
    resolutions representing approximately those of the GATOS Seyfert
    observations. Parameters are as in
      Fig.~\ref{fig:simulated_fwd0.6_thick} but for a ``thin''
      geometry. The image brightness scales and
      contour levels are as in Fig.~\ref{fig:simulated_fwd0.3_thick}. }
              \label{fig:simulated_fwd0.6_thin}
    \end{figure}

    \begin{figure}[!ht]
    \includegraphics[width=9cm]{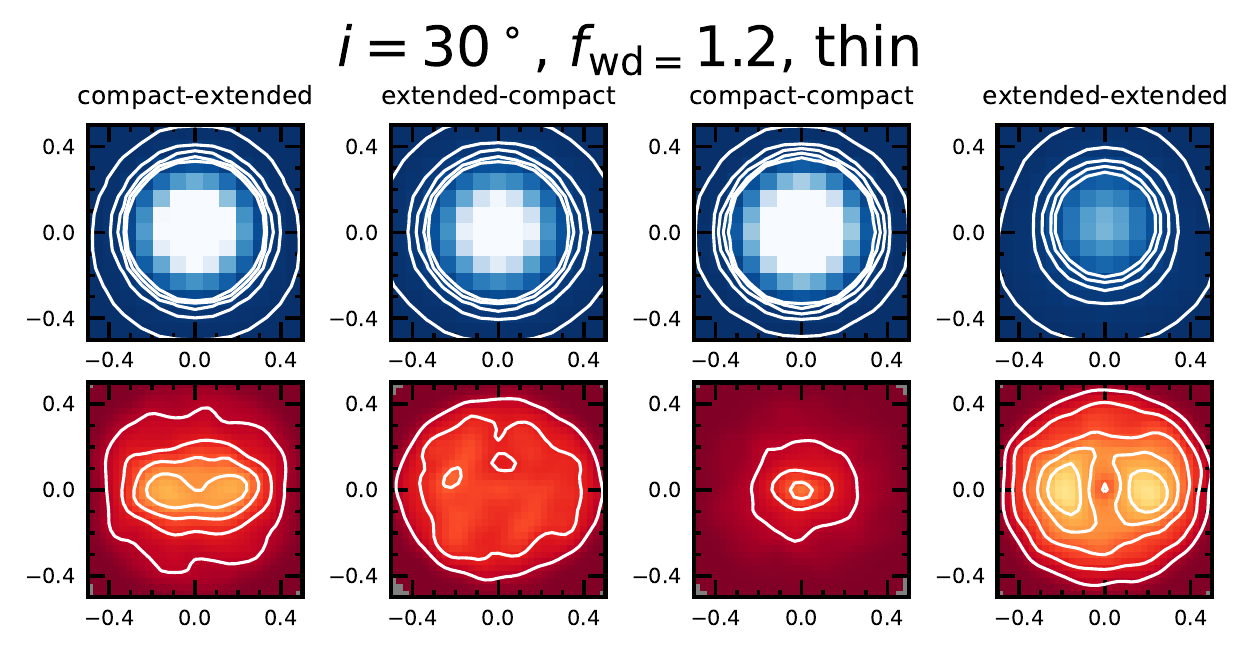}
    \includegraphics[width=9cm]{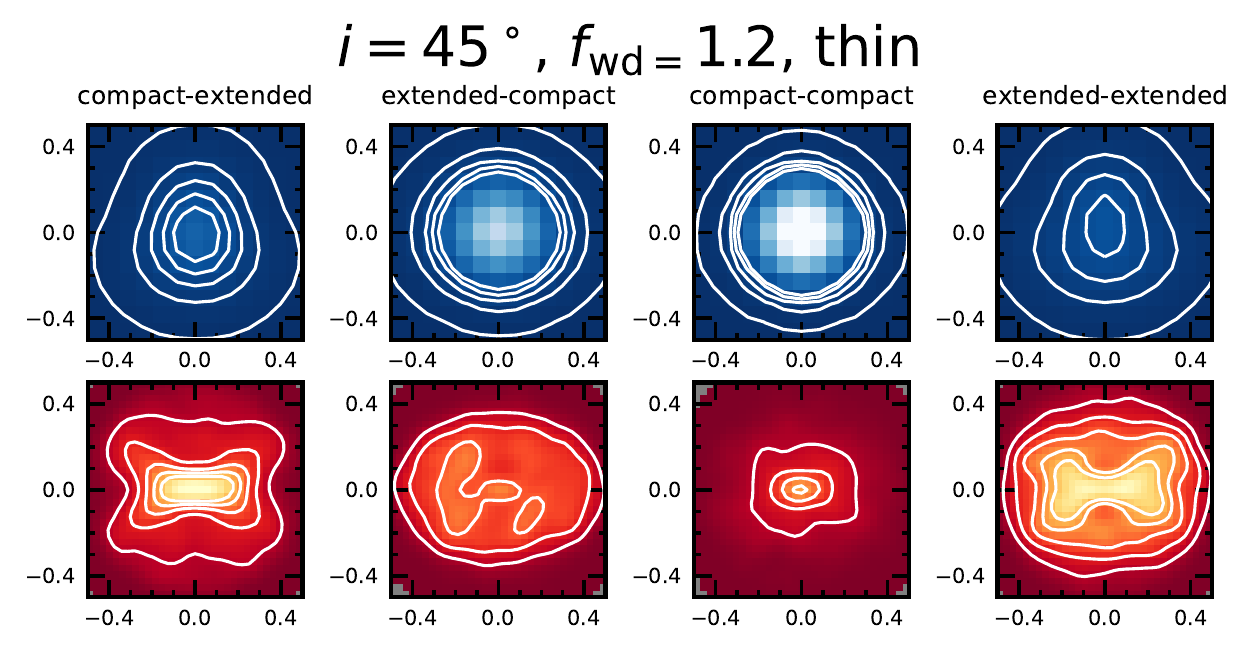}
    \includegraphics[width=9cm]{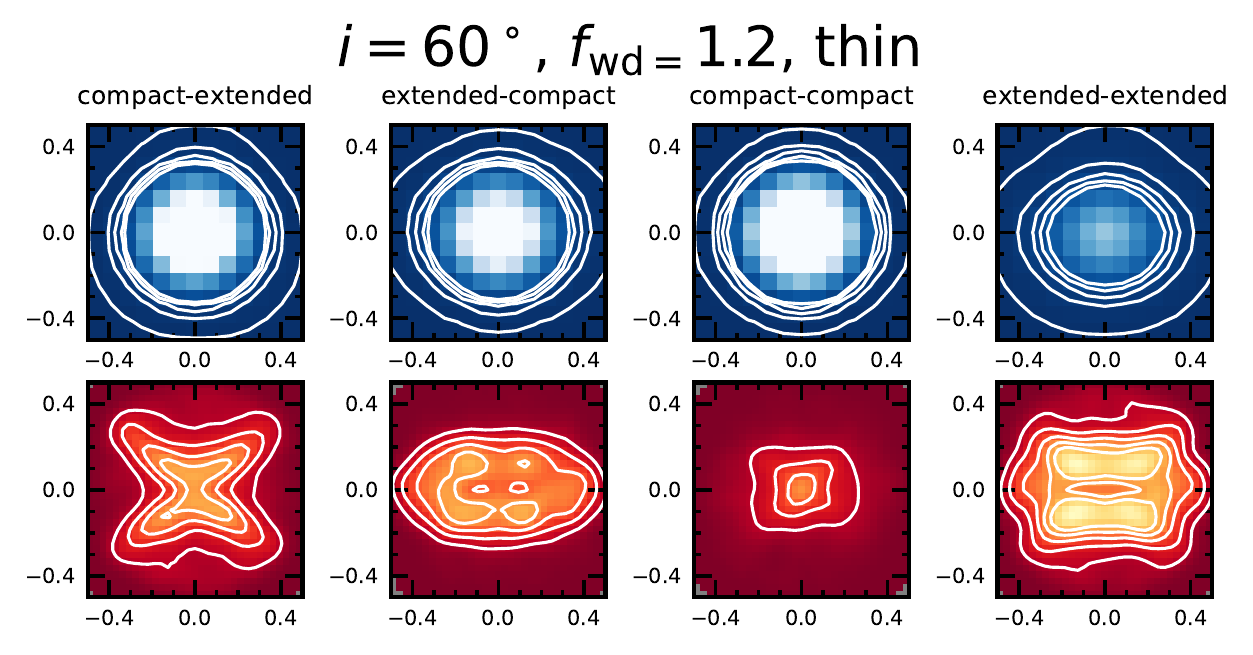}
    \includegraphics[width=9cm]{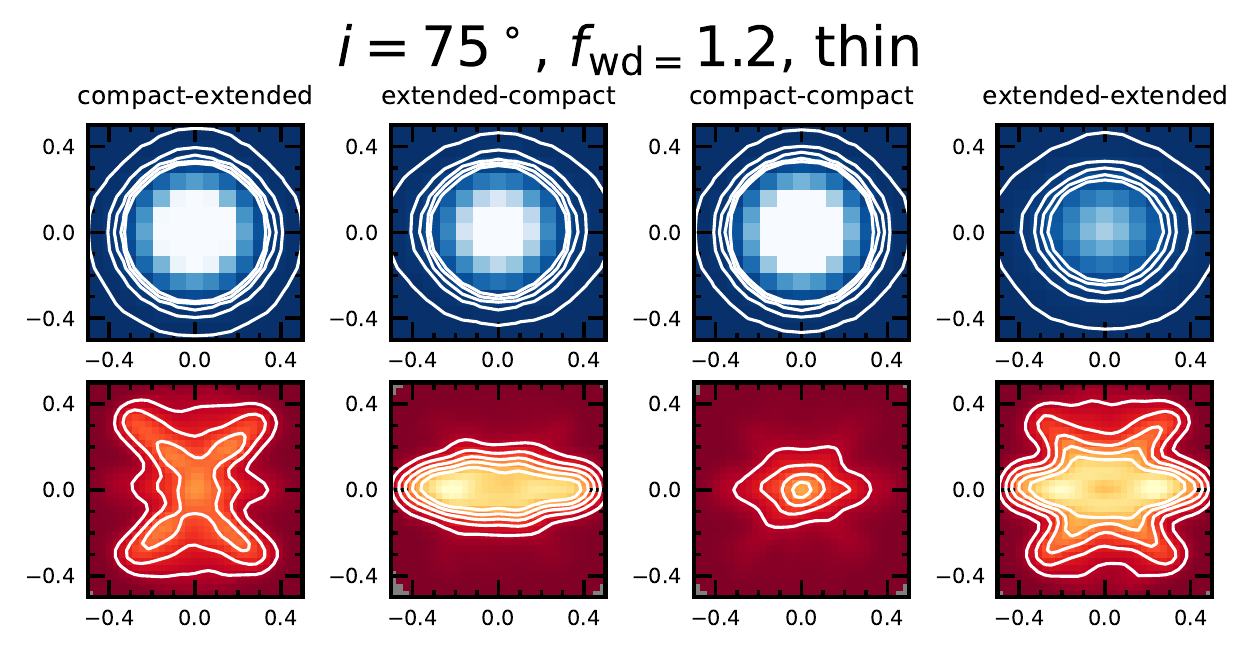}
  \caption{Simulated CAT3D-WIND images  matching for the GATOS Seyferts. Parameters are the
      same as in Fig.~\ref{fig:simulated_fwd0.6_thin} but for a wind-to-disk
    ratio of $f_{\rm wd}=1.2$. The image brightness scales and
      contour levels are as in Fig.~\ref{fig:simulated_fwd0.3_thick}.}
              \label{fig:simulated_fwd1.2}
    \end{figure}

    \begin{figure}[!ht]
  \centering
    \includegraphics[width=5cm]{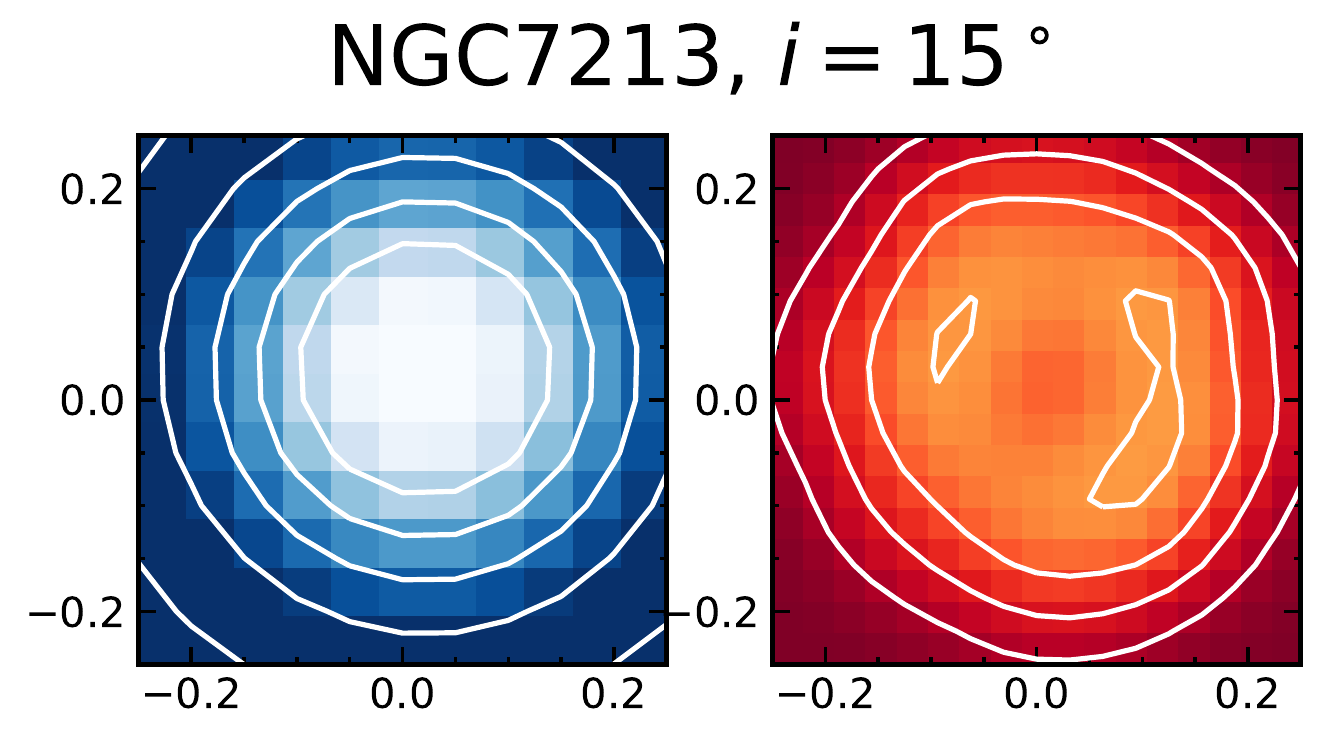}
    \caption{Simulated CAT3D-WIND model images at $8\,\mu$m (left, blue colors)
      and $700\,\mu$m (right, orange colors) with the parameters of
      Fig.~\ref{fig:CAT3D-WINDNGC7213} matching approximately the
      angular resolution of NGC~7213.}
              \label{fig:simulated_NGC7213}
    \end{figure}

\begin{figure*}[ht!]
  \centering
  \includegraphics[width=4.4cm]{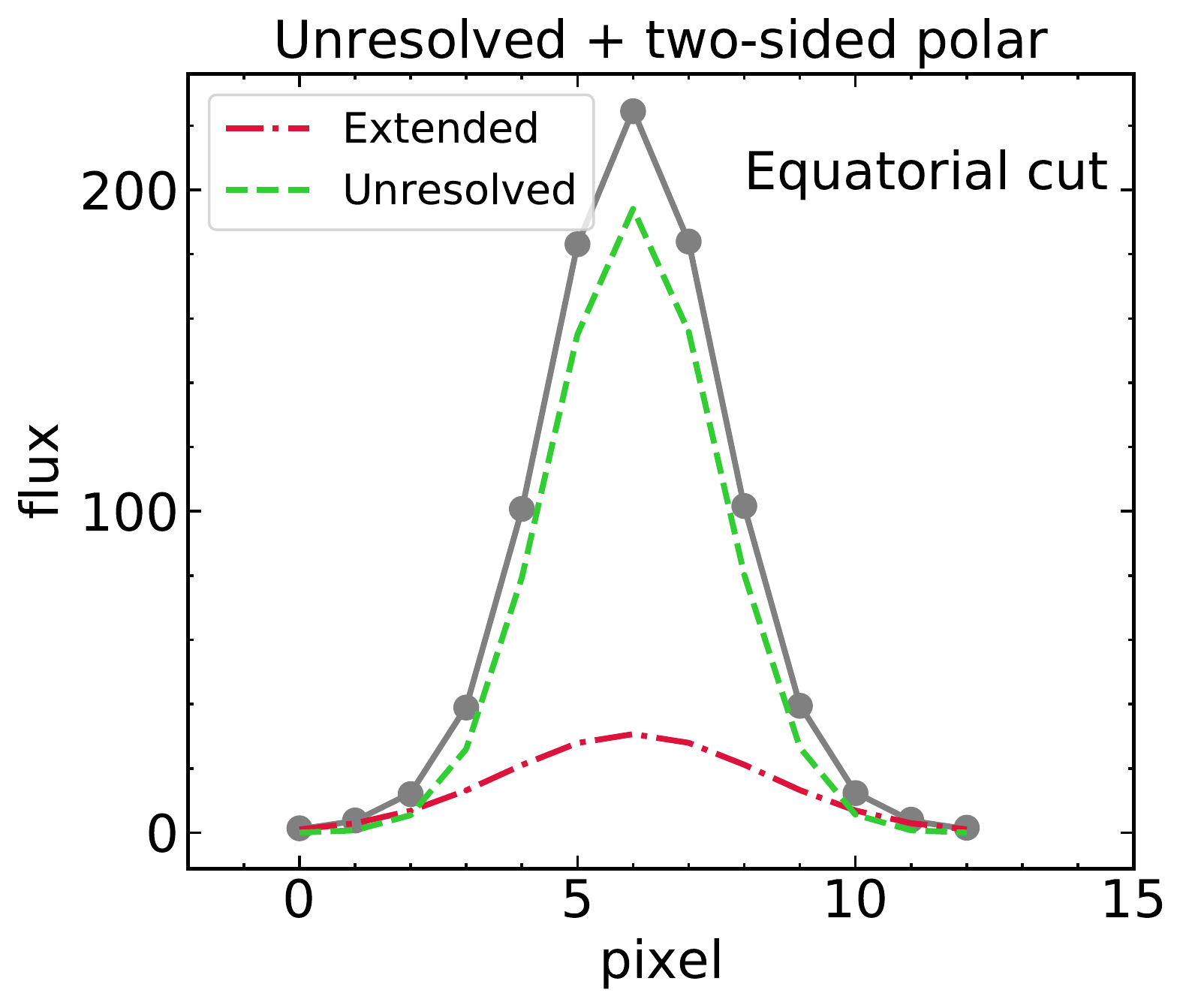}
  \includegraphics[width=4.4cm]{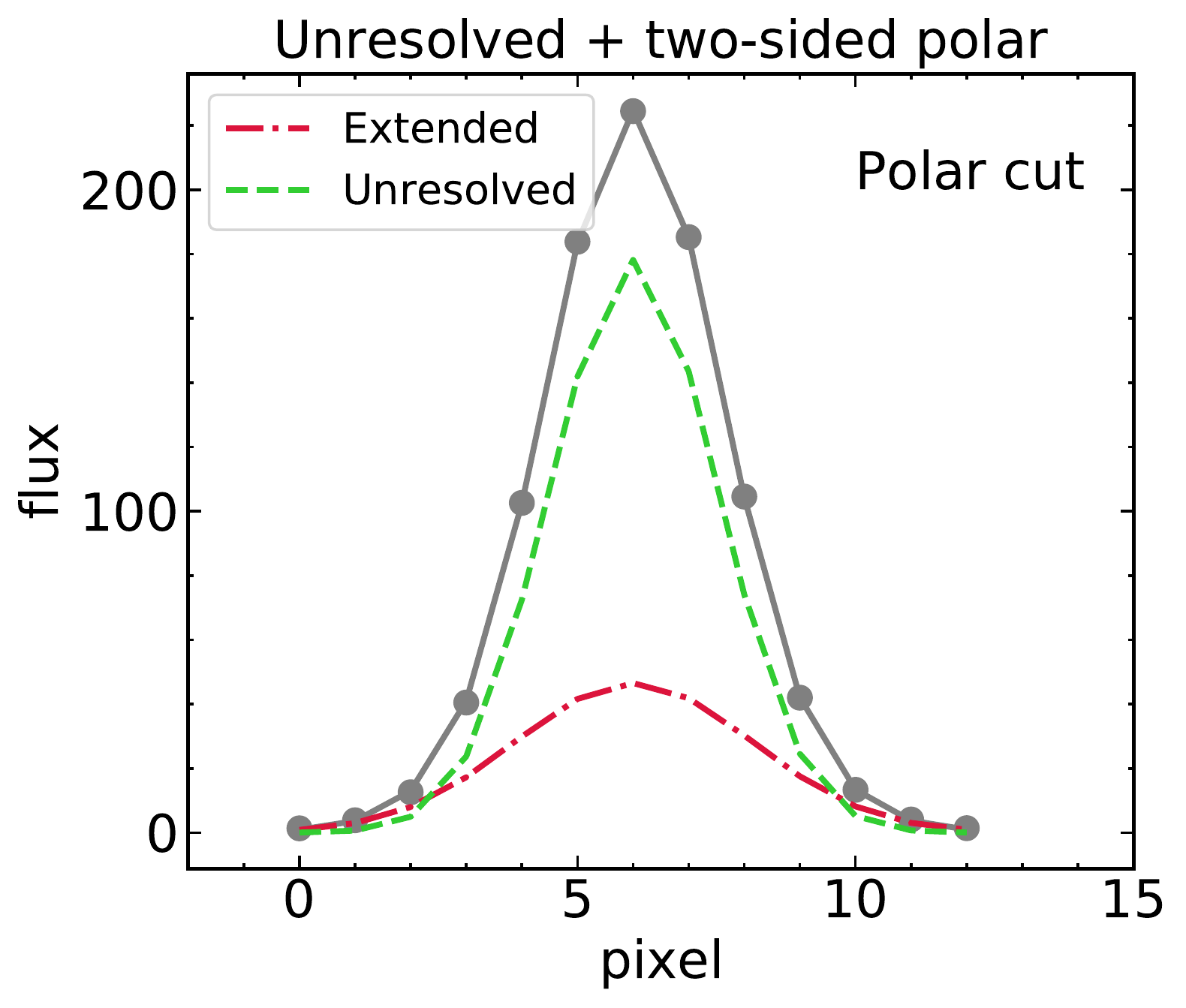}
  \includegraphics[width=7.cm]{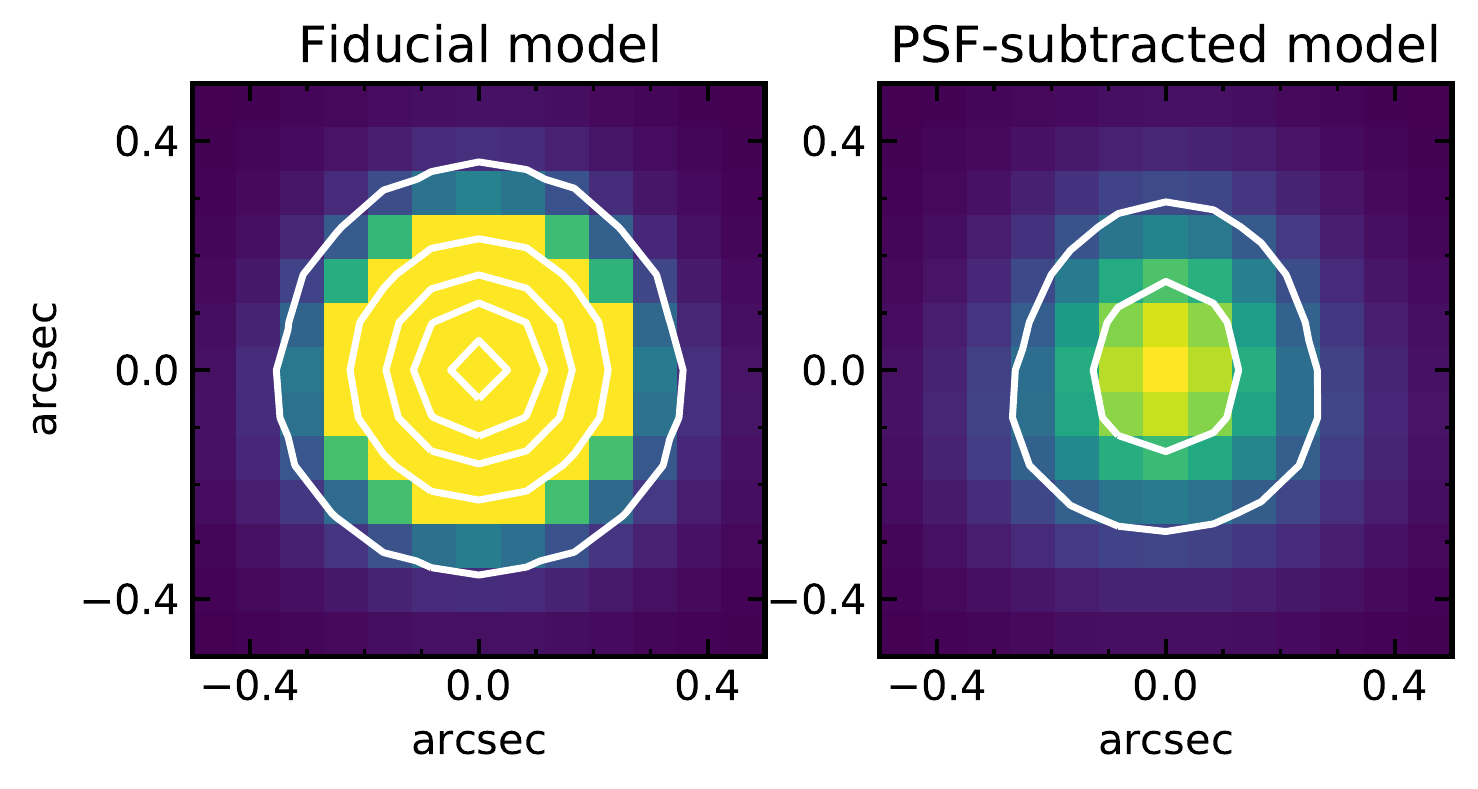}
  \caption{The two most left panels are  equatorial and polar  1D
    profiles for a CAT3D-WIND simulated galaxy image at $8\,\mu$m
    with the typical resolutions of our mid-IR observations. The model parameters are: a
    compact disk - extended wind, ``thin'' geometry with $f_{\rm wd}=0.6$ at
    $i=45^\circ$. The    dashed and dashed-dotted lines are the fitted unresolved and 
    extended components, respectively, and the solid lines the sum of
    both. The two most right panels are the mid-IR fiducial model image and
    the PSF-subtracted model image, respectively. }\label{fig:simulated1Dprofiles_NGC1365}
\end{figure*}

\begin{figure*}[ht!]
  \centering
  \includegraphics[width=4.4cm]{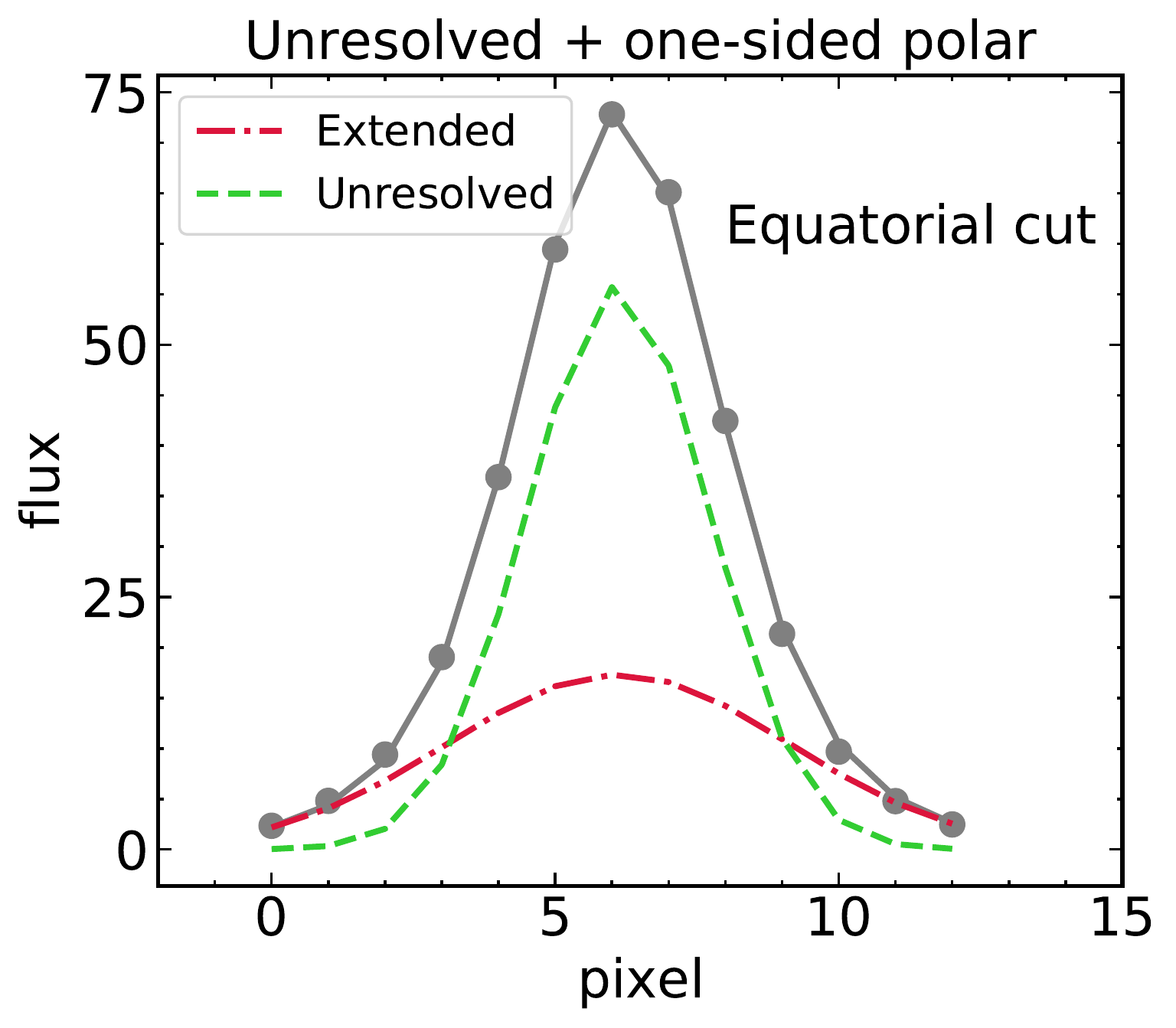}
  \includegraphics[width=4.4cm]{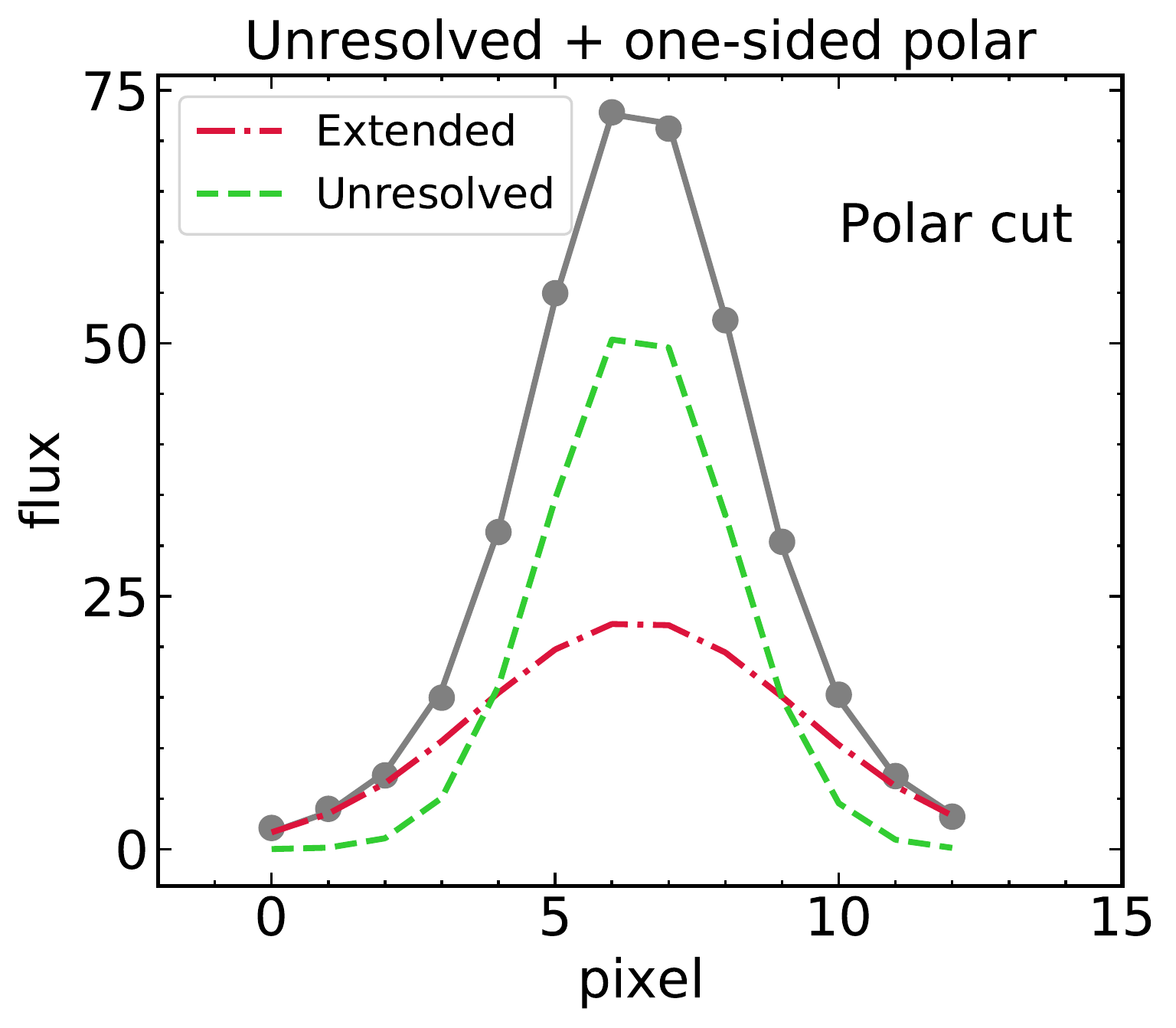}
  \includegraphics[width=7.cm]{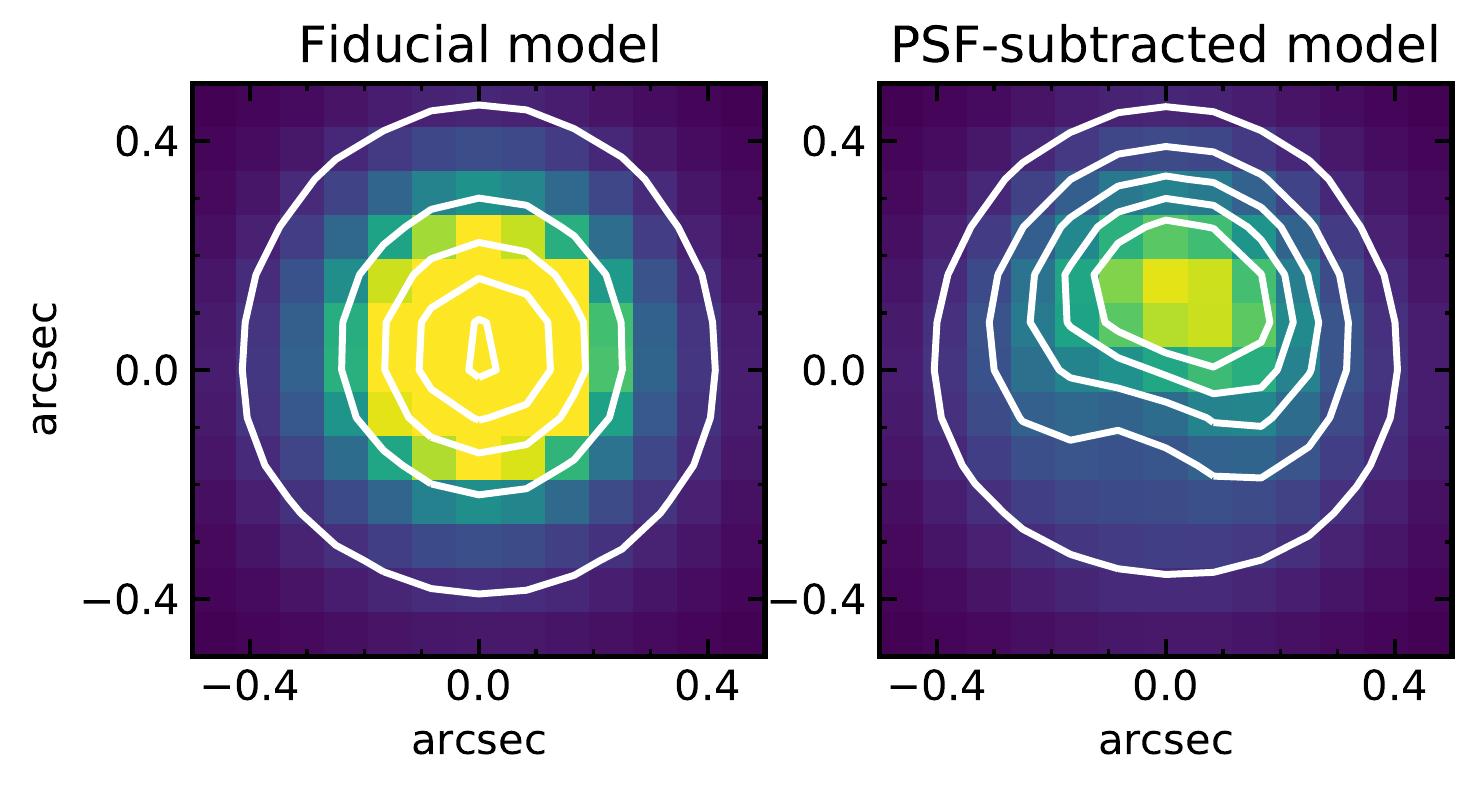}
\caption{Same as Fig.~\ref{fig:simulated1Dprofiles_NGC1365} but for an     
    extended disk - extended wind, ``thick'' geometry model with $f_{\rm wd}=0.3$ at $i=45^\circ$.}\label{fig:simulated1Dprofiles_NGC3227}
\end{figure*}
  
\begin{figure*}[ht!]
  \centering
  \includegraphics[width=4.4cm]{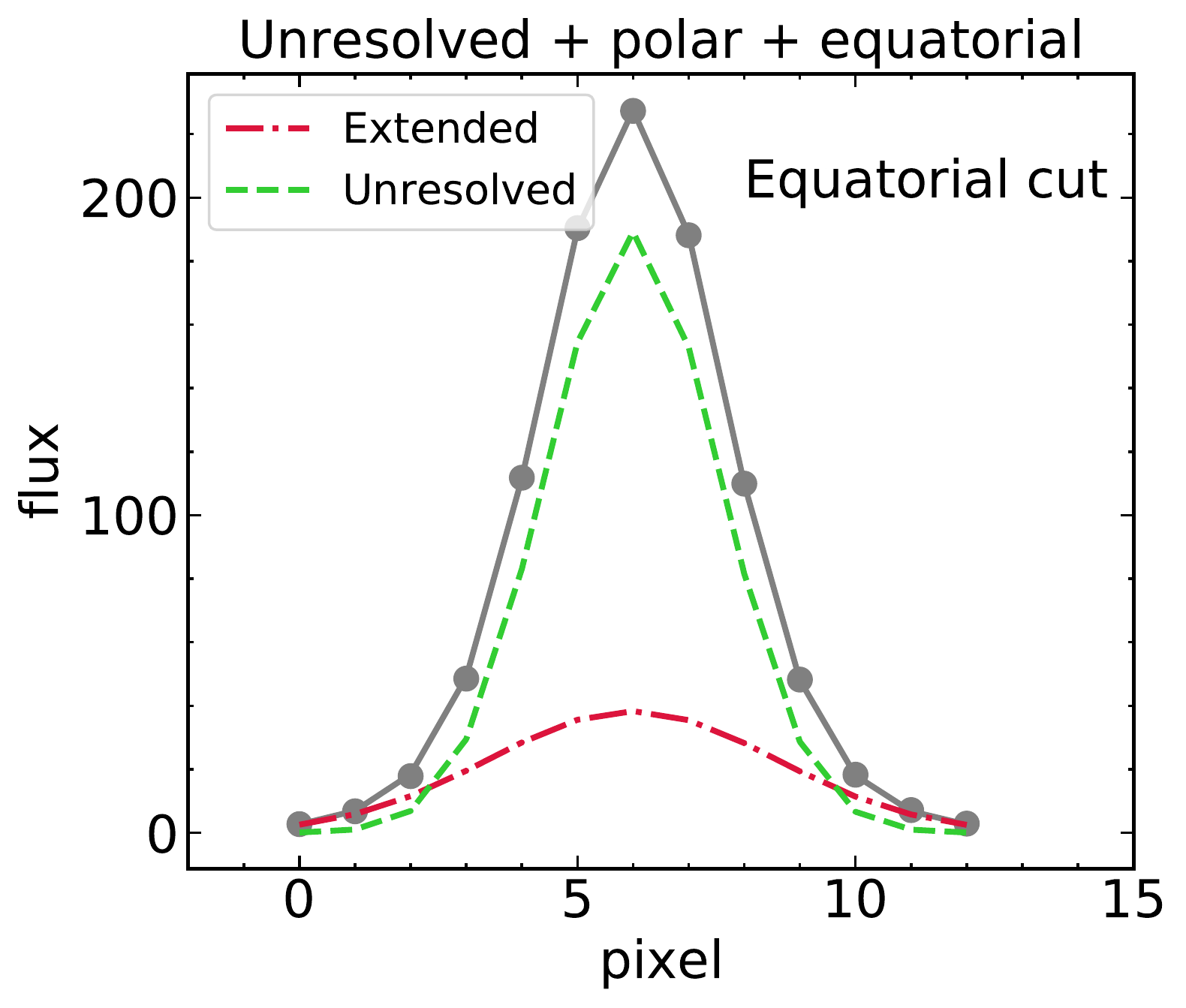}
  \includegraphics[width=4.4cm]{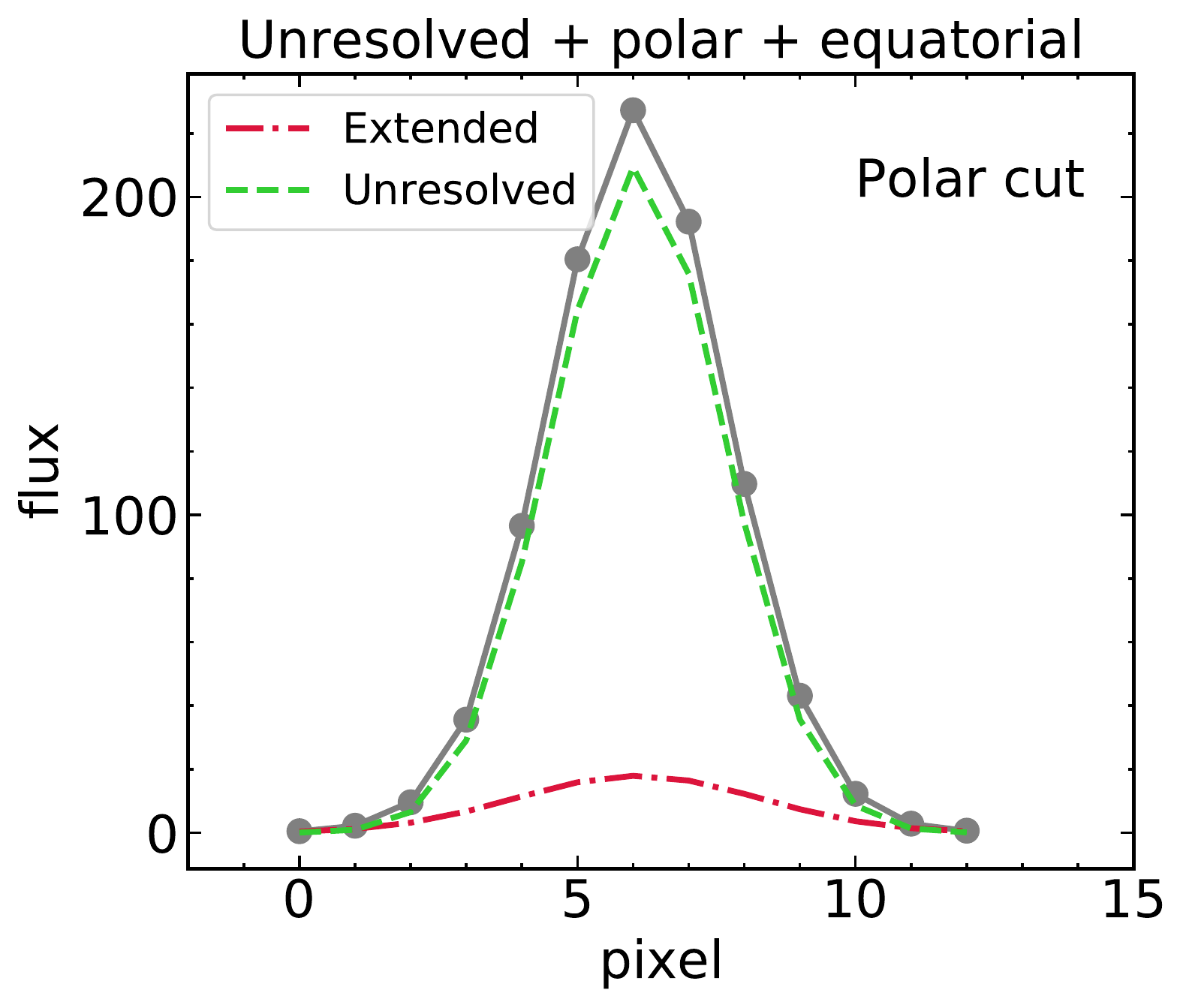}
  \includegraphics[width=7.cm]{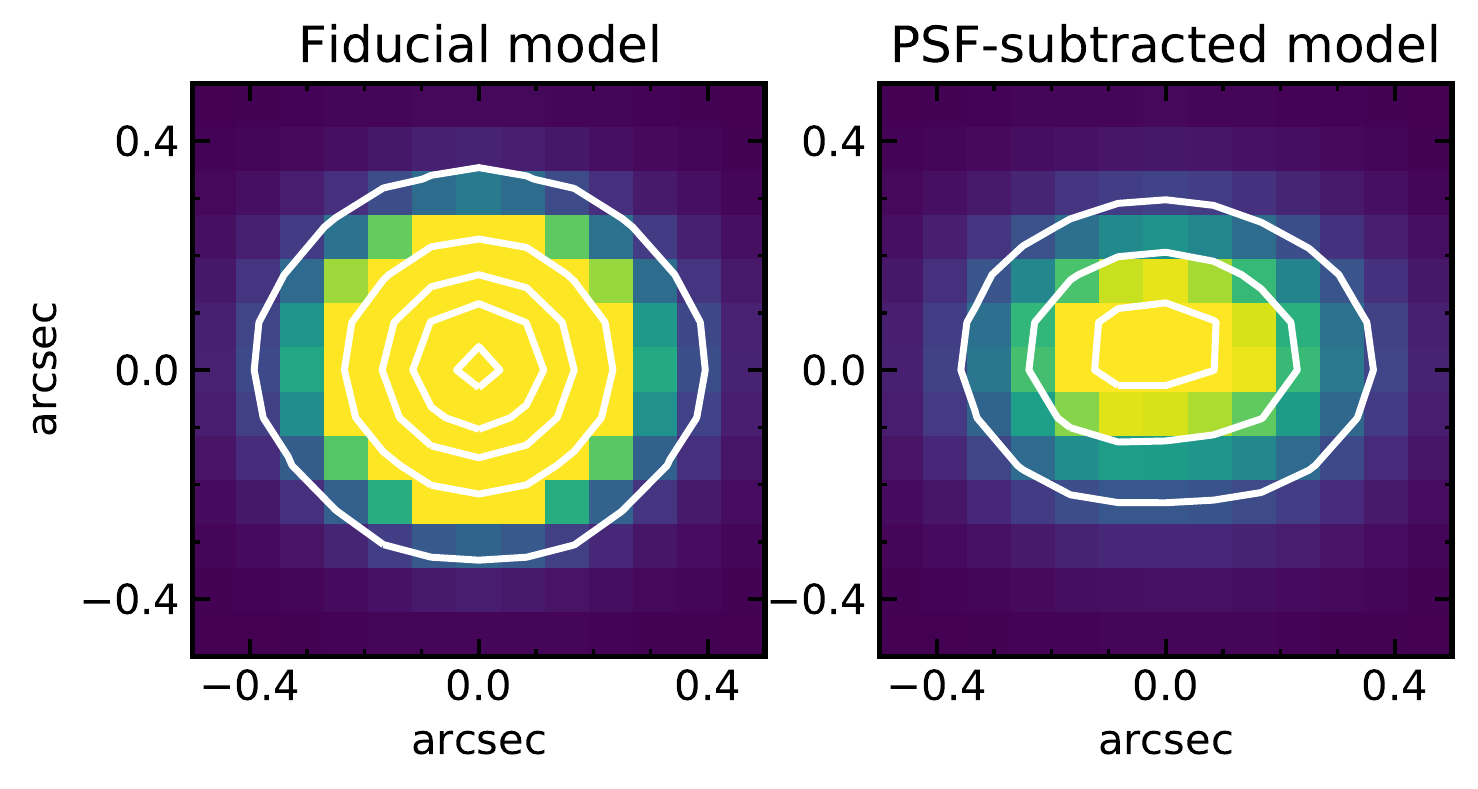}
\caption{ Same as Fig.~\ref{fig:simulated1Dprofiles_NGC1365} but for
  a compact disk - compact wind, ``thick'' geometry model with $f_{\rm wd}=0.3$ at
    $i=75^\circ$.} \label{fig:simulated1Dprofiles_NGC5643}
\end{figure*}

\begin{figure*}[ht!]
  \centering
  \includegraphics[width=4.4cm]{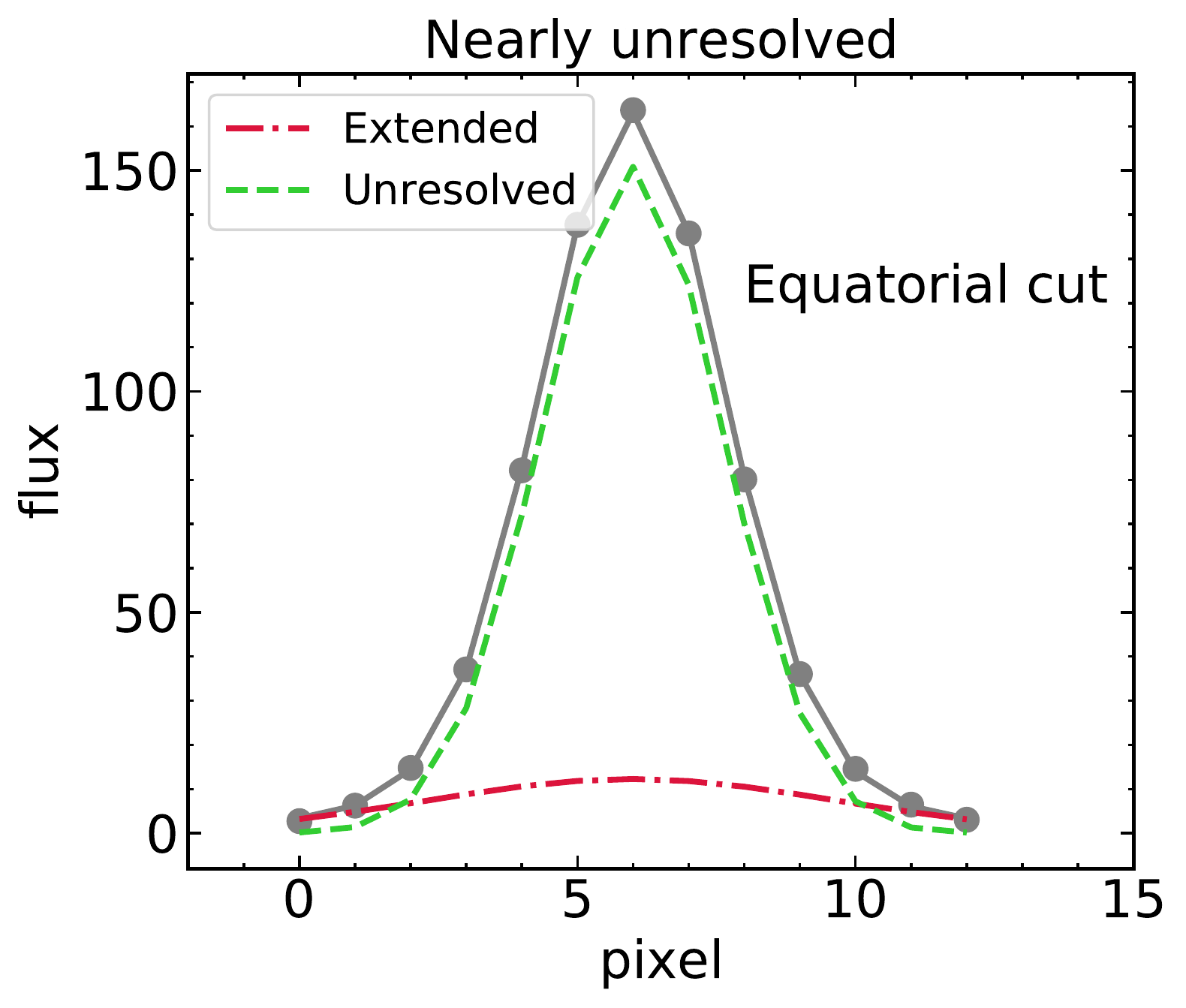}
  \includegraphics[width=4.4cm]{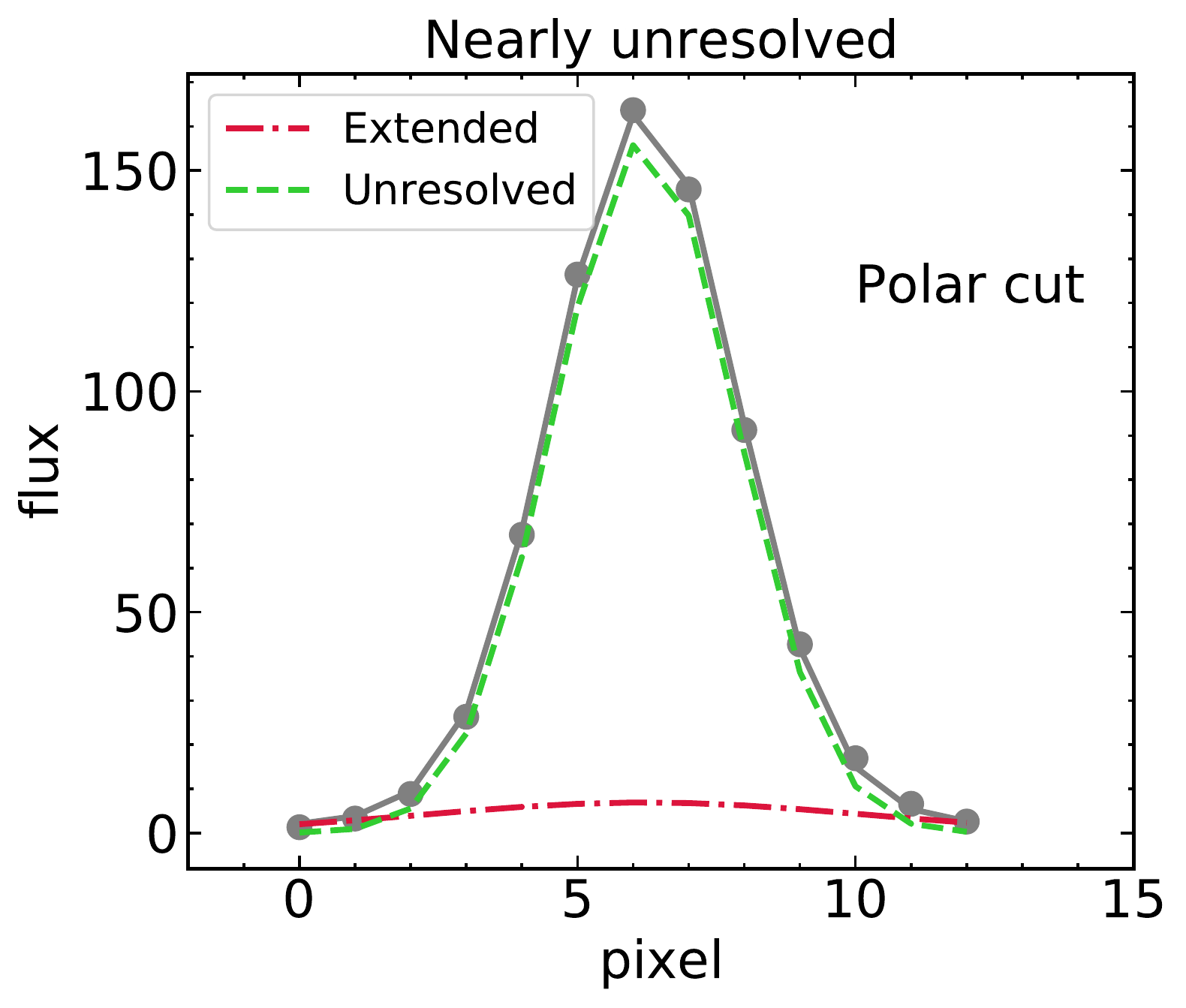}
  \includegraphics[width=7.cm]{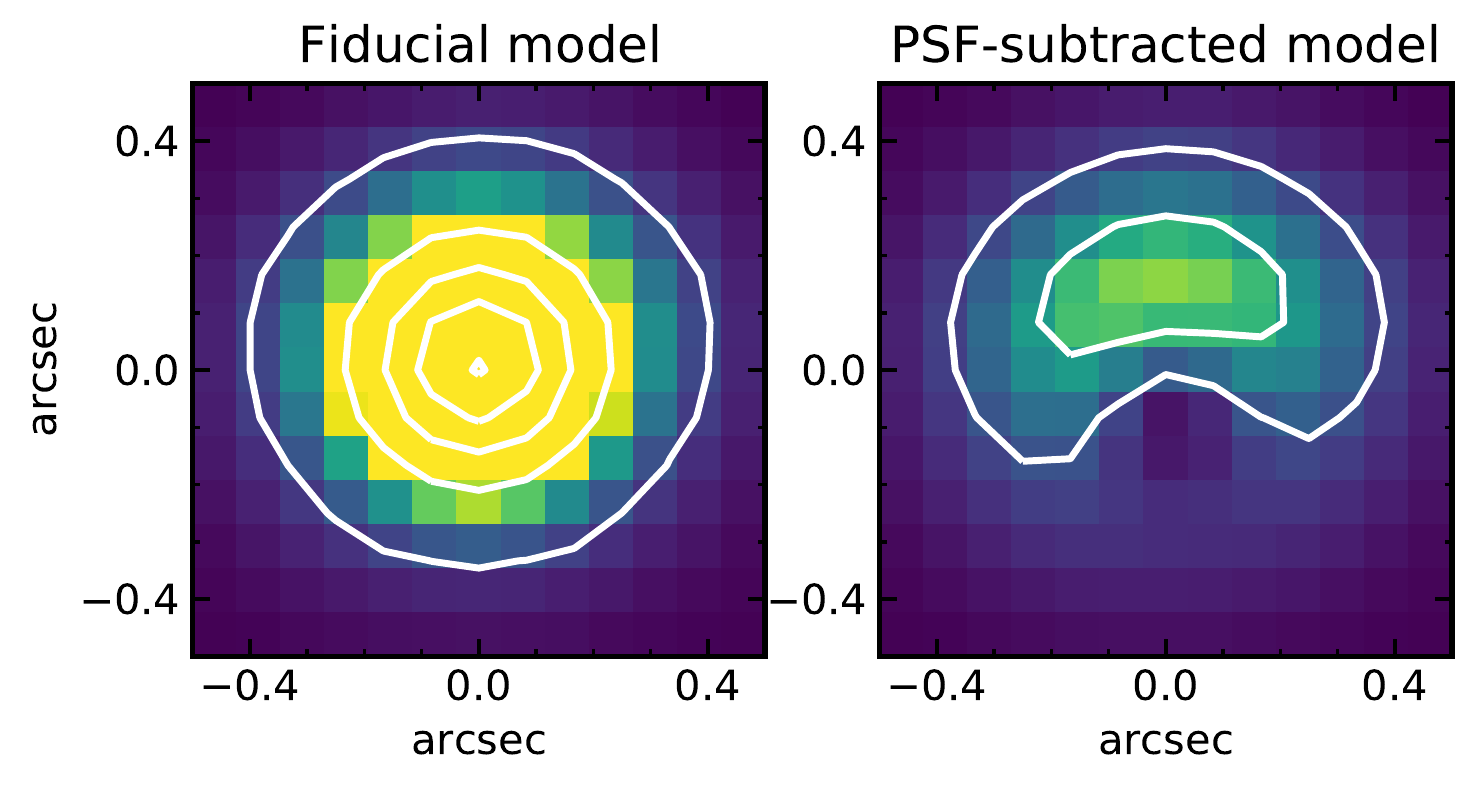}
\caption{Same as Fig.~\ref{fig:simulated1Dprofiles_NGC1365} but for an
  extended disk - compact wind,  ``thick'' geometry model with
    $f_{\rm wd}=0.3$ at $i=60^\circ$.}\label{fig:simulated1Dprofiles_NGC7314}
  
\end{figure*}

The CAT3D-WIND $2000 r_{\rm sub} \times 2000 r_{\rm sub}$ model images cover typical
physical sizes of  $\sim 80\,{\rm pc} \times 80\,{\rm pc}$ or approximately 
angular sizes of $1\arcsec \times 1\arcsec$
for a median distance of 19\,Mpc for our galaxies, excluding NGC~6814
and NGC~7213. For these two galaxies,  the model images have sizes of $\sim 60\,{\rm pc} \times 60\,{\rm pc}$ = $0.5\arcsec \times 0.5\arcsec$ at a distance of
23\,Mpc.
We convolved the $8\,\mu$m and $700\,\mu$m
model images to the typical mid-IR and ALMA far-IR angular
resolutions of $0.3\arcsec$ and $0.1\arcsec$, respectively, using a
Gaussian function. Finally, we resampled the CAT3D-WIND simulated
images to representative pixel sizes of $\sim 0.08$\arcsec/pixel
for the mid-IR (0.0453\arcsec/pixel for NGC~7213)
and 0.03\arcsec/pixel for the far-IR images.

We show a selection of fiducial simulated images for the GATOS
Seyferts. In Figs.~\ref{fig:simulated_fwd0.3_thick} and
\ref{fig:simulated_fwd0.6_thick}, we include models with a ``thick'' geometry and two
values of the wind-to-disk ratio $f_{\rm wd}=0.3$ and $f_{\rm
  wd}=0.6$ (compare with the original resolution model images in
Sect.~\ref{sec:CAT3DWINDmodelimages_THICK}). In Figs.~\ref{fig:simulated_fwd0.6_thin} and 
\ref{fig:simulated_fwd1.2} we display the simulated images for a ``thin'' geometry with
$f_{\rm wd}=0.6$  (compare with the original resolution model images
of Sect.~\ref{sec:CAT3DWINDmodelimages_THIN})  and $f_{\rm wd}=1.2$
(compare with model images in the Appendix), respectively.
In the next two sections, we describe the morphologies of the
fiducial models and make 
qualitative comparisons with our far-IR and
mid-IR observations.

\subsubsection{Mid-IR  model images}\label{sec:resultsMIR}

The fiducial mid-IR simulated model images
do not show a high level of detail compared
to the far-IR. The reasons are twofold. First,
most of the mid-IR emission in many of the geometries studied in this work
comes from the inner part of the disk and wind. Second, the angular
resolution of our mid-IR observations is a
factor of three worse than the ALMA ones.
Nevertheless, there are some morphological differences in the
mid-IR.

The large majority
of the CAT3D-WIND  model parameters investigated  here produces bright
and centrally peaked mid-IR emission, which 
appears unresolved or only slightly resolved at the considered angular
resolution.  This is most noticeable for compact disk 
configurations as well as the case of the specific simulation for
NGC~6814 and NGC~7213 (left panel of Fig.~\ref{fig:simulated_NGC7213}). There is extended mid-IR emission along the polar direction
at intermediate inclinations ($i=45^\circ$) for the extended
wind configurations in both ``thin'' and ``thick'' geometries with
$f_{\rm wd} =0.6$ and $1.2$ (second panels from the top in
Figs.~\ref{fig:simulated_fwd0.6_thick}, \ref{fig:simulated_fwd0.6_thin}, and \ref{fig:simulated_fwd1.2}) and
the case of a mild  extended wind with $f_{\rm wd}=0.3$ (see Fig.~\ref{fig:simulated_fwd0.3_thick}) as
well as at nearly face-on inclinations ($i=30^\circ$) for $f_{\rm wd}
=1.2$ (top panels of Fig.~\ref{fig:simulated_fwd1.2}).
Extended mid-IR emission along the equatorial and/or polar direction of the
disk is produced at high inclinations in extended disk
configurations.

To evaluate the unresolved and extended contribution in the simulated mid-IR model images, we followed a similar
procedure as for the observations in Sect.~\ref{sec:analysis}. We fitted 1D profiles along
the equatorial and polar directions. For the unresolved component we
fixed the Gaussian width to the value
used to convolve the original CAT3D+WIND images. With the scaling
factors derived from these fits, we subtracted the PSF image from the
fiducial mid-IR model images.

Rather than fitting the  individual galaxies,  in
Figs.~\ref{fig:simulated1Dprofiles_NGC1365}-\ref{fig:simulated1Dprofiles_NGC7314}
we show some model examples of the different observed mid-IR morphologies. 
The case of unresolved +
two-sided polar mid-IR emission (Fig.~\ref{fig:simulated1Dprofiles_NGC1365}) is seen in a 
compact disk - extended  wind ``thin'' geometry with a high wind-to-disk ratio ($f_{\rm
  wd}=0.6$, in the figure), at an intermediate inclination of
$i=45^\circ$. These parameters are based on the constraints for
NGC~1365,  including the relatively compact torus seen with ALMA (see Table~\ref{tab:Torus}). The 1D
profiles and PSF-subtracted model image show that the
extended component is more relevant along the polar 
direction.

The unresolved + one-sided polar mid-IR emission (Fig.~\ref{fig:simulated1Dprofiles_NGC3227}) is 
seen mostly at intermediate inclinations ($i=45^\circ$ in the figure)
in the simulated image and profiles, for extended disk - extended wind
configurations
with a ``thick'' geometry. Low to intermediate wind-to-disk values ($f_{\rm
  wd}=0.3$ in the figure) are needed to have an important contribution
from the unresolved mid-IR emission, as seen in our observations of NGC~3227,
and NGC~7582.

The combination of 
unresolved +  equatorial mid-IR emission  (Fig.~\ref{fig:simulated1Dprofiles_NGC5643}) is observed at higher
inclinations ($i=75^\circ$ in the figure) with a ``thick'' geometry
and extended disk and compact wind ($f_{\rm wd} = 0.3$ in the
figure). The PSF-subtracted mid-IR model image shows a morphology  consistent with the observations of
NGC~6300. Finally, we checked models that produce nearly unresolved mid-IR
 morphologies
(Fig.~\ref{fig:simulated1Dprofiles_NGC7314}) such as a ``thick'' geometry
and extended disk and compact wind with $f_{\rm wd} = 0.3$ in the
figure for an inclination of $i=60^\circ$. This would reproduce
the case of NGC~7314 which has a large $870\,\mu$m torus but mostly
unresolved mid-IR emission.

\subsubsection{Far-IR  model images}\label{sec:resultsFIR}

All the simulated far-IR model images are resolved and show a variety of 
morphologies. This is in excellent
agreement with the observational result that the extended ALMA
$870\,\mu$m emission in most GATOS Seyferts dominates
the total  emission at this wavelength (see GB21 for more details). 
At low-to-intermediate inclinations, the far-IR simulated model images with extended
disk components show  ring-like
morphologies. As expected, the compact disk configurations show a more
compact far-IR morphology.
At high inclinations and extended
wind configurations, the characteristic ``X''-shape due to the edge 
brightening in the cone walls, is seen in  
the ``thin'' geometry  and high values of the wind-to-disk cloud ratio
(the two bottom panels of
Figs.~\ref{fig:simulated_fwd0.6_thin} and \ref{fig:simulated_fwd1.2},
that is for $i=60$ and $75^\circ$). It is also seen, although with low
contrast, for 
$f_{\rm wd}= 0.3-0.6$ and ``thick'' geometries
(Figs.~\ref{fig:simulated_fwd0.3_thick} and \ref{fig:simulated_fwd0.6_thick}).
The ALMA $870\,\mu$m images of a few GATOS
Seyferts are suggestive of this ``X''-shape, for instance, 
NGC~5643, NGC~7314, and especially in NGC~7582 (see Figs.~2-4 in
GB21 and Fig.~\ref{fig:ALMAmidIRimages}). NGC~4388 shows mixed/polar
emission at $870\,\mu$m. A compact disk and compact wind
``thick'' configuration with a low wind-to-disk ratio and seen at an
intermediate inclination reproduces well the
observed morphology.

We show the specific simulation for 
NGC~7213 in Fig.~\ref{fig:simulated_NGC7213} (right panel), which is
also valid for NGC~6814.
The far-IR model image shows a
ring-like geometry due to the low  inclination of $i=15^\circ$ and the
$a=-1$ cloud radial distribution in the disk. The ALMA extended $870\,\mu$m
emission of NGC~6814 and 
NGC~7213 (Fig.~3 in GB21) shows centrally peaked emission. As
discussed by GB21,  it is
possible that a large fraction of this emission is not associated with
cold dust. This was based on the discrepancies between the
molecular gas masses derived from the CO(3-2) and the $870\,\mu$m
emission.

\section{Discussion}\label{sec:discussion}
The emerging picture for the central regions of radio-quiet AGNs
has the obscuring {\it torus} as part of a gas flow cycle where gas is
brought in from the host galaxy  and then driven out by the AGN in a
wind \citep[see][]{Elitzur2006}. 
High angular resolution observations gathered over the last 15 years point to a multi-phase
multi-component structure where 
both the disk/torus and the base of the wind can contribute to the AGN
obscuration \citep[see the review by][]{RamosAlmeida2017}.
Recently \cite{Hoenig2019} used simple physical
principles to support this new unifying view in a disk+wind
scenario. Up until now, this view was based on observations of a few
individual AGN. GATOS aims for a comprehensive
study of the nuclear activity and its connection with the host galaxy
in a well-defined sample of nearby Seyfert galaxies.

\subsection{Observational considerations}
One of the main caveats from the analysis of the extended mid-IR emission with
ground-based facilities is the stability of the PSF. Observing conditions can change on scales of a
few minutes and affect the shape of the PSF, which we took from the emission of a
standard star observed close in time. This combined with the modest
sensitivity to diffuse emission due to the high thermal background make it challenging to derive the
morphologies of the
extended mid-IR emission in the vicinity of the unresolved AGN emission.
Nevertheless, we detected extended mid-IR emission in seven of the
twelve Seyfert galaxies analyzed in this work, with sizes between 50
and 160\,pc.

The measured orientations of the extended mid-IR emission are
approximately along the 
polar/NLR direction   and perpendicular to the ALMA-identified tori
(Fig.~\ref{fig:ALMAmidIRimages})  in five  galaxies  (NGC~1365,
NGC~3227,  NGC~5643, and NGC~7582, and tentatively in NGC~4388)
as well as the inner regions of NGC~5506 \citep[see also][]{Asmus2019}.  In NGC~5506,
 NGC~5643, and NGC~6300, there is also extended mid-IR emission along
 the equatorial direction of the torus  and/or the host galaxy,
 adding to the complexity of the dust emission. The extended mid-IR
 emission along the polar direction 
is  likely associated with AGN-driven dusty outflows \citep{Wada2018,
  Williamson2020, Venanzi2020} and/or dust located in
the edges of the NLR and ionization cones. With the existing data, we cannot
distinguish between the two possibilities. We note that contributions
from emission lines in the imaging filters, in particular
[Ne\,{\sc ii}]$12.8\,\mu$m in NGC~5643 and NGC~ 7582, might be partly responsible for the
alignment along the polar direction. However, there are galaxies in our sample with relatively bright
optical NLR emission and faint or
no extended polar mid-IR emission. 

\subsection{The role of the IR radiation pressure}
In the immediate surroundings of an active nucleus, the gas and dust
are subject to the AGN radiation pressure. In particular, the AGN
IR radiation pressure may play a significant role in both maintaining the
vertical extent of the torus and launching dusty outflows
\citep{PierKrolik1992_IRradpressure, Krolik2007, Fabian2008,
  Venanzi2020, Tazaki2020}. We investigated  the observed mid-IR
morphologies of our sample  in the context of the \cite{Venanzi2020}
semi-analytical disk+wind models, which include the AGN radiation pressure,
gravity from the central black hole, and the IR radiation pressure. 
They showed  that dusty outflows are launched more efficiently when
the IR radiation pressure 
is the dominant component. This takes place when the AGN radiation
pressure  balances gravity from the central
black hole.

We found that those Seyferts with 
polar mid-IR emission show intermediate nuclear hydrogen column densities ($\log
N_{\rm H} \simeq 22.5-23\,{\rm cm}^{-2}$) and moderate
Eddington ratios ($\log \lambda_{\rm Edd} \simeq -1\, {\rm to}
-1.75$). As can be seen from Fig.~\ref{fig:dustywinds}, the observed nuclear column densities
and Eddington ratios place them close to the region where the IR
radiation pressure is most effective. 
However, polar outflows are also possible at high nuclear column densities
for sufficiently high Eddington ratios, as is the case of NGC~1068 and
even Circinus.  
The dashed line plotted in  Fig.~\ref{fig:dustywinds} is thus not a hard boundary,
but it rather indicates the combination of Eddington ratios and
column densities where the IR radiation pressure
dominates for individual clouds. The well-studied case of Circinus
illustrates this. Indeed, \cite{Venanzi2020} 
used their simulations and  predicted a disk + wind distribution for 
 this galaxy, in good agreement with both observations
 \citep{Packham2005} and
the radiative transfer modeling done by \cite{Stalevski2017}.

Molecular outflows 
at the  torus  or its vicinity (tens of parsecs) are observed
in some nearby Seyferts
\citep[see][]{Gallimore2016, Izumi2018, 
  AlonsoHerrero2019, GarciaBurillo2019}, and are also  predicted by
hydrodynamic simulations of radiation-driven winds 
\citep{Wada2016, Williamson2020}.
In addition, the AGN in Fig.~\ref{fig:dustywinds}
in the region where outflows are launched also show strong evidence for the clearing of (cold)
molecular gas in the nuclear regions, possibly related to
these nuclear outflows \citep[see GB21 and also][]{AlonsoHerrero2019}.
We also note that in some Seyfert galaxies (for instance, NGC~1068) the radio
jet may also play a role in maintaining or even boosting 
radiation-driven outflows. This is particularly true, when the
molecular outflows occur at 100\,pc scales far from the scales where the IR radiation
of the disk/torus dominates \citep[that is, near the sublimation radius, see][]{Venanzi2020}.

Seyferts with both polar and equatorial mid-IR emission
have moderate-to-low Eddington ratios  ($\log \lambda_{\rm Edd} \simeq -1.3\, {\rm to}
-2.5$),  and high column
densities ($\log N_{\rm H} \gtrapprox 23\,{\rm cm}^{-2}$). As shown by \cite{Venanzi2020}, these
conditions might be more conducive  to equatorial dusty outflows
\citep[see][]{AlonsoHerrero2018} since large
column densities  might suppress the liftup of material.
The galaxies with little or no extended mid-IR
emission are the least luminous in terms of their $2-10\,$keV
luminosities, typically below $\sim 1.5 \times 10^{42}\,{\rm erg\,s}^{-1}$. In Fig.~\ref{fig:dustywinds}
they are in regions of this diagram where dusty outflows
are not expected, either because they are close to the {\it blowout}
region or their Eddington ratios are very low. In the scenario
proposed by GB21 these are galaxies where we would not expect to see
the imprint of AGN feedback from  nuclear outflows.

The predictions of the models used above regarding the behavior of outflows as
a function of location in  the $N_H$-$\lambda_{\rm Edd}$ space should only
be taken as broad characterisations, rather than as strict
laws. \cite{Venanzi2020} represented clouds as indivisible particles,
while \cite{Fabian2008} represented the entire gas distribution as a
thick spherically symmetric shell. However, any optically thick cloud
experiencing radiation pressure that is above the effective Eddington
limit for dusty gas will fragment through the radiative
Rayleigh-Taylor instability into {\it fingers} of low density outflow and
high density inflow \citep{Jacquet2011, Zhang2018}.
Finally, winds do not need to be launched as polar winds to obtain a
polar outflow.  The anisotropy of the
AGN emission, combined with anisotropic extinction from a torus,
causes more equatorial outflows feel less radiation pressure and are
more likely to fail, even if they are injected at the same speed
\citep{Williamson2020}. Thus, non-polar outflows are therefore likely to
evolve into polar outflows. Anisotropy and gas instabilities together
mean that outflows and inflows can exist (and even co-exist) across a
wide range of inclinations, column densities, and Eddington
factors. Further observations and simulations are needed to better
constrain the necessary parameters for a polar outflow. 

\subsection{Disk-wind models}

The CAT3D-WIND models capture qualitatively the
expected dust and gas configurations predicted by simulations of
dusty disks \citep[see Figs.~10-12 of][]{Venanzi2020}
and/or dust in the NLR walls illuminated by the AGN. 
Moreover, the wind-to-disk cloud ratio in the CAT3D-WIND models can
also be associated qualitatively with feedback effects seen observationally in our sample at high AGN
luminosities and/or Eddington ratios (see GB21). 
Out of the large parameter space
of the CAT3D-WIND models, in this work  we selected parameter ranges
based on results from fitting the {\it
  Spitzer}/IRS spectra \citep{GonzalezMartin2019} for several GATOS
Seyferts. Additionally,  we distributed the
dust in the models over characteristic physical sizes of $2000\,r_{\rm
  sub}$. These were
motivated by the ALMA resolved molecular dusty tori of our sample
(GB21) as well as the presence of extended mid-IR components.
However, we emphasize that the dust emission is from a disk+wind
configuration, and any additional dust is not treated in the picture,
but it may exist in reality.

We showed in Sect.~\ref{sec:comparisondatamodel} that to
make meaningful comparisons between the CAT3D-WIND model images and
observations, even if only qualitatively, it is necessary to convolve
and resample the models to the typical angular
resolutions and  pixel sizes.  The level of morphological detail in
the mid- and far-IR as well as the unresolved mid-IR emission
fractions depend on the assumed sublimation radius.  If we adopted 
a factor of
two smaller sublimation radius ($r_{\rm sub}=0.02\,$pc) for the fiducial models, the simulated images would cover
typically $40\,{\rm pc} \times 40\,{\rm pc}$  or
approximately $0.5\arcsec \times 0.5\arcsec$. In those mid-IR models
dominated by the point source, a  larger fraction of
the mid-IR emission would be unresolved.  Detailed morphological
comparisons on a case-by-case basis require
accurate estimates of the sublimation radius as well as
higher angular resolution observations, especially in the mid-IR.

In the fiducial CAT3D-WIND models produced for our sample,  we found
that  the mid-IR emission
comes mainly from the inner part of the disk/cone for
all values of the wind-to-disk cloud ratios
and all inclinations.  Extended  bi-conical and one-sided polar mid-IR 
emission becomes more apparent in the extended-wind configurations at
 $f_{\rm wd}\ge 0.6$. The latter morphology is present mostly at 
intermediate inclinations in the ``thick'' geometry  because of the disk and cone wall
obscuration. Torus models without a wind component also give rise to
similar one-sided mid-IR
morphologies due to self obscuration \citep[see
e.g.,][]{Siebenmorgen2015, LopezRodriguez2018, Nikutta2021}. 

To reproduce the relatively large
contributions from the unresolved component seen in the mid-IR at our
resolution requires  low to
moderate values of the wind-to-dust ratio ($f_{\rm wd} =0.3-0.6$) in
most GATOS Seyferts. We also note, that although clumpiness introduces some inhomogeneities,
the CAT3D-WIND models produce mostly symmetric (with respect to the
wind axis) wind/hollow cone morphologies. 
For several bright nearby AGN the polar extension observed in the
mid-IR  is preferentially tilted towards one side of the ionization
cone and counter-cone, similarly to observations of the NLR. Circinus is one of the
best examples of this \citep[see][]{Packham2005, Stalevski2017}. This could be just a sign
of general inhomogeneity of outflows \citep{Wada2016, Wada2018} or a slight misalignment of
accretion disk and dust disk from where the winds are driven, as
proposed by \cite{Stalevski2017}.

The observed ALMA
 $870\,\mu$m disk-like morphologies and extents of  the GATOS Seyferts are reproduced
 qualitatively with 
the simulated CAT3D-WIND model images, including those with a dominant
equatorial torus emission  as well as  the
small number of galaxies with far-IR  along the polar
direction. The characteristic ``X''-shape associated with
the dusty winds \citep[see e.g., ][]{Wada2016, Williamson2020} is seen at
intermediate to  edge-on inclinations in the extended disk and wind
model configurations. Furthermore, this morphology is appreciated better
for higher values of $f_{\rm wd}$ and thin cone walls, and in the
far-IR rather than the mid-IR. In a few Seyferts in our sample
(NGC~5643, NGC~7314, and NGC~7582),
there is some evidence of this ``X''-shape.

Observationally, the clearest example of an ``X''-shape morphology
is observed in the HCO+(4-3) dense molecular
gas emission of NGC~1068 \citep[see Figs.~11 and 12 of][]
{GarciaBurillo2019}. Moreover, at $432\,\mu$m the torus of NGC~1068 is
resolved in the equatorial direction but it  also shows a
polar elongation \citep{GarciaBurillo2016}, as in the
simulations. The dusty torus/disk may spread over larger
regions (diameter of $\sim 60\,$pc), as traced
by near-IR scattered light \citep{Gratadour2015}.
The sublimation radius of $r_{\rm sub}=0.1-0.2\,$pc and torus diameter
of $20-60\,$pc of NGC~1068 mean that 
in a CAT3D-WIND model the dust should be  
distributed over a region of $\sim 300-600\,r_{\rm sub}$ in diameter, but
mostly concentrated towards the smaller region probed by the ALMA
continuum. The polar component in the mid-IR is more extended and brighter
to the north of the AGN \citep{Tomono2001}. For the estimated $i=70^\circ$
\citep{GarciaBurillo2019}, a model with a compact disk - extended wind
configuration and a ``thin'' geometry reproduces all these
observations (Fig.~\ref{fig:simulated_fwd0.6_thin}). Higher angular
resolution ALMA continuum images are needed for NGC~1068 and other
nearby Seyferts to confirm the ``X''-shape morphologies.

\section{Summary}\label{sec:conclusions} 
In this work we analyzed high-angular ($\sim 0.3\arcsec$)
mid-IR imaging observations of 12 GATOS Seyferts  and compared them
with the ALMA observations from Paper I (GB21).
We assessed the observations in the context of the \cite{Venanzi2020}
semi-analytical models which include the AGN radiation pressure,
gravity from the central black hole, and the IR radiation pressure. Motivated by the observed mid- and far-IR morphologies in our sample, we generated new radiative
transfer CAT3D-WIND models \citep{Hoenig2017} and  $8\,\mu$m,
$12\,\mu$m, and
$700\,\mu$m model images (see Figs.~\ref{fig:CAT3D-WINDcompactdiskextendedwind_thin}-\ref{fig:CAT3D-WINDNGC7213} and Appendix). 
We made an informed choice of the model
parameters (see Table~\ref{tab:modelparameters}) using the
fits of the {\it Spitzer}/IRS spectra   from
\cite{GonzalezMartin2019} for the GATOS
Seyferts in their sample.
Thus, the models are tailored to the
properties of the GATOS Seyferts analyzed in this work, which are
representative of X-ray selected AGN with median luminosities $
L(2-10\,{\rm keV}) =2\times 10^{42}\,{\rm erg\,s}^{-1}$.
We distributed the dust
in a disk + wind geometry 
over a  $2000r_{\rm sub} \times 2000r_{\rm sub}$ region to account for
the relatively large torus sizes observed with ALMA in our sample. We
included compact and extended configurations for the disk and wind,
``thin'' and ``thick'' geometries for the disk and cone walls, and a
range of wind-to-disk cloud ratios. \\

Our main results are the following:

\begin{itemize}

\item
In seven out of the twelve galaxies we detected
extended mid-IR emission  with sizes between 50 and 160\,pc (Fig.~\ref{fig:midIRimages}).
The other five galaxies only show  faint extended
or mostly unresolved mid-IR emission. At our
current mid-IR angular resolutions, however, the unresolved emission
contributes between 60\%  and 100\% of the nuclear
1.5\arcsec$\sim$150\,pc emission in the majority of the sources.

\item
There is a diversity of nuclear extended mid-IR morphologies.   In six galaxies there is
  extended mid-IR emission approximately in the polar direction
  and  perpendicular to the  extended
 $870\,\mu$m component detected with ALMA. The latter
 continuum emission traces the equatorial
 component of the dusty molecular torus in the majority of the GATOS
 Seyferts (GB21). The only
exception is NGC~6300 where the extended mid-IR is along the
equatorial direction of the torus and dust in the
host galaxy. In some galaxies there are other extended mid-IR
components associated with emission in the host galaxy and/or star formation.

\item
The majority of galaxies in our sample with extended mid-IR  emission show
intermediate-to-high nuclear column densities ($\log N_H$(ALMA)$\sim$
22.5-23.9\,${\rm cm}^{-2}$) and moderate Eddington ratios ($\log \lambda_{\rm Edd}
\simeq -1.9$ to $-1.2$) and
have $L(2-10{\rm keV})> 1.5 \times 10^{42}\,{\rm erg\,s}^{-1}$.
According to the simulations of \cite{Venanzi2020}, 
these conditions are favorable to launching dusty
winds (Fig.~\ref{fig:dustywinds}), both in the polar and equatorial direction. 

\item

At our mid-IR 0.3\arcsec \, resolutions, the fiducial CAT3D-WIND
models show unresolved or only slightly resolved mid-IR emission in a significant number of model
configurations
(Figs.~\ref{fig:simulated_fwd0.3_thick}-\ref{fig:simulated_NGC7213}). Polar
mid-IR emission is observed at intermediate inclinations, extended
wind configurations, and intermediate-to-high wind-to-disk cloud ratios.
To reproduce the observed mid-IR unresolved/extended fractions
in our galaxies,   low-to-moderate wind-to-disk ratios are needed (up to $f_{\rm
  wd} \sim 0.6$).

\item

  At our ALMA 0.1\arcsec \,
  resolution, the fiducial CAT3D-WIND model images show resolved far-IR morphologies,
trace well the extent of the dust distribution,
and reproduce the observed morphologies. In the models, the
``X-''shape morphologies associated with 
a strong and extended wind component are seen at this resolution for
high inclinations and are more apparent in the far-IR.  
While there is some ALMA observational evidence for these in our
sample, higher angular resolution data are needed.

\end{itemize}

In conclusion, this work together with GB21 provide observational
support for the disk/torus+wind scenario.
Dusty nuclear winds with the accompanying polar dust emission and
possibly with on-going clearing of the nuclear regions might be common in
Seyferts with high Eddington ratios and/or AGN
luminosities \citep[see also][]{Ricci2017, GonzalezMartin2019}. 
At  moderate Eddington ratios if the nuclear column densities are
high, the IR radiation pressure
uplift  may be suppressed and only equatorial dusty outflows could
occur \citep{Venanzi2020}. GB21 also showed that (radio-quiet)
AGN with low Eddington ratios and/or luminosities might
be dominated by the disk/torus component, and have higher
covering factors \citep{Ezhikode2017, Ricci2017, GonzalezMartin2019}. These are
predicted to show
little or no polar dust emission, as confirmed by our observations.  

In the near future, the Mid-Infrared Instrument
\citep[MIRI,][]{Rieke2015, Wright2015} onboard 
the James Webb Space Telescope will 
provide  observations at angular resolutions
(FWHM$\sim 0.3\arcsec$ at $8\,\mu$m) similar to those of the
ground-based mid-IR imaging observations analyzed in this
work. MIRI high sensitivity and spectral coverage, and especially the
stability of the PSF will enable for the separation of  the emitting
components in the nuclear regions of AGN.  The next generation of near and mid-IR
instruments on 30-40\,m-class ground-based telescopes
\citep{Packham2018, Brandl2021, Davies2021, Thatte2021}  will allow observations of
nearby Seyfert galaxies at angular resolutions in the tens of
milli-arcsecond range that are comparable to the best resolutions currently
achieved with  ALMA. All these instruments will provide an
unprecedented IR view of the
dust emission and gas cycle in the immediate surroundings of nearby AGN.
Finally, the comparison between  models and observations will benefit from accurate estimates of the
sublimation radii, dust composition, and the inclusion of the
circum-nuclear environment.



\begin{acknowledgements}

We thank D. Asmus, M. Villar-Mart\'{\i}n, and M. Venanzi for interesting discussions.  AA-H and SG-B acknowledge support
through grant PGC2018-094671-B-I00 (MCIU/AEI/FEDER,UE). AA-H, AL, and MP-S  work
was done under project No. MDM-2017-0737 Unidad de Excelencia "Mar\'{\i}a
de Maeztu"- Centro de Astrobiolog\'{\i}a (INTA-CSIC). 
SG-B thanks support  from  the  research project PID2019-106027GA-C44
from the Spanish Ministerio de Ciencia e Innovaci\'on.
SFH acknowledges support by the EU Horizon 2020 framework programme via the ERC Starting Grant DUST-IN-THE-WIND (ERC-2015-StG-677117).
IGB acknowledges support from STFC through grant ST/S000488/1.
CRA acknowledges financial support from the Spanish Ministry of Science, Innovation
and Universities (MCIU) under grant with reference RYC-2014-15779, from
the European Union's Horizon
2020 research and innovation programme under Marie Sk\l odowska-Curie
grant agreement No 860744 (BiD4BESt), from the State
Research Agency (AEI-MCINN) of the Spanish MCIU under grants
"Feeding and feedback in active galaxies" with reference
PID2019-106027GB-C42 and "Quantifying the impact of quasar feedback on galaxy evolution (QSOFEED)" with reference EUR2020-112266. CRA also acknowledges support from the
Consejería de Econom\' ia, Conocimiento y Empleo del Gobierno de
Canarias and the European Regional Development Fund (ERDF) under grant with reference ProID2020010105 and from IAC project
P/301404, financed by the Ministry of Science and Innovation, through
the State Budget and by the Canary Islands Department of Economy,
Knowledge and Employment, through the Regional Budget of the Autonomous
Community.
OG-M acknowledges support from UNAM PAPIIT IN105720.
AJB has received funding from the European Research Council (ERC)
under the European Union's Horizon 2020 Advanced Grant 789056 ``First
Galaxies’’.
AA-H and AJB acknowledge support from a Royal Society International Exchange Grant.
BG-L acknowledges support from the Spanish Agencia Estatal de
Investigaci\'on del Ministerio de Ciencia e Innovaci\'on (AEI-MCINN)
under grant with reference PID2019-107010GB-I00.
TI is supported by Japan Society for the Promotion of Science (JSPS) KAKENHI grant No. JP20K14531.
AL acknowledges the support from Comunidad de Madrid through the
Atracción de Talento Investigador Grant 2017-T1/TIC-5213, and
PID2019-106280GB-I00 (MCIU/AEI/FEDER,UE).
MPS acknowledges support from the Comunidad de Madrid through the Atracción de Talento Investigador Grant 2018-T1/TIC-11035 and PID2019-105423GA-I00 (MCIU/AEI/FEDER,UE).
DR acknowledges support from the Oxford Fell Fund and STFC through grant ST/S000488/1.
DJR acknowledges support from the STFC (ST/T000244/1).
MS is supported by the Ministry of Education, Science and
Technological Development of the Republic of Serbia through the
contract no. 451-03-9/2021-14/200002 and the Science Fund of the
Republic of Serbia, PROMIS 6060916, BOWIE.

Based on observations made with the Gran Telescopio Canarias (GTC),
installed at the Spanish Observatorio del Roque de los Muchachos of
the Instituto de Astrof\'{\i}sica de Canarias, in the island of La
Palma. Based on observations collected at the European Organisation
for Astronomical Research in the Southern Hemisphere. Based on
observations obtained at the international Gemini Observatory, a
program of NSF’s NOIRLab, which is managed by the Association of
Universities for Research in Astronomy (AURA) under a cooperative
agreement with the National Science Foundation. on behalf of the
Gemini Observatory partnership: the National Science Foundation
(United States), National Research Council (Canada), Agencia Nacional
de Investigaci\'{o}n y Desarrollo (Chile), Ministerio de Ciencia,
Tecnolog\'{i}a e Innovaci\'{o}n (Argentina), Minist\'{e}rio da
Ci\^{e}ncia, Tecnologia, Inova\c{c}\~{o}es e Comunica\c{c}\~{o}es
(Brazil), and Korea Astronomy and Space Science Institute (Republic of
Korea). This paper makes use of the following ALMA data:
ADS/JAO.ALMA\#2017.1.00082.S and\#2018.1.00113.S. ALMA is a partnership
of ESO (representing its member states), NSF (USA) and NINS (Japan),
together with NRC (Canada) and NSC and ASIAA (Taiwan), in cooperation
with the Republic of Chile. The Joint ALMA Observatory is operated by ESO,
AUI/NRAO and NAOJ. The National Radio Astronomy Observatory is a facility
of the National Science Foundation operated under cooperative agreement
by Associated Universities, Inc. This research has made use of the NASA/IPAC Extragalactic Database (NED),
which is operated by the Jet Propulsion Laboratory, California Institute of Technology,
under contract with the National Aeronautics and Space Administration.
This research made use of 
Astropy,\footnote{http://www.astropy.org} a community-developed core
Python package for Astronomy \citep{astropy:2013, astropy:2018}.  
IRAF is distributed by the National Optical Astronomy Observatory, which is operated by the Association of Universities for Research in Astronomy (AURA) under a cooperative agreement with the National Science Foundation.
\end{acknowledgements}

\bibliographystyle{aa} 
   \bibliography{bibliography} 

 \appendix
\section{Additional CAT3D-WIND model images}

\begin{figure}[!ht]
    \includegraphics[width=9cm]{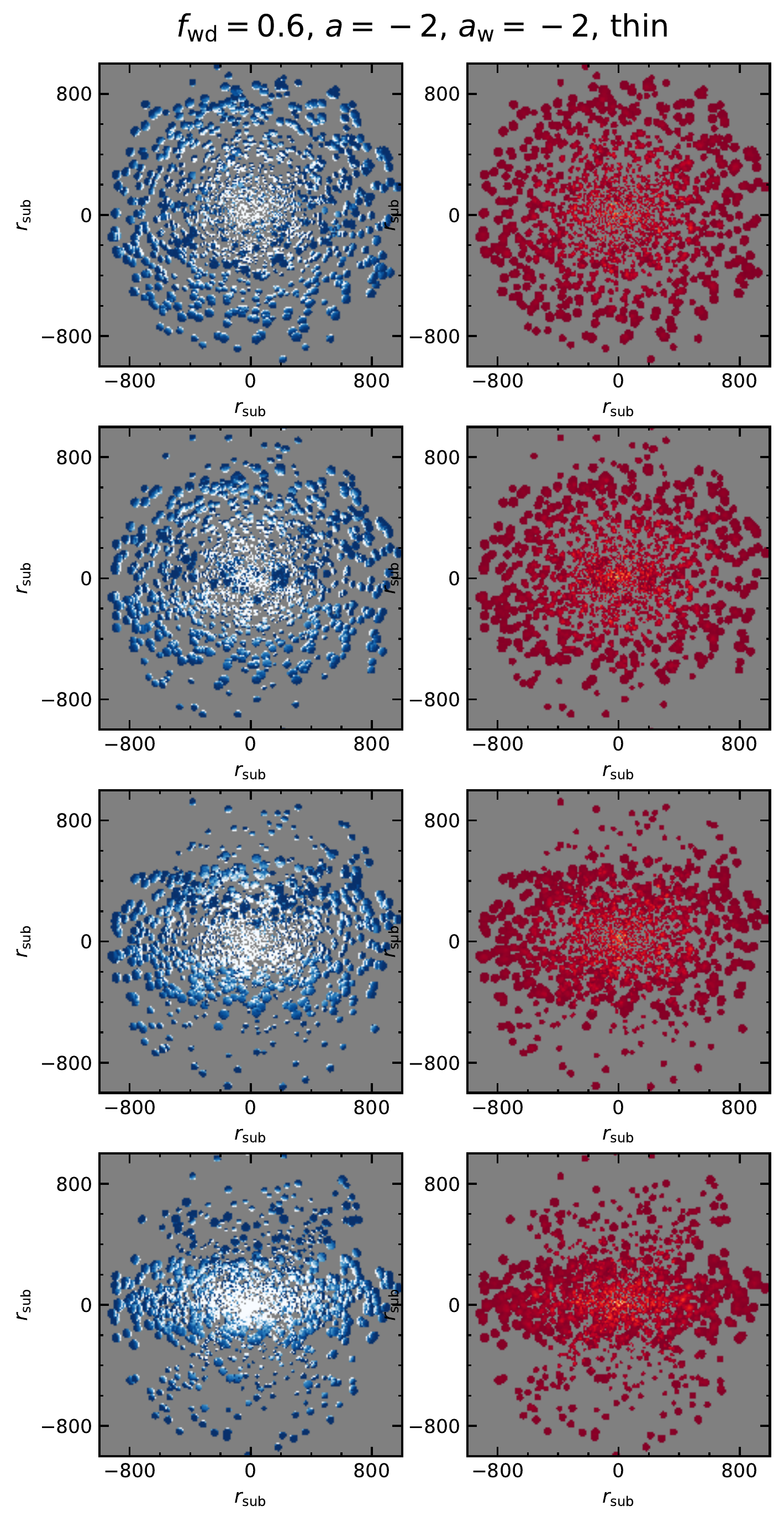}
  \caption{CAT3D-WIND   model images for a ``thin'' geometry and
    $f_{\rm wd}=0.6$ at
    $8\,\mu$m (in blue colors, left panels) and
    $700\,\mu$m (in orange colors, right panels) for a compact disk - wind
    ($a=-2$, $a_w=-2$). From top to bottom the inclinations
are   $30 ^\circ$ (nearly face-on disk), $45 ^\circ$, $60 ^\circ$, and $75^\circ$ (nearly edge-on disk). 
 }
              \label{fig:CAT3D-WINDcompactdiskcompactwind_thin}%
    \end{figure}

\begin{figure}[!ht]
    \includegraphics[width=9.2cm]{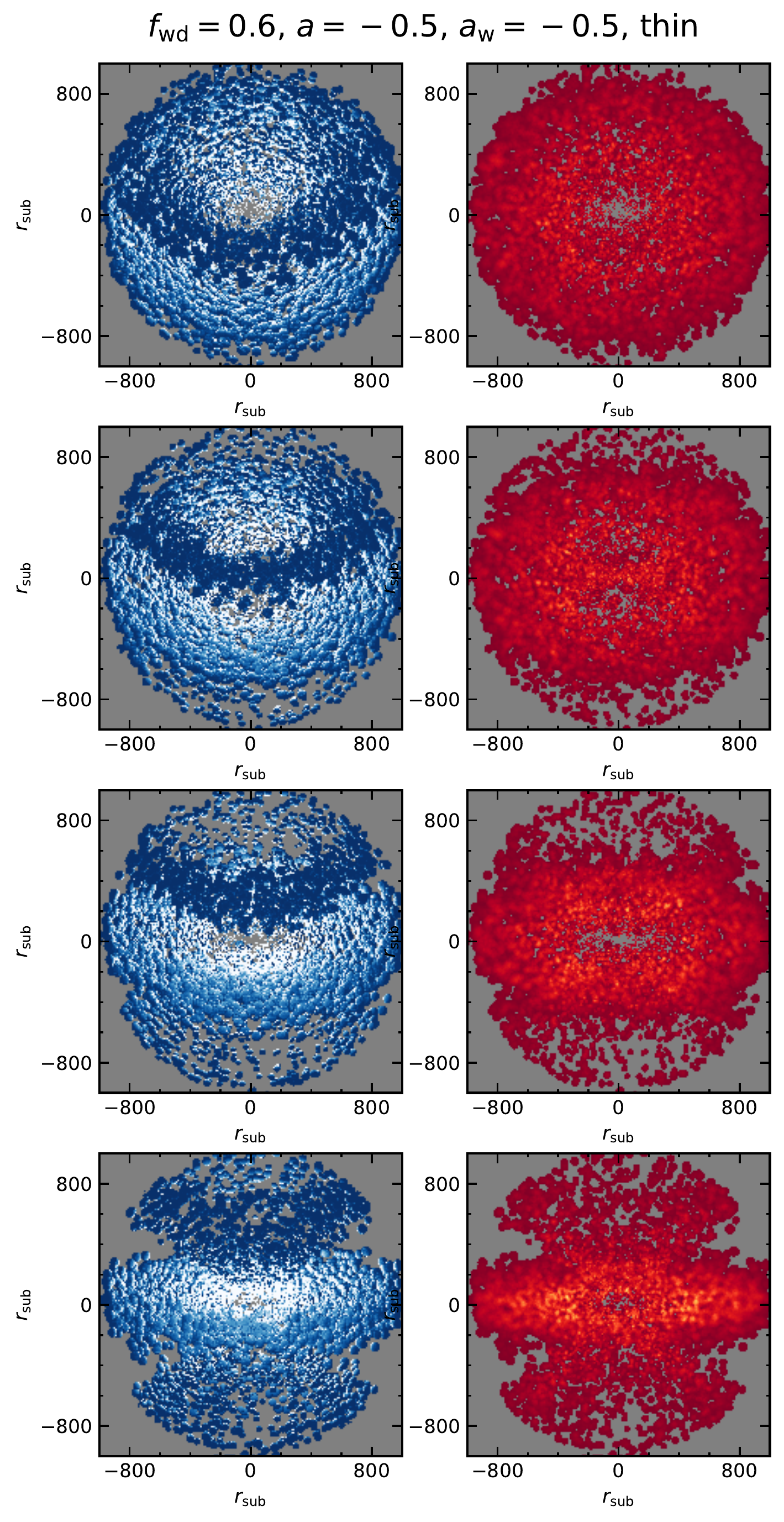}
  \caption{CAT3D-WIND model images for a ``thin'' geometry and $f_{\rm
      wd}=0.6$
    at $8\,\mu$m (in blue colors, left panels) and
    $700\,\mu$m (in orange colors, right panels) for an extended disk - wind
    ($a=-0.5$, $a_w=-0.5$). From top to bottom the inclinations
are   $30^\circ$ (nearly face-on disk), $45 ^\circ$, $60^\circ$, and $75^\circ$ (nearly edge-on disk). 
 }
              \label{fig:CAT3D-WINDextendeddiskextendedwind_thin}%
    \end{figure}

\begin{figure}
    \includegraphics[width=9cm]{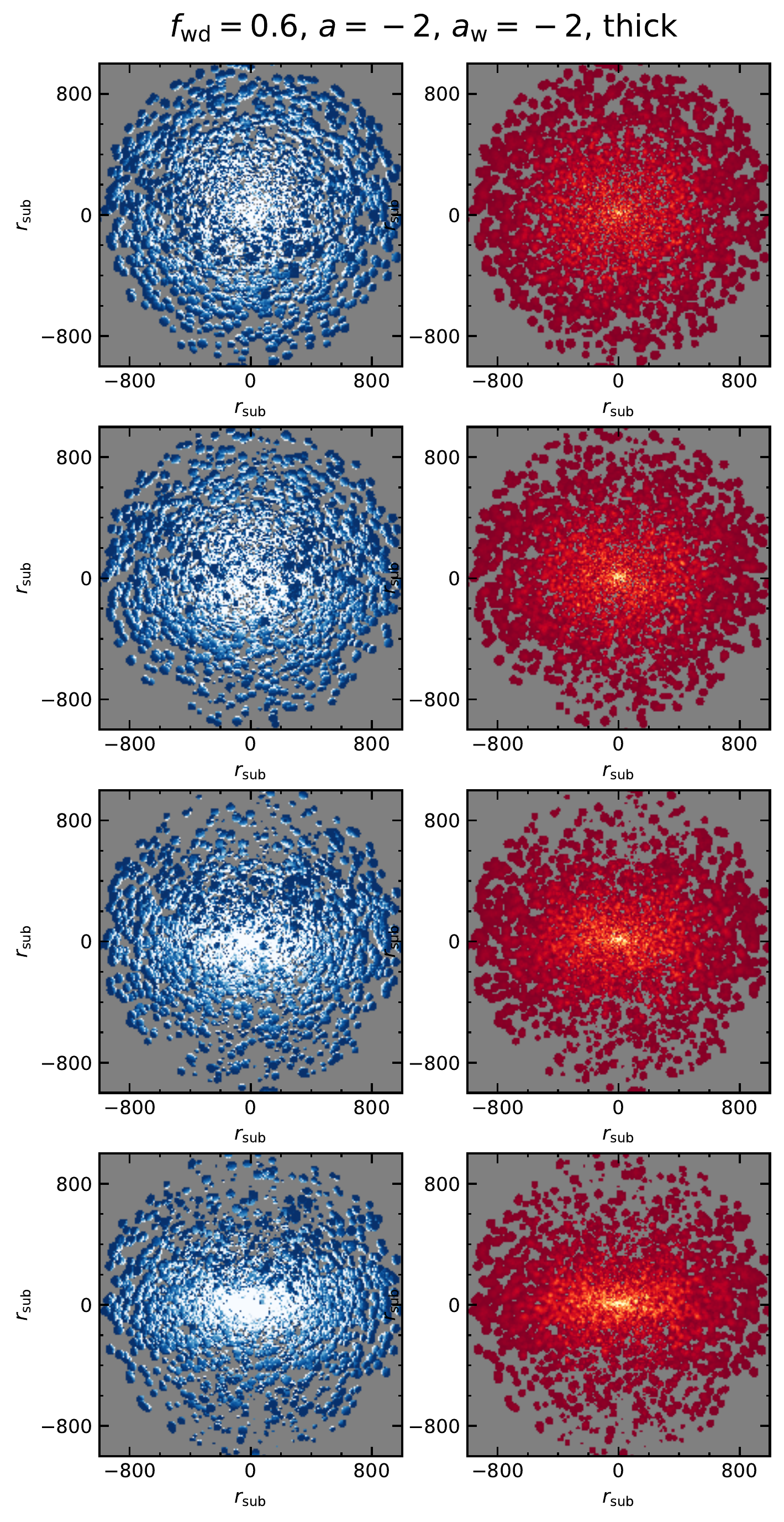}
  \caption{CAT3D-WIND model images  for a ``thick'' geometry and
    $f_{\rm wd}=0.6$ at $8\,\mu$m (in blue colors, left panels) and
    $700\,\mu$m (in orange colors, right panels) for a compact disk - wind
    ($a=-2$, $a_w=-2$). From top to bottom the inclinations
are   $30 ^\circ$ (nearly face-on disk), $45 ^\circ$, $60 ^\circ$, and $75^\circ$  (nearly edge-on disk). 
 }
              \label{fig:CAT3D-WINDcompactdiskcompactwind}%
    \end{figure}

\begin{figure}
    \includegraphics[width=9cm]{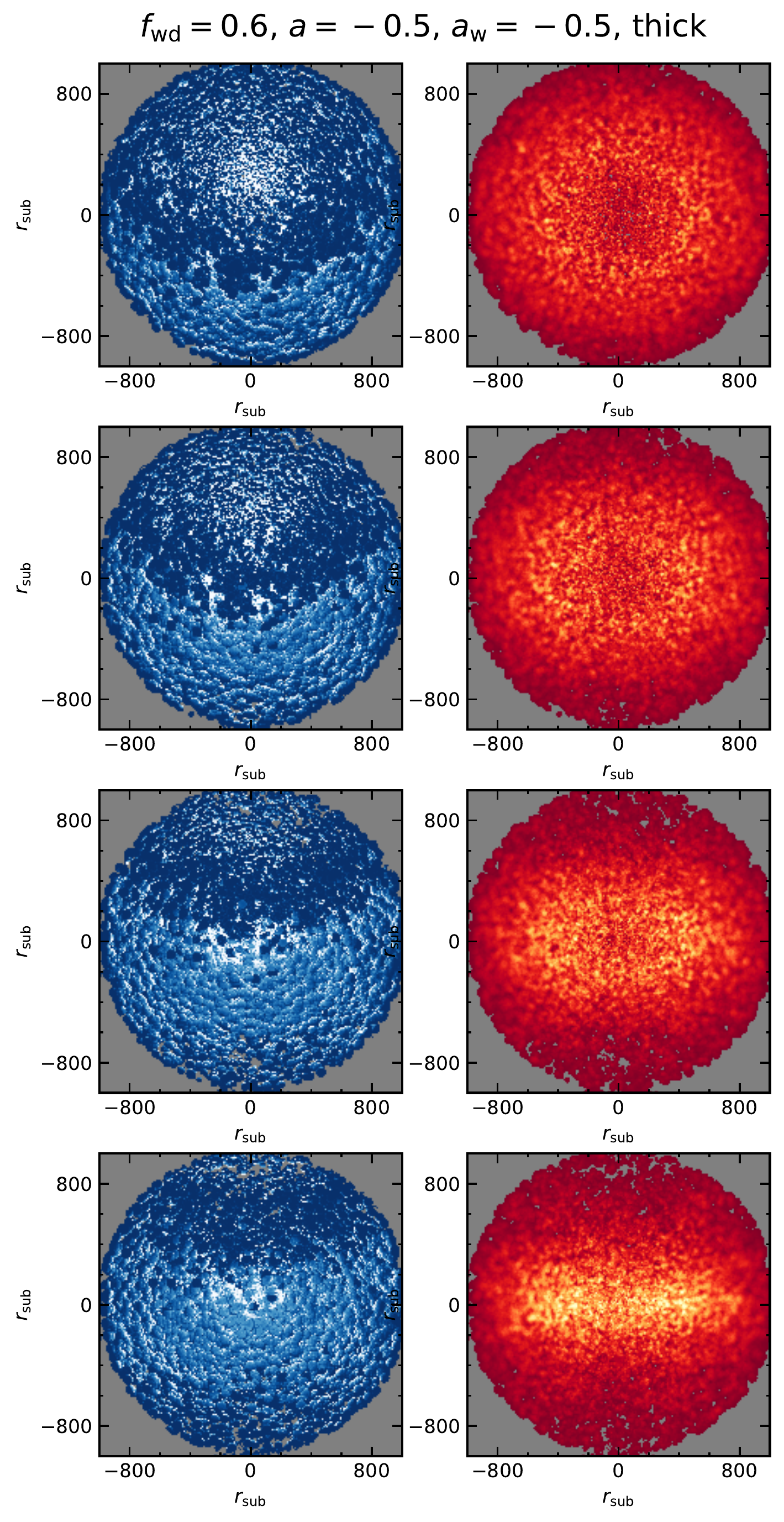}
  \caption{CAT3D-WIND  model images  for a ``thick'' geometry and
    $f_{\rm wd}=0.6$ at $8\,\mu$m (in blue colors, left panels) and
    $700\,\mu$m (in orange colors, right panels) for an extended disk - wind
    ($a=-0.5$, $a_w=-0.5$). From top to bottom the inclinations
are   $30 ^\circ$ (nearly face-on disk), $45 ^\circ$, $60 ^\circ$, and $75^\circ$ (nearly edge-on disk). 
 }
              \label{fig:CAT3D-WINDextendeddiskextendedwind}%
    \end{figure}

\begin{figure}
    \includegraphics[width=9cm]{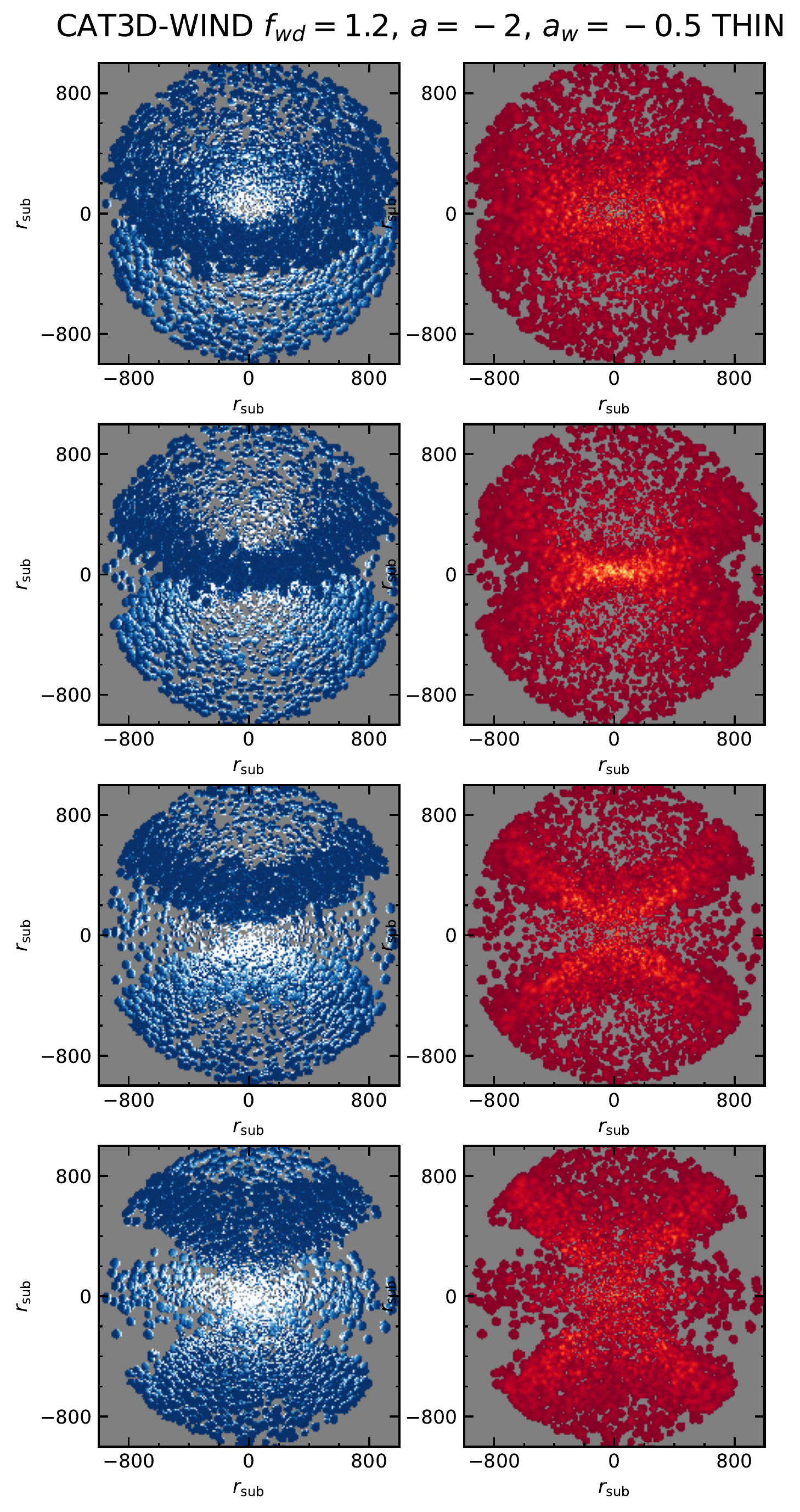}
  \caption{CAT3D-WIND model images  for a ``thin'' geometry and
    $f_{\rm wd}=1.2$ at $8\,\mu$m (in blue colors, left panels) and
    $700\,\mu$m (in orange colors, right panels) for a compact disk
    - extended wind configuration. From top to bottom the inclinations
are   $30 ^\circ$ (nearly face-on disk), $45 ^\circ$, $60 ^\circ$, and $75^\circ$  (nearly edge-on disk). 
 }
              \label{fig:CAT3D-WINDcompactdiskextendedwind_fwd1.2}%
    \end{figure}

\begin{figure}
    \includegraphics[width=9cm]{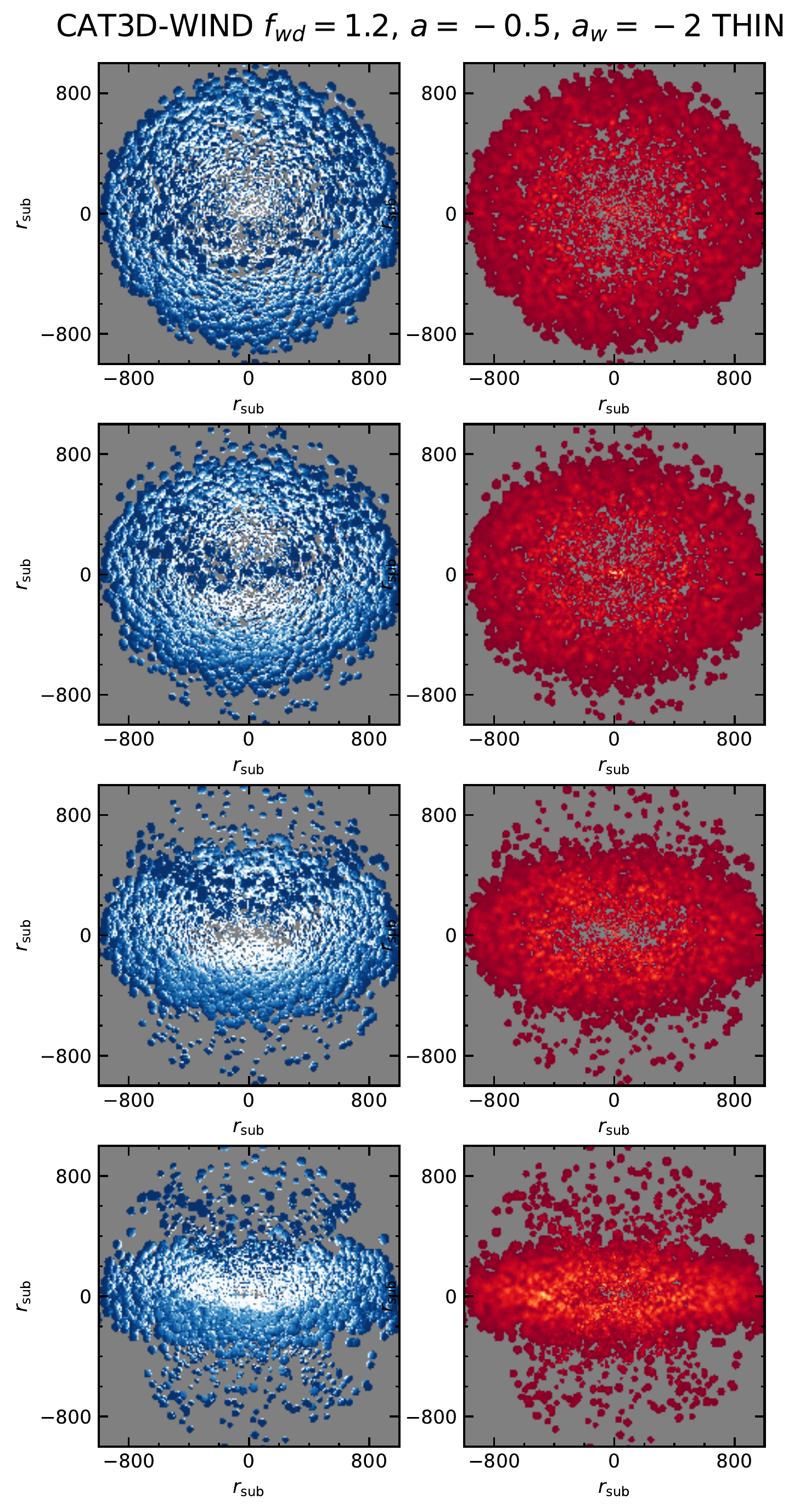}
  \caption{CAT3D-WIND model images  for a ``thin'' geometry and
    $f_{\rm wd}=1.2$ at $8\,\mu$m (in blue colors, left panels) and
    $700\,\mu$m (in orange colors, right panels) for an extended disk
    - compact wind configuration . From top to bottom the inclinations
are   $30 ^\circ$ (nearly face-on disk), $45 ^\circ$, $60 ^\circ$, and $75^\circ$  (nearly edge-on disk). 
 }
              \label{fig:CAT3D-WINDextendeddiskcompactwind_fwd1.2}%
    \end{figure}

\begin{figure}
    \includegraphics[width=9cm]{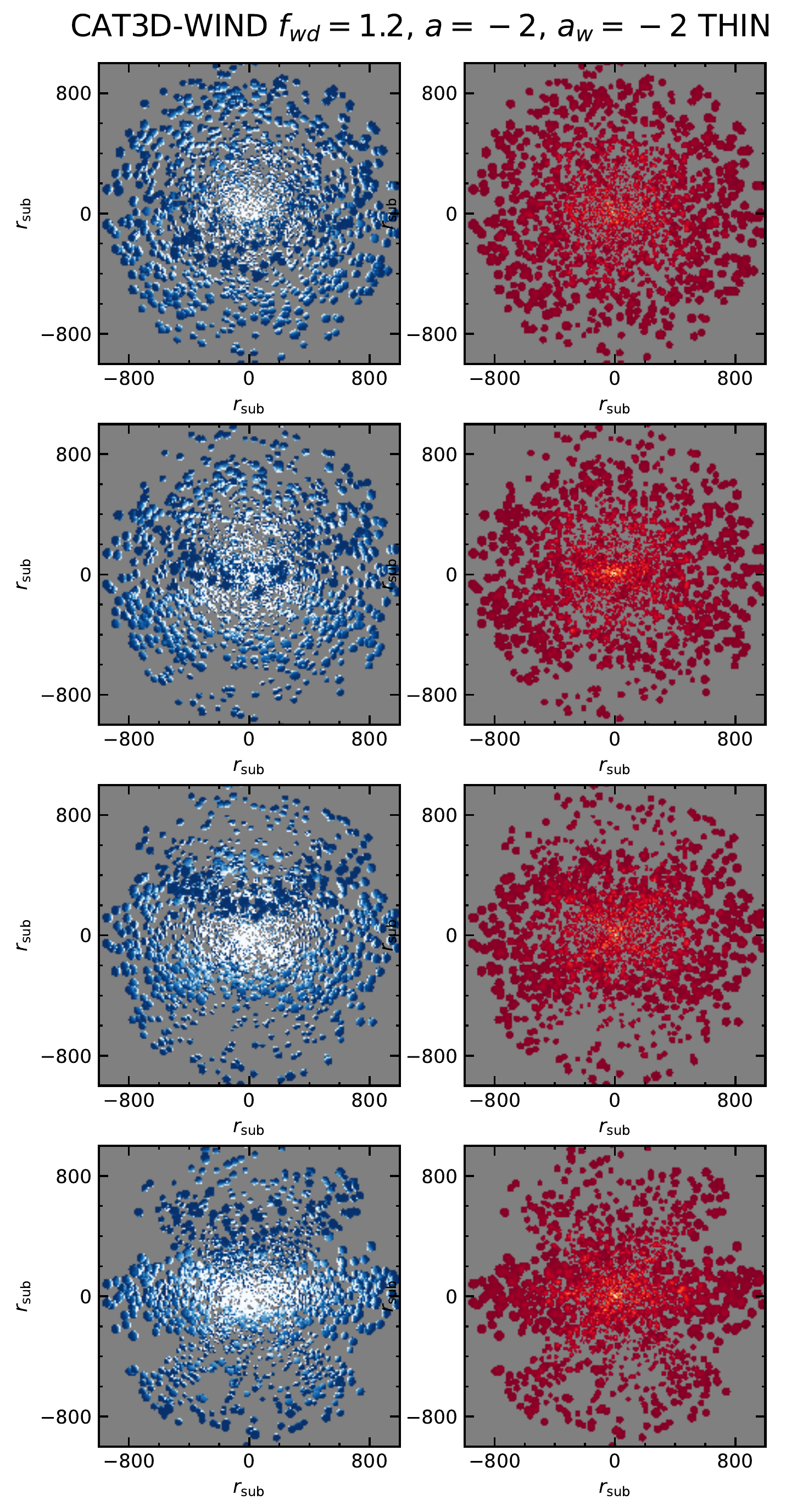}
  \caption{CAT3D-WIND model images  for a ``thin'' geometry and
    $f_{\rm wd}=1.2$ at $8\,\mu$m (in blue colors, left panels) and
    $700\,\mu$m (in orange colors, right panels) for a compact disk -
    compact wind configuration . From top to bottom the inclinations
are   $30 ^\circ$ (nearly face-on disk), $45 ^\circ$, $60 ^\circ$, and $75^\circ$  (nearly edge-on disk). 
 }
              \label{fig:CAT3D-WINDcompactdiskcompactwind_fwd1.2}%
    \end{figure}

\begin{figure}
    \includegraphics[width=9cm]{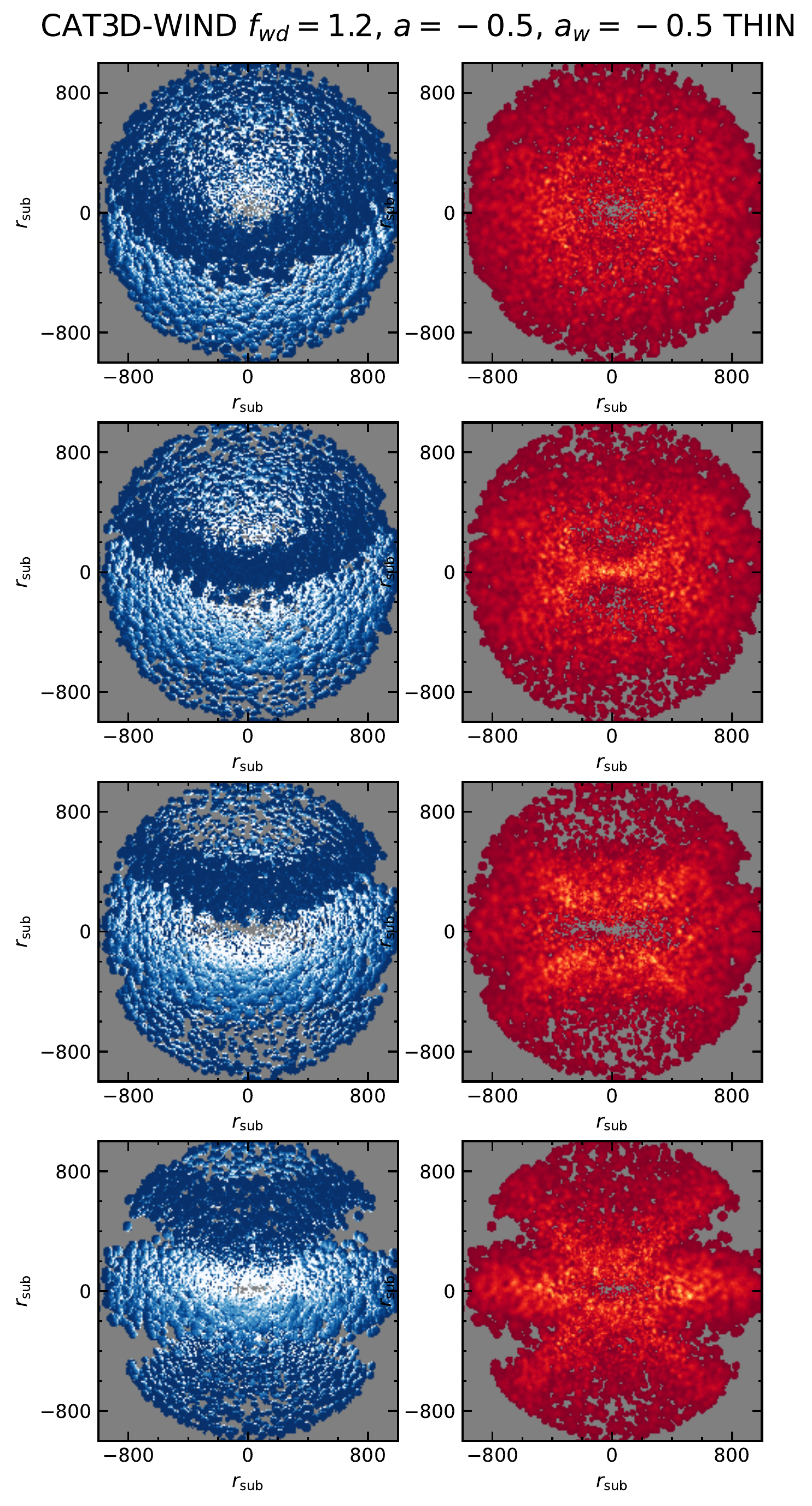}
  \caption{CAT3D-WIND model images  for a ``thin'' geometry and
    $f_{\rm wd}=1.2$ at $8\,\mu$m (in blue colors, left panels) and
    $700\,\mu$m (in orange colors, right panels) for an extended disk
    - extended wind configuration. From top to bottom the inclinations
are   $30 ^\circ$ (nearly face-on disk), $45 ^\circ$, $60 ^\circ$, and $75^\circ$  (nearly edge-on disk). 
 }
              \label{fig:CAT3D-WINDextendeddiskextendedwind_fwd1.2}%
    \end{figure}

   We show in this appendix additional CAT3D-WIND mid- and far-IR  model images for
   the compact disk - compact wind and extended disk - extended wind
   configurations for the ``thin'' and ``thick'' geometries and
   $f_{\rm wd}=0.6$ in
   Figs.~\ref{fig:CAT3D-WINDcompactdiskcompactwind_thin}-\ref{fig:CAT3D-WINDextendeddiskextendedwind}, respectively.
   We also show for the ``thin'' geometry and the
   four disk and wind configurations the mid- and far-IR model images
   for a wind-to-disk ratio of  $f_{\rm wd}=1.2$ in
   Figs.~\ref{fig:CAT3D-WINDcompactdiskextendedwind_fwd1.2}-\ref{fig:CAT3D-WINDextendeddiskextendedwind_fwd1.2}. All these images were discussed briefly in
   Sects.~\ref{sec:CAT3DWINDmodelimages_THIN} and \ref{sec:CAT3DWINDmodelimages_THICK}.

\end{document}